%% file: adv_poly_article.tex
\documentclass[fleqn]{svmult}

\usepackage{svaips}		
\usepackage{makeidx}		
\usepackage{graphicx}		
\usepackage{multicol}		

\makeindex             

\usepackage{times}

\usepackage{url}

\usepackage[numbers,sort&compress]{natbib}

\input macros.tex

\begin{document}
\title*{Lattice Boltzmann simulations of soft matter systems}
\titlerunning{Lattice Boltzmann simulations of soft matter systems}
\author{Burkhard D\"unweg\inst{1} and
Anthony J. C. Ladd\inst{2} }
\institute{
Max Planck Institute for Polymer Research \\
Ackermannweg 10, 55128 Mainz, Germany \\
\textit{duenweg@mpip-mainz.mpg.de}
\and
Chemical Engineering Department \\
University of Florida, Gainesville, FL 32611-6005, USA \\
\textit{tladd@che.ufl.edu}
}

\maketitle

\begin{abstract}

  This article concerns numerical simulations of the dynamics of
  particles immersed in a continuum solvent. As prototypical systems, we
  consider colloidal dispersions of spherical particles and
  solutions of uncharged polymers. After a brief explanation of the
  concept of hydrodynamic interactions, we give a general overview
  of the various simulation methods that have been developed to
  cope with the resulting computational problems. We then focus on the
  approach we have developed, which couples a system
  of particles to a lattice-Boltzmann model representing the
  solvent degrees of freedom. The standard D3Q19 lattice-Boltzmann
  model is derived and explained in depth, followed by a detailed
  discussion of complementary methods for the coupling of solvent and
  solute. Colloidal dispersions are best described in terms
  of extended particles with appropriate boundary conditions at the
  surfaces, while particles with internal degrees of freedom are easier to
  simulate as an arrangement of mass points with frictional coupling to
  the solvent. In both cases, particular care has been taken to
  simulate thermal fluctuations in a consistent way.
  The usefulness of this methodology is illustrated by studies from
  our own research, where the dynamics of colloidal and polymeric
  systems has been investigated in both equilibrium and nonequilibrium
  situations.

  \keywords{
    Soft matter,
    colloidal dispersions,
    polymer solutions,
    lattice Boltzmann,
    Chapman Enskog,
    boundary conditions,
    force coupling,
    Brownian motion,
    fluctuation-dissipation theorem,
    hydrodynamic interactions,
    hydrodynamic screening
  }

\end{abstract}
\dominitoc
%

\section{Introduction} \label{sec:intro}
\input intro.tex

\section{Particle-fluid systems} \label{sec:partfluid}
\input partfluid.tex

\section{The fluctuating lattice-Boltzmann equation} \label{sec:flbe}
\input fluct_lbe.tex

\section{Coupling the LB fluid to soft matter} \label{sec:coupling}
\input coupling.tex

\section{Applications with hydrodynamic interactions} \label{sec:apps}
\input apps.tex

\section*{Acknowledgements}

We thank E. Wajnryb --- IPPT (Institute of Fundamental Research),
Warsaw --- for access to the multipole code used in Sec.
\ref{sec:coupling}. A. J. C. Ladd thanks the Alexander von Humboldt
Foundation for supporting his stay at the Max Planck Institute for
Polymer Research by a Humboldt Research Award.

\bibliographystyle{huepfer}
\bibliography{bibs/Polymers,bibs/Suspensions,bibs/LG_LB,%
bibs/MD,bibs/Misc,bibs/Porous_Media,%
bibs/erice,bibs/stochlb}

\printindex
\end{document}

%% file: macros.tex
\newcommand{\fullwidth}   {9.5cm}
\newcommand{\halfwidth}   {5.0cm}
\newcommand{\smallwidth}  {4.0cm}
\newcommand{\boldbox}[1] {\mbox{\boldmath${#1}$}}
\newcommand{\mod}[1]     {{\left \vert {#1} \right \vert}}
\newcommand{\fracn}[2]   {{\textstyle\frac{#1}{#2}}}

\newcommand{\bc}         {{\vec c}}

\newcommand{\be}         {{\vec e}}
\newcommand{\bj}         {{\vec j}}

\newcommand{\bm}         {{\vec m}}
\newcommand{\bn}         {{\vec n}}
\newcommand{\bp}         {{\vec p}}

\newcommand{\br}         {{\vec r}}

\newcommand{\bu}         {{\vec u}}

\newcommand{\bx}         {{\vec x}}

\newcommand{\bF}         {{\vec F}}

\newcommand{\bR}         {{\vec R}}

\newcommand{\bU}         {{\vec U}}

\newcommand{\bnu}        {{\boldbox{\nu}}}

\newcommand{\bzeta}      {{\boldbox{\zeta}}}

\newcommand{\hh}         {{\fracn{1}{2}h}}

\newcommand{\Dd}         {{\partial}}

\newcommand{\cf}         {{\it c.f.}}
\newcommand{\ie}         {{\it i.e.}}
\newcommand{\eg}         {{\it e.g.}}
\newcommand{\etal}       {{\it et al.}}

\newcommand{\tensor}[1]  {\tens {#1}}

%% file: intro.tex
The term ``soft condensed matter'' generally refers to materials which
possess additional ``mesoscopic'' length scales between the atomic
and the macroscopic scales~\cite{chaikin_lubensky,%
  russel_saville_schowalter,lyklema,evans_wennerstrom,dGbook,%
  doi:86,grosberg,phil_nelson}. While simple fluids are 
characterized by the atomic size ($3 \times 10^{-10} m$),
soft--matter systems contain one or more additional length scales,
typically of order $10^{-9} m - 10^{-6} m$. There are many
examples of matter with mesoscale structure, including
suspensions, gels, foams, and emulsions; all of these are
characterized by viscoelastic behavior, which means a response that is 
fluid--like on long time scales but solid--like on shorter time
scales.  The two prototype systems considered in this article are colloidal
dispersions of hard particles, where the additional length scale is
provided by the particle size, and polymer systems, where the length
scale is the size of the macromolecule. The main difference between
these two systems is the presence of internal degrees of
freedom in the polymer, such that a statistical description is
necessary even on the single--molecule level. Many additional
soft--matter systems exist.  For example, dispersions may not only
contain spherical particles, but rather rod--like or disk--like
objects. For polymers, there are many possible molecular
architectures; in addition to simple linear chains, there are
rings, stars, combs, bottle--brush polymers and dendrimers.
Polymers may also self--assemble into two--dimensional membranes, either
free or tethered, which are of paramount biological
importance in cell membranes, vesicles, and red blood cells.

Strongly non-linear rheology is characteristic of soft matter.
In simple fluids, it is difficult to observe any
deviations from Newtonian behavior, which is well described by
the hydrodynamic equations of motion with linear transport
coefficients that depend only on the thermodynamic state. Indeed,
Molecular Dynamics simulations~\cite{Ald70} have revealed that a
hydrodynamic description is valid down to astonishingly
small scales, of the order of a few collisions of an individual molecule.
This means that one would have to probe the system with very
short wave lengths and very high frequencies, which are typically not
accessible to standard experiments (with the exception of neutron
scattering \cite{Mon89}), and even less in everyday life. However, in
soft--matter systems microstructural components (particles and polymers
for example) induce responses that depend very much on
frequency and length scale. These systems are often referred to as
``complex fluids''.

The nonlinear rheological properties of soft matter pose a
substantial challenge for theory~\cite{bird}.  Therefore, the study
of simple model systems is often the only way to make systematic progress.
Numerical simulations allow us to follow the dynamics of model systems
without invoking the uncontrolled approximations that are usually
required by purely analytical methods. Simulations can be used to isolate
and investigate the influence of microstructure, composition, external
perturbation and geometry in ways that cannot always be duplicated
in the laboratory.  In particular, simulations can provide
a well--defined test bed for theoretical ideas, allowing them to be
evaluated in a simpler and more rigorous environment than is possible
experimentally. Finally they can provide more detailed
and direct information on the particle dynamics and structure than is
typically possible with experimental measurements.

In this article we will focus on systems which comprise particles,
with or without internal degrees of freedom, suspended in a simple
fluid. We will first outline the necessary ingredients for a
theoretical description of the dynamics, and in particular explain the
concept of hydrodynamic interactions (HI). Starting from this
background, we will provide a brief overview of the various simulation
approaches that have been developed to treat such systems. All of
these methods are based upon a description of the solute in terms of
particles, while the solvent is taken into account by a simple
(but sufficient) model, making use of the fact that it can be
described as a Newtonian fluid. Such methods are often referred
to as ``mesoscopic''. We will then describe and derive in some
detail the algorithms that have been developed by us to couple a
particulate system to a lattice--Boltzmann fluid. The usefulness of
these methods will then be demonstrated by applications to colloidal
dispersions and polymer solutions. Some of
the material presented here is a summary of previously published work.

%% file: partfluid.tex
\subsection{Coarse--grained models}
\label{sec:partfluid:coarsegrain}

The first step towards understanding systems of particles suspended in
a solvent is the notion of \emph{scale separation}. Colloidal
particles are much larger (up to $10^{-6} m$) than solvent molecules,
and their relaxation time (up to $10 s$), given by the time the
particle needs to diffuse its own size, is several orders of magnitude
larger than the corresponding solvent time scale ($10^{-12} s$) as
well. A similar separation of scales holds for solute with internal
degrees of freedom, like polymer chains, membranes, or vesicles, as
far as their diffusive motion or their global conformational
reorganization is concerned. However, these systems contain a
hierarchy of length and time scales, which can be viewed as a spectrum
of internal modes with a given wavelength and relaxation time. For the
long--wavelength part of this spectrum, the same scale separation
holds, but when the wavelength is comparable with the solvent size
molecular interactions become important.  However, on scales of
interest, these details can usually be lumped into a few parameters.
The solute is then modeled as a system of ``beads'' interacting with
each other via an effective potential.  The beads should be viewed as
collections of atomic--scale constituents, either individual atoms or
functional groups, which have been combined into a single effective
unit in a process known as ``coarse--graining''.  On the scale of the
beads, the solvent may be viewed as a hydrodynamic continuum,
characterized by its shear viscosity and temperature.  The flow of
soft matter is usually isothermal, incompressible, and inertia--free
(zero Reynolds number).  Then, the most natural parameter to describe
solute--solvent coupling in polymeric systems is the Stokes friction
coefficient of the individual beads, or (in the case of anisotropic
subunits) the corresponding tensorial generalization. The effective
friction coefficient is the lumped result of a more detailed, or
``fine--grained'', description at the molecular scale, as is the
bead--bead potential, which should be viewed as a potential of mean
force. In order to make contact with experimental systems, these
parameters must be calculated from more microscopic theories or
simulations, or deduced from experimental data.

Further simplification may arise from the type of scientific question
being addressed by the coarse--grained model. If the interest is not
in specific material properties of a given chemical species, but
rather in generic behavior and mechanisms, as is the case for the
examples we will discuss in this article, then the details of the
parameterization are less important than a model that is both
conceptually simple and computationally efficient. The standard model
for particulate suspensions is a system of hard spheres, while for
polymer chains the Kremer--Grest model~\cite{grest:1986} has proved to
be a valuable and versatile tool. Here, the beads interact via a
purely repulsive Lennard--Jones (or WCA \cite{wca}) potential,
\begin{equation} \label{eq:ulennardjones}
   V_{LJ} (r) = \left\{
   \begin{array}{l  r}
     4 \varepsilon \left[ { \left( {\sigma \over r} \right) }^{12}  -
                          { \left( {\sigma \over r} \right) }^{6} +
                          { 1 \over 4} \right]  & 
                          \hspace{2cm} r \le 2^{1/6} \sigma , \\
     0                                          & 
                          \hspace{2cm} r \ge 2^{1/6} \sigma , \\
   \end{array}                               \right.
\end{equation}
while consecutive beads along the chain are connected via a
FENE (``finitely extensible nonlinear elastic'') potential,
\begin{equation} \label{eq:ufene}
V_{ch} (r) = - {k \over 2} R_0^2
\ln \left( 1 - { {r^2} \over {R_0^2} } \right), 
\end{equation}
with typical parameters $R_0 / \sigma = 1.5$, $k \sigma^2 / \varepsilon
= 30$.  Chain stiffness can be incorporated by an
additional bond--bending potential, while more complicated
architectures (like stars and tethered membranes) require
additional connectivity. Poor solvent quality is modeled
by adding an attractive part to the non--bonded interaction, while
Coulomb interactions (however without local solvent polarization)
can be included by using charged beads.

There are two further ingredients which any good soft--matter model
should take into account: On the one hand, \emph{thermal
fluctuations} are needed in order to drive Brownian motion and
internal reorganization of the conformational degrees of freedom.
There are however special situations where thermal noise can be
disregarded, and ideally the simulation method should be flexible
enough to be able to turn the noise both on and off. On the other
hand, \emph{hydrodynamic interactions}, which will be the subject of
the next subsection, need to be taken into account in most circumstances.

\subsection{Hydrodynamic interactions}
\label{sec:partfluid:hydrodynint}

The term ``hydrodynamic interactions'' describes the \emph{dynamic}
correlations between the particles, induced by diffusive momentum
transport through the solvent. The physical picture is the same,
whether the particle motion is Brownian (\ie\ driven by thermal noise)
or the result of an external force (\eg\ sedimentation or
electrophoresis). The motion of particle $i$ perturbs the surrounding
solvent, and generates a flow. This signal spreads out diffusively, at
a rate governed by the kinematic viscosity of the fluid $\eta_{kin} =
\eta / \rho$ ($\eta$ is the solvent shear viscosity and $\rho$ is its
mass density). On interesting (long) time scales, only the transverse
hydrodynamic modes \cite{Han86} remain, and the fluid may be
considered as incompressible. The viscous momentum field around a
particle diffuses much faster than the particle itself, so that the
Schmidt number
\begin{equation}
Sc = \frac{\eta_{kin}}{D} 
\end{equation}
is large. In a molecular fluid, $D$ is the diffusion coefficient of
the solvent molecules and $Sc \sim 10^2 - 10^3$, while for soft
matter, where $D$ is the diffusion coefficient of the polymer or
colloid, $Sc$ is much larger, up to $10^6$ for micron size
colloids. The condition $Sc \gg 1$ is an important restriction on the
dynamics, which good mesoscopic models should satisfy. Consider two
beads $i$ and $j$, separated by a distance $r_{ij}$. The momentum
generated by the motion of $i$ with respect to the surrounding fluid
reaches bead $j$ after a ``retardation time'' $\tau \sim r_{ij}^2 /
\eta_{kin}$. After this time the motion of $j$ becomes
\emph{correlated} with that of bead $i$.  However, during the
retardation time the beads have traveled a small distance $\sim
\sqrt{D \tau} \sim r_{ij} / \sqrt{Sc}$, which is negligible in
comparison with $r_{ij}$. Therefore, it is quite reasonable to
describe the Brownian motion of the beads neglecting retardation
effects, and consider their random displacements to be
\emph{instantaneously} correlated.

These general considerations suggest a Langevin description
(stochastic differential equation) for the the time
evolution of the bead positions $\vec r_i$:
\begin{equation} \label{eq:langevin_bead_positions}
\frac{d}{dt} r_{i \alpha} = \sum_j \mu_{i \alpha, j \beta}
F_{j \beta}^c + \Delta_{i \alpha} ,
\end{equation}
where $\alpha$, $\beta$ are Cartesian indexes and the Einstein summation
convention has been assumed. $\vec F_j^c$ is the conservative force
acting on the $j$th bead, while the mobility tensor $\tensor \mu_{ij}$
describes the velocity response of particle $i$ to $\vec F_j^c$;
both $\vec F_j^c$ and $\tensor \mu_{ij}$ depend on the
configuration of the $N$ particles, $\vec r^N$. The random
displacements (per unit time) $\Delta_{i \alpha}$ are Gaussian white noise
variables~\cite{risken,gardiner:1985} satisfying the
fluctuation--dissipation relation
\begin{eqnarray}
&& \left< \Delta_{i \alpha} \right> = 0, \\
&& \left< \Delta_{i \alpha} (t) \Delta_{j \beta} (t^\prime) \right> =
2 k_B T \mu_{i \alpha, j \beta} \delta ( t - t^\prime ) ;
\end{eqnarray}
here $T$ is the temperature and $k_B$ denotes the Boltzmann
constant. In general, the mobility tensor, $\mu_{i \alpha, j \beta}$
depends on particle position, in which case the stochastic integral
should be evaluated according to the Stratonovich calculus. This is
because during a finite time step a particle samples different
mobilities as its position changes. Numerically simpler is the Ito
calculus~\cite{risken,gardiner:1985}, which uses the mobility at the
beginning of the time step. In this case an additional term needs to
be added to Eq.~\ref{eq:langevin_bead_positions},
\begin{equation} \label{eq:langevin_bead_positions_ito}
\frac{d}{dt} r_{i \alpha} = \sum_j \mu_{i \alpha, j \beta}
F_{j \beta}^c + k_B T \sum_j 
\frac{\partial \mu_{i \alpha, j \beta}}{\partial r_{j \beta}}
+\Delta_{i \alpha}.
\end{equation}
In cases where the divergence of the mobility vanishes, $\partial \mu_{i \alpha, j \beta}/\partial r_{j \beta} = 0$, the Ito and Stratonovich interpretations coincide.

Processes generated by Eq.~\ref{eq:langevin_bead_positions} (Stratonovich) or
Eq.~\ref{eq:langevin_bead_positions_ito} (Ito)
sample trajectories from a probability distribution in the configuration space of the beads, $P \left( \vec r^N, t \right)$, which evolves according to a Fokker-Planck equation (Kirkwood diffusion equation):
\begin{equation}
\frac{\partial}{\partial t} P \left( \vec r^N, t \right) =
{\cal L} P \left( \vec r^N, t \right) ,
\end{equation}
with the Fokker--Planck operator
\begin{equation}\label{eq:FP1}
{\cal L} = 
\sum_{ij} \frac{\partial}{\partial r_{i \alpha}}
\mu_{i \alpha, j \beta} \left( k_B T
\frac{\partial}{\partial r_{j \beta}}
- F_{j \beta}^c \right) .
\end{equation}
Writing the forces as the derivative of a potential,
\begin{equation}
F_{i \alpha}^c = - \frac{\partial}{\partial r_{i \alpha}}
V \left( \vec r^N \right) ,
\end{equation}
it follows that the Boltzmann distribution, $P \propto \exp \left( - V
  / k_B T \right)$, is the stationary solution of the Kirkwood
diffusion equation; in other words the model satisfies the fluctuation-dissipation relation.

The mobility tensor can be derived from Stokes--flow hydrodynamics.
Consider a set of spherical particles, located at positions $\vec r_i$
with radius $a$, surrounded by a fluid with shear viscosity
$\eta$. Each of the particles has a velocity $\vec v_i$, which, as a
result of stick boundary conditions, is identical to the local fluid
velocity on the particle surface. The resulting fluid motions generate
hydrodynamic drag forces $\vec F_i^d$, which at steady state are
balanced by the conservative forces, $\vec F_i^d + \vec F_i^c = 0$.
The commonly used approximation scheme is a systematic multipole
expansion, similar to the analogous expansion in
electrostatics~\cite{Maz82,%
  cichocki_felderhof_multipole,Lad90b,sangani_mo,%
  Cic00}.  For details, we refer the reader to the original
literature~\cite{Maz82}, where the contributions from rotational
motion of the beads are also considered.  As a result of the linearity
of Stokes flow, the particle velocities and drag forces are linearly
related,
\begin{equation}
v_{i \alpha} = -\sum_{j}
\mu_{i \alpha, j \beta} F_{j \beta}^d =
\sum_{j}
\mu_{i \alpha, j \beta} F_{j \beta}^c .
\end{equation}
Since $\tensor \mu_{ij}$ describes the velocity response of particle
$i$ to the force acting on particle $j$, it must be identical to
the mobility tensor appearing in the Langevin equation.

In general the mobility matrix is a function of all the
particle coordinates, but to leading order, it is pairwise additive:
\begin{equation}\label{eq:HI-RP}
\tensor \mu_{i j} = \frac{\delta_{ij} \tensor 1} {6 \pi \eta a}
+ \frac{\left(1 - \delta_{ij}\right)} {8 \pi \eta r_{ij}}
\left( \tensor 1 + \frac{\vec r_{ij} \vec r_{ij}}{r_{ij}^2} \right)
+ \frac{\left(1 - \delta_{ij}\right)a^2}{12 \pi \eta r_{ij}^3}
\left( \tensor 1 - 3 \frac{\vec r_{ij} \vec r_{ij}}{r_{ij}^2} \right),
\end{equation}
where $\vec r_{ij} = \vec r_i - \vec r_j$, and $\tensor 1$ denotes the
unit tensor. The hydrodynamic interaction is long--ranged and
therefore has a strong influence on the collective dynamics of
suspensions and polymer solutions. This approximate form for the
mobility matrix follows from the assumption that the force density on
the sphere surface is constant. A point multipole expansion, by
contrast, generates the Oseen ($1/r_{ij}$) interaction at lowest
order~\cite{Hap86}, and can lead to non-positive-definite mobility
matrices~\cite{Rot69}. Thus the simplest practical form for the
hydrodynamic interaction is the Rotne-Prager tensor~\cite{Rot69} given
in Eq. \ref{eq:HI-RP}. Both the Oseen and Rotne-Prager mobilities are
divergence free, and therefore there is no distinction between Ito and
Stratonovich interpretations.  However, at higher orders in the
multipole expansion, the divergence is non-zero
\cite{wajnryb_divergence}.

\subsection{Computer simulation methods and models}
\label{sec:partfluid:simuls}

In this section we briefly summarize the Brownian dynamics algorithm and
its close cousin Stokesian Dynamics. We then outline the motivation and
development of several mesoscale methods, some of which are reviewed 
elsewhere in this series.

\subsubsection{Brownian Dynamics}
\label{sec:partfluid:simuls:browniandyn}

Brownian dynamics is conceptually the most straightforward
approach~\cite{Ott96,kroger_review}. Starting from the
Langevin equation for the particle coordinates,
Eq.~\ref{eq:langevin_bead_positions}, and discretizing the time into
finite length steps $h$, gives a first-order (Euler) update for the
particle positions,
\begin{equation} \label{eq:brownian_dynamics_euler_update}
r_{i \alpha} (t + h) = 
r_{i \alpha} (t) + \sum_j \mu_{i \alpha, j \beta} F_{j \beta}^c h 
+ \sqrt{2 k_B T h} \sum_j \sigma_{i \alpha, j \beta} q_{j \beta} .
\end{equation}
Here $q_{i \alpha}$ are random variables with
\begin{eqnarray}
&& \left< q_{i \alpha} \right> = 0 \\
&& \left< q_{i \alpha} q_{j \beta} \right> =
\delta_{ij} \delta_{\alpha \beta} ,
\end{eqnarray}
while the matrix $\sigma_{i \alpha, j \beta}$ satisfies the
relation
\begin{equation}
\sum_{k} \sigma_{i \alpha, k \gamma}
                \sigma_{j \beta, k \gamma} =
\mu_{i \alpha, j \beta} .
\end{equation}
Note that in Eq. \ref{eq:brownian_dynamics_euler_update} we have
assumed a divergence--free mobility tensor

Although the number of degrees of freedom has been minimized, this
approach is computationally intensive, and imposes severe limitations
on the size of the system that can be studied.  Since every particle
interacts with every other particle, the calculation of the mobility
matrix scales as $O(N^2)$, where $N$ is the number of Brownian
particles. In addition, the covariance matrix for the random
displacements requires a Cholesky decomposition of the mobility
matrix, which scales as $O(N^3)$~\cite{Erm78}.  The computational
costs of Brownian dynamics are so large that even today one cannot
treat more than a few hundred Brownian particles~\cite{boliu}.

``Stokesian Dynamics''~\cite{Bra88b} is an improved version of
Brownian dynamics, in which the mobility tensor takes into account
short--range (lubrication) contributions to the hydrodynamic forces.
It also improves the far-field interactions by including
contributions from torques and stresslets, although still higher
moments are needed for accurate results in concentrated
suspensions~\cite{Lad90b}. Stokesian Dynamics is even more
computationally intensive than Brownian dynamics; the determination
of the mobility tensor is already an $O(N^3)$ process.

However, there have been two important improvements in efficiency.
First, Fixman~\cite{Ott96,kroger_review,Fix86,jendrejack00} has
proposed an approximation to $\sigma_{i \alpha, j \beta}$
by a truncated expansion in Chebyshev polynomials, which has a more
favorable scaling than Cholesky decomposition. Second, the long-range
hydrodynamic interactions can be calculated by
Fast Fourier Transforms~\cite{Sie01,BanchioBrady03,%
  Saintillanetal05,Hernandez-Ortizetal07,menghigdon1,menghigdon2}, or
hierarchical multipole expansions~\cite{sangani_mo}. Accelerated
Brownian Dynamics and Stokesian Dynamics algorithms scale
close to linearly in the number of particles, and their full potential
is not yet explored.  However, it should be noted that all these methods are
based upon an efficient evaluation of the Green's function
for the Stokes flow, which depends on the global boundary
conditions.  For planar boundaries, solutions are
available~\cite{Bla71,Lir76,cichocki_jones_physica,Cic00}, but a
more general shape requires a numerical calculation of the Green's
function between a tabulated set of source
and receiver positions~\cite{jendrejack03}.

\subsubsection{Mesoscale methods}

In view of the computational difficulties associated with Brownian Dynamics,
several ``mesoscale'' methods have been developed recently. The central
idea is to \emph{keep} the solvent degrees of freedom, but to describe
them in a simplified fashion, such that only the most salient features
survive. As we have already seen, it is in principle sufficient to
describe the solvent as a Navier--Stokes continuum, \emph{or} by some
suitable model which behaves like a Navier--Stokes continuum on
sufficiently large length and time scales. At least
asymptotically, the solvent dynamics must be described by the
equations
\begin{eqnarray}\label{eq:NS}
&& \partial_t \rho + \partial_\alpha \left( \rho u_\alpha \right) = 0 ,
\nonumber \\
&& \partial_t \left( \rho u_\alpha \right) +
   \partial_\beta \left( \rho u_\alpha u_\beta \right) +
   \partial_\alpha p =
   \partial_\beta \sigma_{\alpha \beta} +
   \partial_\beta \sigma_{\alpha \beta}^f + f_\alpha ,
\end{eqnarray}
where $\rho$ is the mass density, $\rho \bu$ the momentum density, $p$
the thermodynamic pressure, $\vec f$ an external force density applied
to the fluid, $\tensor \sigma$ the viscous stress tensor, and $\tensor
\sigma^f$ the fluctuating (Langevin) stress~\cite{Lan59}, whose
statistical properties will be discussed in later chapters of this
article. The viscous stresses are characterized by the shear and bulk
viscosities, $\eta$ and $\eta_v$, which we will assume to be
constants, independent of thermodynamic state and flow conditions:
\begin{equation} \label{eq:NSStress}
\sigma_{\alpha \beta} = \eta 
\left( \partial_\alpha u_\beta + \partial_\beta u_\alpha
- \frac{2}{3} \partial_\gamma u_\gamma \delta_{\alpha \beta}
\right)
+ \eta_v \partial_\gamma u_\gamma \delta_{\alpha \beta} .
\end{equation}

The advantage of such approaches is their spatial locality, resulting
in favorable $O(N)$ scaling, combined with ease of implementation
and parallelization. The disadvantage is the introduction of
additional degrees of freedom, and of additional (short) time scales
which are not of direct interest. The coupling between solvent and
solute varies from method to method. However, in all cases one takes
the masses and the momenta of the solute particles explicitly into
account, and makes sure that the total momentum is conserved.

Lattice models (Navier--Stokes, lattice Boltzmann) simulate a
discretized field theory in which thermal fluctuations can be added,
but also avoided if desired. Particle methods (Molecular Dynamics,
Dissipative Particle Dynamics, Multi--Particle Collision Dynamics)
simulate a system of interacting mass points, and therefore thermal
fluctuations are always present. The particles may have size and
structure or they may be just point particles. In the former case, the
finite solvent size results in an additional potential of mean force
between the beads. The solvent structure extends over unphysically
large length scales, because the proper separation of scale between
solute and solvent is not computationally realizable. In dynamic
simulations of systems in thermal equilibrium~\cite{ahlrichs:01},
solvent structure requires that the system be equilibrated with the
solvent in place, whereas for a structureless solvent the solute
system can be equilibrated by itself, with substantial computational
savings~\cite{ahlrichs:01}. Finally, lattice models have a
(rigorously) known solvent viscosity, whereas for particle methods the
existing analytical expressions are only approximations (which however
usually work quite well).

These considerations suggest that lattice methods are somewhat more
flexible and versatile for soft--matter simulations.
On the other hand, the coupling between solvent and immersed particles
is less straightforward than for a pure particle system. The
coupling between solid particles and a lattice--based fluid model
will be discussed in detail in Sec. \ref{sec:coupling}.

\subsubsection{Molecular Dynamics (MD)}
\label{sec:partfluid:simuls:md}

Molecular Dynamics is the most fundamental approach to soft--matter
simulations. Here the solute particles are immersed in a bath of solvent
molecules and Newton's equations of motion are solved numerically. In this
case, it is impossible to make the solvent structureless -- a
structureless solvent would be an ideal gas of point particles, which
never reaches thermal equilibrium.
Furthermore, the model interaction potentials are stiff and
considerable simulation time is spent following the motion of the
solvent particles in their local ``cages''. These disadvantages are so
severe that nowadays MD is rarely applied to soft--matter systems
of the type we are discussing in this article.

\subsubsection{Dissipative Particle Dynamics (DPD)}
\label{sec:partfluid:simuls:dpd}

Dissipative Particle Dynamics, which has become quite popular in the
soft--matter
community~\cite{koelman:1992,koelman:1993,pep:1995,pep:1995:2,groot:1997,%
  pep:1998,ignacio,gibson,besold,besold2,mikko,shardlow,thoso}, was
developed to address the computational limitations of MD.
A very soft interparticle potential, representing coarse--grained
aggregates of molecules, enables a large time step to be used.
Furthermore, a momentum--conserving Galilean--invariant
thermostat is included, representing the degrees of freedom that have been
lost in the coarse--graining process. Practically, these two parts
are unrelated, such that it is legitimate to apply the DPD thermostat to
a standard MD system. The DPD thermostat is
consistent with macroscopic isothermal thermodynamics. Since this
already introduces interparticle collisions, it is
possible to run DPD using an ideal gas solvent and still achieve
thermal equilibrium..

The key innovation in DPD is to apply the thermostat to particle
\emph{pairs}. A frictional damping is applied to the \emph{relative}
velocities between each neighboring pair, and a corresponding
random force is added in a pairwise fashion also, such that
Newton's third law holds exactly. The implementation is as follows. We
define two functions $\zeta(r) \ge 0$, the relative friction
coefficient for particle pairs with interparticle distance $r$, and
$\sigma(r) \ge 0$, characterizing the strength of the stochastic force
applied to the same particle pair. The fluctuation--dissipation theorem
requires that
\begin{equation} \label{eq:FDTforDPD}
\sigma^2(r) = k_B T \zeta(r) .
\end{equation}
The functions have compact support, so that only near neighbors need be
taken into account.

The frictional force on particle $i$ is determined by projecting the
relative velocities onto the interparticle separation
($\hat r_{ij} = \vec r_{ij} / \left\vert \vec r_{ij} \right\vert$):
\begin{equation}
\vec F^{d}_i = - \sum_j \zeta(r_{ij})
\left[ \left( \vec v_i - \vec v_j \right) \cdot \hat r_{ij} \right]
\hat r_{ij} ,
\end{equation}
which conserves momentum exactly, $\sum_i \vec F^{d}_i = 0$.
Similarly, the stochastic forces are directed along the interparticle
separation, again so that momentum is conserved pair--by--pair,
\begin{equation}
\vec F^{f}_i =  \sum_j \sigma(r_{ij}) \, \eta_{ij} (t) \, \hat r_{ij} .
\end{equation}
The noise $\eta_{ij}$ satisfies the relations $ \eta_{ij} =
\eta_{ji}$, $\left< \eta_{ij} \right> = 0$, and
\begin{equation}
\left< \eta_{ij} (t) \eta_{kl} (t^\prime) \right> =
2 ( \delta_{ik} \delta_{jl} + \delta_{il} \delta_{jk} )
\delta(t - t^\prime),
\end{equation}
such that different pairs are statistically independent and
$\sum_i \vec F^{f}_i = 0$. The equations of
motion for a particle of mass $m_i$ and momentum ${\vec p}_i$ are:
\begin{eqnarray}
\frac{d}{dt} \vec r_i & = & \frac{1}{m_i} \vec p_i , \\
\frac{d}{dt} \vec p_i & = & \vec F_i^c + \vec F^{d}_i + \vec F^{f}_i .
\end{eqnarray}
Exploiting the relation between this stochastic differential equation
and its Fokker--Planck equation, it can be shown that
the fluctuation--dissipation theorem holds~\cite{pep:1995}, and that the
method therefore simulates a canonical ensemble. DPD can be extended
to thermalize the perpendicular component of the interparticle
velocity as well, thereby allowing more control over the transport
properties of the model~\cite{pep:1998,junghans}.

\subsubsection{Multi--Particle Collision Dynamics (MPCD)}
\label{sec:partfluid:simuls:srd}

This method
\cite{malevanets_kapral,ihle_mpcd1,ihle_mpcd2,kikuchi,ihle_mpcd3}
works with a system of ideal--gas particles and therefore has no
artificial depletion forces. Free streaming of the particles,
\begin{equation}
\vec r_i (t + h) = \vec r_i (t) + h \vec v_i (t) ,
\end{equation}
alternates with momentum and energy conserving collisions, which
are implemented via a Monte Carlo procedure:
\begin{itemize}
\item Sub--divide the simulation volume into a regular array of cells.
\item For each cell, determine the set of particles residing in
      it. For one particular cell, let these particles be
      numbered $i = 1, \ldots, n$. Then in
      each box:
\item Determine the local center--of--mass velocity:
      \begin{equation}
        \vec v_{CM} = \frac{1}{n} \sum_{i = 1}^{n} \vec v_i .
      \end{equation}
\item For each particle in the cell, perform a Galilean
      transformation into the local center--of--mass system:
      \begin{equation}
        \tilde{\vec v}_i = \vec v_i - \vec v_{CM} .
      \end{equation}
\item Within the local center--of--mass system, rotate
      all velocities within the cell by a random rotation
      matrix $\tensor R$:
      \begin{equation}
        \tilde{\vec v}_i^\prime = \tensor R \tilde{\vec v_i} .
      \end{equation}
\item Transform back into the laboratory system:
      \begin{equation}
        \vec v_i^\prime = \tilde{\vec v}_i^\prime + \vec v_{CM} .
      \end{equation}
\end{itemize}
By suitable random shifts of the cells relative to the fluid, it is
possible to recover strict Galilean invariance~\cite{ihle_mpcd1,ihle_mpcd2}. MPCD results in hydrodynamic behavior on large length and time
scales, and is probably the simplest and most efficient particle
method to achieve this.

\subsubsection{Lattice Boltzmann (LB)}
\label{sec:partfluid:simuls:lb}

Here one solves the Boltzmann equation, known from the kinetic theory
of gases, in a fully discretized fashion. Space is discretized into
a regular array of lattice sites, time is discretized,
and velocities are chosen such that one time step will connect only
nearby lattice sites. Free streaming along the lattice links
alternates with local on--site collisions. Care must be taken to
restore isotropy and Galilean invariance in the hydrodynamic limit, and
asymptotic analysis is an indispensable tool in this process. Further
details will be provided in the following chapters.

\subsubsection{Navier-Stokes}
\label{sec:partfluid:simuls:navstokes}

It is possible to start from a discrete representation of Eqs.~\ref{eq:NS}
but this has not been particularly popular in soft--matter simulations,
due to the difficulty of including thermal fluctuations
(but see Ref.~\cite{Sha04}). Finite-difference methods
share many technical similarities with LB and are roughly comparable
in terms of computational resources.
However, to our knowledge, no detailed benchmark comparisons are
available as yet. In order to be competitive with LB, we believe that the
solver must (i) make sure that mass and momentum are conserved within
machine accuracy, as is the case for LB, and (ii) \emph{not} work
in the incompressible limit, in order to avoid the costly
non--local constraints imposed by the typical
Poisson solver for the pressure. 
The incompressible limit is an approximation, which
eliminates the short time scales associated with wave--like motion.
However, in soft matter the solute particles must be simulated on
short inertial time scales, which requires that the solvent is simulated
on rather short time
scales as well. For this reason, we believe that enforcing an
incompressibility constraint does not pose a real advantage, and it is
instead preferable to allow for finite compressibility, such
that one obtains an explicit and local algorithm. This idea is
analogous to the Car--Parrinello
method~\cite{carparrinello}, where the Born--Oppenheimer
constraint is also discarded, in favor of 
an approximate but adequate separation of time scales. For
simulations of soft--matter systems coupled to a Navier--Stokes
background, see Refs.~\cite{hofler_schwarzer,schwarzer,%
kalthoff_schwarzer,wachmann_granular,delgado1,delgado2,delgado3,%
defabritiis,giupponi,delgado4}.

%% file: fluct_lbe.tex
The motivation for the development of lattice gas cellular
automata (LGCA) \cite{Fri86,Fri87} was to apply a highly simplified
Molecular Dynamics (MD) to simulations of hydrodynamic flows.
In LGCA, particles move along the links of a regular lattice, typically
cubic or triangular. Each lattice direction is encoded with a label $i$
and a vector $h \bc_i$ connects neighboring pairs of sites. During each
time step $h$, all particles with a direction $i$ are displaced $h \bc_i$
to an adjacent lattice site; thus $\bc_i$ is the (constant) velocity
of particles of type $i$.  Interparticle interactions are reduced to
collisions between particles on the
same lattice site, such that the conservation laws for mass and
momentum are satisfied; in single speed LGCA models \cite{Fri86,Fri87},
mass conservation implies energy conservation as well.
The lattice-Boltzmann (LB) method \cite{sauro,McN88,Hig89,Ben92} was
developed to reduce the thermal noise in LGCA, which requires extensive
averaging to obtain statistically significant results.

The LB model preserves the structural simplicity of LGCA, but
substitutes an ensemble-averaged collision operator for the
detailed microscopic dynamics of the LGCA. The hydrodynamic flow fields
develop without thermal noise, but the underlying connection with
statistical mechanics is lost (Sec. \ref{sec:fluct_lbe:general}).
The LB model turns out to be more flexible than LGCA, and there is
now a rich literature that includes
thermal \cite{alexander:93,ihle_thermal:00,%
lallemand_thermal:03, guo:07,ansumali:05,prasianakis:07} and 
multiphase flows, involving both liquid--gas coexistence and
multicomponent mixtures \cite{Gun91,shanchen:93,julia,%
julia_noise:99,lishinonideal,lishimixture,guo_zhao:05,arcidiacono:06,%
halliday:07,li_wagner:07}. 
In the present article, we will consider only single--phase flows of a
single solvent species, such that we can describe the dynamics in
terms of a single particle type. The algorithm can be summarized by
the equation
\begin{equation} \label{eq:GLG}
\nu_i (\vec r + \vec c_i h, t + h) =
\nu_i^\star (\vec r, t) = \nu_i (\vec r, t) +
{\Delta}_i \left( \bnu(\vec r, t) \right) ,
\end{equation}
where $\nu_i (\vec r, t)$ is
the number of particles that, at the discrete time $t$ just prior
to collision, reside at the lattice site $\vec r$, and have
velocity $\vec c_i$; $\nu_i^\star (\vec r, t)$ indicates the velocity
distribution immediately after collision. The difference
${\Delta}_i$ between the pre- and post-collision states is
called the ``collision operator'' and depends on the complete set of
populations at the site $\bnu(\vec r, t)$.
The left hand side of Eq. \ref{eq:GLG}
describes the advection of the populations along the links connecting
neighboring lattice sites. The 
velocity set $\vec c_i$ is chosen such that each new position
$\vec r + \vec c_i h$ is again at a lattice site; $\vec c_i = 0$
is possible.

For simplicity and computational efficiency,
the number of velocities should be small. Therefore the set of
velocities, $\vec c_i$, is typically limited to two or three neighbor
shells, chosen to be compatible with the symmetry of the lattice.
In two dimensions a single shell of 6 neighbors is sufficient for
hydrodynamic flows, but a single set of cubic lattice vectors
leads to anisotropic momentum diffusion, even at large spatial scales.
Thus LB models employ a judicious mixture of
neighboring shells, suitably weighted so that isotropy is
recovered. We use the classification scheme introduced by
Qian \etal\ \cite{qian}: for instance D2Q9 refers to an LB model on a
square lattice in two dimensions, using nine velocities
(zero, four nearest neighbors, four next--nearest neighbors), while
D3Q19 indicates a three--dimensional model on a simple cubic lattice
with nineteen velocities (zero, six nearest neighbors, twelve
next--nearest neighbors).

\subsection{Fluctuations}
\label{sec:fluct_lbe:general}

The difference between lattice gas and lattice
Boltzmann lies in the nature of the $\nu_i$. In a lattice gas
$\nu_i$ is a Boolean variable (\ie\ only the values zero and one are
allowed), while in the lattice-Boltzmann equation it is a positive
real--valued variable. In Sec. \ref{sec:fluct_lbe:statmech}
we will consider the case where $\nu_i$ is a large
positive integer, a conceptual model we call a ``Generalized Lattice Gas''
(GLG). Thinking of these models as a simplified Molecular Dynamics,
and considering fluctuations in $\nu_i$, it becomes clear what the
key difference between LGCA and the LB equation is. We define a
dimensionless ``Boltzmann number'', $Bo$, by the fluctuations
in $\nu_i$ at a single site,
\begin{equation}
Bo = \frac{ \left( \left< \nu_i^2 \right> -
\left< \nu_i \right>^2 \right)^{1/2}  }{\left< \nu_i \right>} ,
\end{equation}
where $\left< \ldots \right>$ denotes the ensemble average.
One could define Boltzmann numbers for other observables, but they
would all produce similar values. The important point is that $Bo$ tells
us how coarse--grained the model is, compared to microscopic MD:
$Bo \sim 1$ (the maximum value) corresponds to a fully microscopic model
where fluctuations are of the same order as the mean. This is exactly the
case for lattice gas cellular automata, which should therefore be
viewed as a simplified, but not coarse--grained MD. Conversely,
deterministic LB algorithms, at sufficiently small Reynolds numbers,
and with time--independent driving forces,
bring the system to a stationary state
with well--defined values for the $\nu_i$. In other words,
they are characterized by $Bo = 0$, which is the minimal value,
corresponding to entirely deterministic physics.

Originally, LGCA and LB algorithms were developed to simulate
macroscopic hydrodynamics. Here, a large Boltzmann number (order 1) is
undesirable, since the hydrodynamic behavior is only revealed after
extensive sampling. For many macroscopic applications a deterministic
LB simulation at $Bo = 0$ is hence entirely appropriate. In reality,
however, the Boltzmann number is finite, since the spatial domain in
the physical system corresponding to a single lattice site is also
finite. In soft--matter applications the spatial scales are so small
that these fluctuations do need to be taken into account, although in
many cases $Bo$ is fairly small. This suggests it would be
advantageous to introduce small thermal fluctuations into the LB
algorithm, in a controlled fashion, by means of a \emph{stochastic}
collision operator \cite{Lad94,Lad94a,Lad01}. The
fluctuation--dissipation relation can be satisfied by enforcing
consistency with fluctuating hydrodynamics \cite{Lan59} on large
length and time scales. An important refinement is to thermalize the
additional degrees of freedom that are not directly related to
hydrodynamics \cite{Adh05}, which leads to equipartition of
fluctuation energy on all length scales. A comprehensive understanding
of these approaches in terms of the statistical mechanics of LB
systems has been achieved only recently \cite{DSL}.

The number variables, $\nu_i$, can be connected to the hydrodynamic fields,
mass density $\rho(\vec r, t)$, momentum density $\vec j (\vec r, t)$, and
fluid velocity $\vec u(\vec r, t)$ ($\vec j = \rho \vec u)$, by
introducing the mass of an LB particle, $m_p$, and the mass density
parameter
\begin{equation} \label{eq:massden}
\mu = \frac{m_p}{b^3} ;
\end{equation}
here $b$ is the lattice spacing and a three--dimensional
lattice has been assumed. We then use the mass densities
of the individual populations,
\begin{equation}
n_i (\vec r, t) = \mu \nu_i (\vec r, t) ,
\end{equation}
to re--write the LB equation as
\begin{equation} \label{eq:LBE}
n_i (\vec r + \vec c_i h, t + h) =
n_i^\star (\vec r, t) =
n_i (\vec r, t) + \Delta_i \left( \bn(\vec r, t) \right) .
\end{equation}
The mass and momentum densities, $\rho$ and $\vec j$, are 
moments of the $n_i$'s with respect to the velocity vectors,
\begin{eqnarray}
&&\rho (\vec r, t)  =  \sum_i n_i (\vec r, t) , \\
&&\vec j (\vec r, t)  =  \sum_i n_i (\vec r, t) \vec c_i ,
\end{eqnarray}
and therefore, the collision operator
must satisfy the constraints of mass and momentum conservation,
\begin{equation}
\sum_i \Delta_i = \sum_i \Delta_i \vec c_i = 0 .
\end{equation}

The LB algorithm has both locality and conservation laws built in, but
two important symmetries have been lost. The system will in general
exhibit cubic anisotropy, due to the underlying lattice symmetries,
and violate Galilean invariance, due to the finite number of
velocities. Isotropy can be restored in the large--scale limit by a
careful choice of velocities and collision operator; however, the
broken Galilean invariance restricts the method to flows with $u \ll
c_i$. Since the speed of sound $c_s$, the maximum velocity with which
any signal can travel through the system, is of the order of the
$c_i$, the condition actually means low Mach number ($Ma$) flow,
\begin{equation}
Ma = u / c_s \ll 1.
\end{equation}
In soft--matter applications, variations in fluid density are small and
there is a universal equation of state characterized by the pressure at
the mean fluid density and temperature, $p_0 = p(\rho_0,T)$, and the speed
of sound $c_s = \left( \partial p /\partial \rho \right)^{1/2}$
\cite{Lan59},
\begin{equation} \label{eq:EOS}
p = p_0 + \left ( \rho - \rho_0 \right) c_s^2.
\end{equation}
Within an unimportant constant ($p_0 - \rho_0 c_s^2$), Eq. \ref{eq:EOS}
can be replaced by the relation
\begin{equation} \label{eq:EOS2}
p = \rho c_s^2 ,
\end{equation}
which fits well to the linear structure of
Eq. \ref{eq:LBE}. The value of $c_s$ is immaterial except that it
establishes a time-scale separation between sound propagation and
viscous diffusion of momentum. For this reason, a model where $c_s$ is
unphysically small may be used, so long as the dimensionless number
$C_{\eta} = \rho c_s l / \eta$ is sufficiently large; here $l$ is a
characteristic length in the system and $\eta$ is the shear viscosity
of the fluid. For polymers and colloids, $C_{\eta} \sim 10-1000$, but
values of $C_{\eta}$ in excess of 10 lead to quantitatively similar
results.

The simplest equation of state of the form of Eq. \ref{eq:EOS2} is an
ideal gas, 
\begin{equation}
p = \frac{\rho}{m_p} k_B T ,
\end{equation}
where $T$ is the absolute temperature and $k_B$ Boltzmann's
constant. Comparison with Eq. \ref{eq:EOS2} yields
\begin{equation}
k_B T = m_p c_s^2 .
\end{equation}
The temperature is then determined by choosing values for the
discretization parameters $b$ and $h$ ($c_s \sim b/h$), and the
LB particle mass $m_p$. The parameter $m_p$ controls the noise level in
stochastic LB simulations \cite{DSL}: the smaller $m_p$ (at fixed
$c_s$), the smaller the temperature (or the noise level). This makes
physical sense, since small $m_p$ means that a fixed amount of mass
$\rho b^3$ is distributed onto many particles, and therefore the
fluctuations are small.

In this section we will study the connection between the LB equation,
Eq. \ref{eq:LBE}, and the equations of fluctuating hydrodynamics
\cite{Lan59},
\begin{eqnarray}
\label{eq:navstokes1}
&&\partial_t \rho + \partial_\alpha j_\alpha = 0 , \\
\label{eq:navstokes2}
&&\partial_t j_\alpha + 
\partial_\beta \left( \rho c_s^2 \delta_{\alpha \beta}
+ \rho u_\alpha u_\beta \right)
= \partial_\beta \sigma_{\alpha \beta} + 
\partial_\beta \sigma^f_{\alpha \beta} .
\end{eqnarray}
The Greek indexes denote Cartesian components, $\delta_{\alpha
\beta}$ is the Kronecker delta, and the Einstein summation convention is
implied. The viscous stress has a Newtonian constitutive law,
\begin{equation}
\label{eq:macrostressviscous}
\sigma_{\alpha \beta} = \eta_{\alpha \beta \gamma \delta} 
\partial_\gamma u_\delta ,
\end{equation}
and for an isotropic fluid
\begin{equation}
\label{eq:macrostressviscous2}
\eta_{\alpha \beta \gamma \delta} = \eta
\left(
\delta_{\alpha \gamma} \delta_{\beta \delta} + 
\delta_{\alpha \delta} \delta_{\beta \gamma} -
\frac{2}{3} \delta_{\alpha \beta} \delta_{\gamma \delta} 
\right)
+ \eta_v \delta_{\alpha \beta} \delta_{\gamma \delta} ,
\end{equation}
with shear and bulk viscosities $\eta$ and $\eta_v$. 
The fluctuating stress tensor, $\sigma^f_{\alpha \beta}$,
is a Gaussian random variable characterized by zero mean,
$\left< \sigma^f_{\alpha \beta} \right> = 0$,
and a covariance matrix
\begin{equation} \label{eq:macrostressfluct}
\left< \sigma^f_{\alpha \beta}  \left(\vec r, t \right)
       \sigma^f_{\gamma \delta} \left(\vec r^\prime, t^\prime \right) 
\right> = 
2 k_B T \eta_{\alpha \beta \gamma \delta}
\delta \left( \vec r - \vec r^\prime \right)
\delta \left( t - t^\prime \right) .
\end{equation}
In the limit that $T \rightarrow 0$, $\sigma^f_{\alpha\beta}$ vanishes, and
the Navier-Stokes equations are recovered.

We begin our analysis with a general description of the dynamics
of the lattice-Boltzmann equation, based on a Chapman-Enskog
expansion (Sec. \ref{sec:fluct_lbe:chapman_enskog}). Then we
consider the equilibrium distribution for the D3Q19 model
(Sec. \ref{sec:fluct_lbe:equil_pops}), followed by deterministic
(Sec. \ref{sec:fluct_lbe:determ_coll}) and stochastic
(Sec. \ref{sec:fluct_lbe:noise}) collision operators. Finally we consider
the connection of the fluctuating LB model to statistical mechanics
(Sec. \ref{sec:fluct_lbe:statmech}) and the effects of external forces
(Sec. \ref{sec:fluct_lbe:forces}).

\subsection{Chapman--Enskog expansion}
\label{sec:fluct_lbe:chapman_enskog}

The Navier--Stokes description of a fluid is more
coarse--grained than the original LB equation, and to connect
the microscopic scales with the hydrodynamic scales we follow a
standard asymptotic analysis \cite{hinch}. We first
introduce a dimensionless scaling parameter $\varepsilon \ll 1$ and write
\begin{equation}
\vec r_1 = \varepsilon \vec r .
\end{equation}
The idea is to measure spatial positions with a ruler that has such a
coarse scale that details at the lattice level are not
resolved. The position $\vec r_1$ then corresponds to the number
read off from this coarse--grained ruler; for example instead of
talking about $1000$ nanometers, we talk about one micrometer. For
two points to be distant on the hydrodynamic scale, it is not sufficient
that $\left\vert \Delta \vec r \right\vert$ is large, but
rather that $\left\vert \Delta \vec r_1 \right\vert$ is large.
However, from the perspective of practical computation, the degree
of coarse graining is never as extensive as implied by our analysis;
the calculations would take far too long. Instead there is usually
only a few grid points separating the lattice scale from the smallest
hydrodynamic scale. Surprisingly the LB method can be quite accurate,
even in these circumstances \cite{Lad94a,Ngu02}.

In a similar way, we can also introduce a coarse--grained clock
for the time variable, and write
\begin{equation}
t_1 = \varepsilon t .
\end{equation}
The fact that we choose the same factor $\varepsilon$ for both
space and time is related to the typical scaling of wave--like
phenomena, where the time scale of a process is \emph{linearly}
proportional to the corresponding length scale. However, hydrodynamics 
also includes diffusion of momentum, where the time scale is
proportional to the \emph{square} of the length scale. These
processes occur on a much longer time scale, and to capture
the slow dynamics we introduce a second clock that is even more
coarse--grained,
\begin{equation}
t_2 = \varepsilon^2 t .
\end{equation}
We can therefore distinguish between ``short times'' on the hydrodynamic
scale, characterized by $t_s = t_1 / \varepsilon$,
and ``long times'', where $t_l = t_2 / \varepsilon^2$.
Both $t_s$ and $t_l$ are implicitly large on the lattice scale, with the
hydrodynamic limit being reached as $\varepsilon \to 0$.  But once again,
practical computation limits the separation between the time scales
$h$, $t_s$, and $t_l$ to one or two orders of magnitude each.

In the ``multi--time scale'' analysis,
the LB population densities may be considered to be functions of the
coarse--grained position and times, $\vec r_1$, $t_1$, and $t_2$;
$n_i \equiv n_i \left( \vec r_1, t_1, t_2 \right)$. When the
algorithm proceeds by one time step, $t \rightarrow t+h$, 
$t_1 \rightarrow t_1 + \varepsilon h$, and
$t_2 \rightarrow t_2 + \varepsilon^2 h$. The LB
equation in terms of the coarse-grained variables is then,
\begin{equation} \label{eq:scaledLBE}
n_i (\vec r_1 + \varepsilon \vec c_i h, t_1 + \varepsilon h, 
t_2 + \varepsilon^2 h) - n_i (\vec r_1, t_1, t_2)
= \Delta_i \left( \bn(\vec r_1, t_1, t_2) \right) .
\end{equation}
%

The population densities are slowly varying functions of
coarse-grained variables, and we may obtain hydrodynamic behavior by a
Taylor expansion of $n_i$ (Eq. \ref{eq:scaledLBE}) to second order in
powers of $\varepsilon$:
\begin{equation}
n_i \left( \bx + \delta \bx \right) =
n_i \left( \bx \right)
+ \sum_k \frac{\partial n_i}{\partial x_k} \delta x_k
+ \frac{1}{2} \sum_{kl} \frac{\partial^2 n_i}{\partial x_k \partial x_l}
\delta x_k \delta x_l + \ldots ,
\end{equation}
where we use $\bx$ to indicate the coarse-grained variables,
$\left[\vec r_1, t_1, t_2\right]$. Since the distribution function
itself depends on the degree of coarse--graining, we must take the
$\varepsilon$ dependence of the $n_i$ and $\Delta_i$ into account as
well:
\begin{eqnarray} \label{eq:niexpans}
&&n_i = n_i^{(0)} + \varepsilon n_i^{(1)} + O(\varepsilon^2) ,
\\ \label{eq:deliexpans}
&&\Delta_i = \Delta_i^{(0)} + \varepsilon \Delta_i^{(1)} + 
	\varepsilon^2 \Delta_i^{(2)} + O(\varepsilon^3) .
\end{eqnarray}
The conservation laws for mass and momentum must hold
independently of the value of $\varepsilon$, and thus
at every order $k$:
\begin{equation} \label{eq:delicons}
\sum_i \Delta_i^{(k)} = \sum_i \Delta_i^{(k)} \vec c_i = 0 .
\end{equation}
Inserting these expansions into
Eq. \ref{eq:scaledLBE}, and collecting terms at different orders
of $\varepsilon$, we obtain:
\begin{itemize}
\item at order $\varepsilon^0$,
\begin{equation} \label{eq:eps0LBE}
\Delta_i^{(0)} = 0 ;
\end{equation}
\item at order $\varepsilon^1$,
\begin{equation} \label{eq:eps1LBE}
\left( \partial_{t_1} + \bc_i \cdot \Dd_{\br_1} \right)
n_i^{(0)} = h^{-1} \Delta_i^{(1)} ;
\end{equation}
\item at order $\varepsilon^2$,
\begin{equation} \label{eq:eps2LBE}
\partial_{t_2} n_i^{(0)}
+ \frac{h}{2}
  \left( \partial_{t_1} + \bc_i \cdot \Dd_{\br_1} \right)^2 n_i^{(0)}
+ \left( \partial_{t_1} + \bc_i \cdot \Dd_{\br_1} \right) n_i^{(1)}
= h^{-1} \Delta_i^{(2)} .
\end{equation}
\end{itemize}
Subsequently, it will prove useful to eliminate the second occurrence of
$n_i^{(0)}$ from Eq. \ref{eq:eps2LBE}, by using  Eq. \ref{eq:eps1LBE}:
\begin{equation} \label{eq:eps2LBEb}
\partial_{t_2} n_i^{(0)}
+ \frac{1}{2}
  \left( \partial_{t_1} + \bc_i \cdot \Dd_{\br_1} \right) 
  \left( n_i^{\star(1)} +  n_i^{(1)} \right) = h^{-1} \Delta_i^{(2)} ,
\end{equation}
where $n_i^\star = n_i + \Delta_i$ is the post-collision population in
direction $i$.

The multi--time--scale expansion of Eq. \ref{eq:scaledLBE} is based on
the physical time-scale separation between collisions ($t \sim h$), sound
propagation ($t \sim h/\varepsilon$), and momentum diffusion
($t \sim h/\varepsilon^2$).
Eqs. \ref{eq:eps0LBE}--\ref{eq:eps2LBE} make the implicit assumption
that these three relaxations can be considered separately, which 
allows the collision operator at order $k+1$ to be calculated
from the distribution functions at order $k$. In essence, the collision
dynamics at order $k+1$ is slaved to the lower-order distributions.
The zeroth--order collision operator must be a function of $\vec n^{(0)}$
only, 
\begin{equation}  \label{eq:deli0}
\Delta_i^{(0)} = \Delta_i ( \bn^{(0)} ),
\end{equation}
which, in conjunction with Eq. \ref{eq:eps0LBE}, shows
that $\bn^{(0)}$ is a collisional invariant; thus we can associate
$\bn^{(0)}$ with the equilibrium distribution $\bn^{eq}$ \cite{cha60}.
In order to avoid spurious conserved quantities, the equilibrium
distribution should be a function of local values of the conserved
variables, $\rho$ and $\vec j$, only. In a homogeneous system,
with fixed mass and momentum densities,
$\bn^{eq}(\rho, \vec j) = \bn^{(0)}(\rho, \vec j)$ is stationary in time.
A stochastic collision operator (see Sec. \ref{sec:fluct_lbe:noise})
cannot satisfy Eq. \ref{eq:eps0LBE} and therefore must enter the
Chapman-Enskog expansion at order $\varepsilon$.
 
From Eq. \ref{eq:niexpans} we can derive
analogous $\varepsilon$ expansions for $\rho$ and $\vec j$,
\begin{eqnarray}
&&\rho  =  \rho^{(0)} + \varepsilon \rho^{(1)} 
+ \varepsilon^2 \rho^{(2)} + O(\varepsilon^3) , \\
&&{\vec j}  =  {\vec j}^{(0)} + \varepsilon {\vec j}^{(1)} 
+ \varepsilon^2 {\vec j}^{(2)} + O(\varepsilon^3) .
\end{eqnarray}
However, inserting these expansions into 
$n_i^{(0)} (\rho, \vec j)$, shows that
\begin{eqnarray} \label{eq:masscons}
&&0  =  \rho^{(1)} = \rho^{(2)} = \ldots , \\ \label{eq:momcons}
&&0  =  {\vec j}^{(1)} = {\vec j}^{(2)} = \ldots ;
\end{eqnarray}
otherwise $n_i^{(0)}$ would have contributions of order $\varepsilon$
and above, in contradiction to Eq. \ref{eq:eps0LBE}. The mass and momentum
densities can therefore be defined as moments of the equilibrium
distribution as well,
\begin{eqnarray} \label{eq:equilmass}
&&\sum_i n_i^{eq}  =  \rho , \\ \label{eq:equilmom}
&&\sum_i n_i^{eq} \vec c_i  =  \vec j.
\end{eqnarray}

We can analyze the dynamics of the LB model on large length and time scales
by taking moments of Eqs. \ref{eq:eps1LBE} and \ref{eq:eps2LBEb} with
respect to the LB velocity set ${\vec c}_i$. From the zeroth moment,
$\sum_i \cdots$, we obtain the continuity equation on the $t_1$
time scale (Eq. \ref{eq:delicons}),
\begin{equation} \label{eq:eps1continuity}
\partial_{t_1} \rho + \partial_{1\alpha} j_{\alpha} = 0 ,
\end{equation}
and incompressibility on the $t_2$ time scale
(Eqs. \ref{eq:delicons}, \ref{eq:masscons}, and \ref{eq:momcons})
\begin{equation} \label{eq:eps2continuity}
\partial_{t_2} \rho = 0 .
\end{equation}
In Eq. \ref{eq:eps1continuity} we have used the shorthand notation
$\partial_{1\alpha}$ for the $\alpha$ component of the spatial derivative
$\partial_{{\vec r}_1}$.

The first moment, $\sum_i \cdots c_{i\alpha}$, leads to momentum
conservation equations on both time scales
(Eqs. \ref{eq:delicons}, \ref{eq:masscons}, and \ref{eq:momcons}):
\begin{eqnarray} \label{eq:eps1momentum}
&&\partial_{t_1} j_\alpha + \partial_{1\beta} \pi_{\alpha\beta}^{(0)}
= 0 ,\\ \label{eq:eps2momentum}
&&\partial_{t_2} j_\alpha + 
\frac{1}{2} \partial_{1\beta} \left( 
\pi_{\alpha \beta}^{\star(1)} + \pi_{\alpha \beta}^{(1)}
\right) = 0 ,
\end{eqnarray}
where $\pi_{\alpha\beta}$ is the momentum flux or second moment,
\footnote{There is a notational inconsistency in Ref. \cite{DSL}.
In Eqs. 71 and 73 of that paper the superscript $neq$ should be replaced
by a superscript 1, and in Eq. 79 $Q_{\alpha\beta}$ should be
 $Q_{\alpha\beta}^1$.}
\begin{equation} \label{eq:definestress}
\pi_{\alpha \beta} = \sum_i n_i c_{i \alpha} c_{i \beta}.
\end{equation}
Momentum is conserved on both the $t_1$ and $t_2$ time scales,
because, in the hydrodynamic limit, the coupling between acoustic and
diffusive modes is very weak. First, sound waves propagate with
negligible viscous damping; then the residual pressure field in a nearly
incompressible fluid relaxes by momentum diffusion. We can write the
conservation laws on each time scale separately, as in Eqs.
\ref{eq:eps1momentum} and \ref{eq:eps2momentum}, or combine them
into a single equation in the lattice--scale variables
$\vec r = \vec r_1/\varepsilon$ and
$t = t_1/\varepsilon = t_2/\varepsilon^2$. The hydrodynamic fields
depend on $\vec r$ and $t$ parametrically, through their dependence on the
coarse-grained variables $\vec r_1$, $t_1$, $t_2$. Using $\partial_\alpha$
for a component of $\partial_{\vec r}$, we have
\begin{eqnarray}
&&\partial_\alpha = \varepsilon \partial_{1\alpha}, \\
&&\partial_t = \varepsilon \partial_{t_1} + \varepsilon^2 \partial_{t_2} .
\end{eqnarray}
The combined equations for the mass and momentum densities on the
lattice space and time scales are then:
\begin{eqnarray}
\label{eq:chapenskresult1}
&&\partial_t \rho + \partial_\alpha j_\alpha  =  0 , \\
\label{eq:chapenskresult2}
&&\partial_t j_\alpha + 
\partial_\beta \pi^{eq}_{\alpha \beta}
+ \frac{1}{2} \partial_\beta \left( 
\pi_{\alpha \beta}^{\star neq} + \pi^{neq}_{\alpha \beta}
\right)  =  0 ,
\end{eqnarray}
where from Eq. \ref{eq:niexpans},
$\pi_{\alpha\beta}^{eq} = \pi_{\alpha\beta}^{(0)}$ and
$\pi_{\alpha\beta}^{neq} = \varepsilon \pi_{\alpha\beta}^{(1)}$.

Finally, we can derive a relation between the pre-collision and
post-collision momentum fluxes, $\pi_{\alpha\beta}$ and
$\pi_{\alpha\beta}^\star$, by taking the second moment of
Eq. \ref{eq:eps1LBE}:
\begin{equation} \label{eq:eps1stress}
\partial_{t_1} \pi_{\alpha \beta}^{(0)} + \partial_{1\gamma}
\Phi_{\alpha \beta \gamma}^{(0)} = h^{-1}
\left( \pi_{\alpha \beta}^{\star(1)} - \pi_{\alpha \beta}^{(1)} \right) ,
\end{equation}
where $\Phi_{\alpha\beta\gamma}$ is the third moment of the distribution,
\begin{equation} \label{eq:definethirdmoment}
\Phi_{\alpha \beta \gamma} = \sum_i n_i c_{i \alpha} c_{i \beta}
c_{i \gamma} .
\end{equation}
We note that $\pi_{\alpha\beta}^{eq}$ is a collisional invariant and
therefore remains unchanged by the collision process. In terms of the
lattice variables,
\begin{equation} \label{eq:pre2post}
\pi_{\alpha \beta}^{\star neq} = \pi_{\alpha \beta}^{neq} +
h \left(\partial_t \pi_{\alpha \beta}^{eq} + \partial_{\gamma}
\Phi_{\alpha \beta \gamma}^{eq}\right).
\end{equation}

Equation \ref{eq:chapenskresult1} shows that continuity 
(Eq. \ref{eq:navstokes1}) is automatically satisfied by any LB model.
The Navier--Stokes equation (Eq. \ref{eq:navstokes2}) will be satisfied,
\emph{if} we succeed in ensuring that the Euler stress $\rho
c_s^2 \delta_{\alpha \beta} + \rho u_\alpha u_\beta$, the Newtonian
viscous stress, $\sigma_{\alpha \beta}$ (Eq. \ref{eq:macrostressviscous}),
and the fluctuating stress $\sigma^f_{\alpha \beta}$
(Eq. \ref{eq:macrostressfluct}) are given correctly
by the sum of the momentum fluxes in Eq. \ref{eq:chapenskresult2}.
Since $\pi^{eq}_{\alpha \beta}$ depends \emph{only} on
$\rho$ and $\vec j$, it must be identified with the Euler stress:
\begin{equation} \label{eq:equileulerstress}
\pi_{\alpha \beta}^{eq} = 
\rho c_s^2 \delta_{\alpha \beta}
+ \rho u_\alpha u_\beta .
\end{equation}
The viscous stress and fluctuating stresses must then
be contained in $( \pi_{\alpha \beta}^{\star neq} +
\pi^{neq}_{\alpha \beta} ) /2$.

This is about as far as we can go in complete generality.
In order to proceed further we need to consider specific equilibrium
distributions and collision operators. The results of this
subsection suggest the following approach towards constructing
an LB method which (asymptotically) simulates the
fluctuating Navier--Stokes equations:
\begin{itemize}
\item Find a set of equilibrium populations $n_i^{eq}$
      such that
      \begin{itemize}
      \item
         \begin{equation}
            \sum_i n_i^{eq} = \rho ;
         \end{equation}
      \item
         \begin{equation}
            \sum_i n_i^{eq} \vec c_i = \vec j ;
         \end{equation}
      \item
         \begin{equation}
            \sum_i n_i^{eq} c_{i \alpha} c_{i \beta}
             = \rho c_s^2 \delta_{\alpha \beta}
               + \rho u_\alpha u_\beta .
         \end{equation}
      \end{itemize}
\item Find a collision operator $\Delta_i$ with the properties:
      \begin{itemize}
      \item
         \begin{equation}
            \sum_i \Delta_i = 0 ;
         \end{equation}
      \item
         \begin{equation}
            \sum_i \Delta_i \vec c_i = 0 ;
         \end{equation}
      \item The nonequilibrium momentum flux
		 $(\pi_{\alpha \beta}^{\star neq} + 
         \pi^{neq}_{\alpha \beta} ) /2$ must be connected
         with the sum of viscous and fluctuating stresses.
      \end{itemize}
\end{itemize}
In the following subsections, we will follow this procedure for the
three-dimensional D3Q19 model.

\subsection{D3Q19 model I: Equilibrium populations}
\label{sec:fluct_lbe:equil_pops}

Early LB models~\cite{McN88,Hig89,Ben92} inherited their equilibrium
distributions from LGCA, along with macroscopic manifestations of
the broken Galilean invariance: an incorrect advection velocity and a
velocity dependent pressure. Subsequently a new equilibrium distribution
was proposed that restored Galilean invariance at the macroscopic
level \cite{McN92,qian}, but with the loss of the connection to statistical
mechanics. The idea was to ensure that the first few moments of $n_i^{eq}$
matched those derived from the Maxwell--Boltzmann distribution for a
dilute gas \cite{cha60},
\begin{equation} \label{eq:MBeq}
n \left({\vec c} | \rho, {\vec u}, T \right) 
= \rho \, \left( \frac{m_p}{2 \pi k_B T} \right)^{3/2}
\exp\left(- \frac{m_p}{2 k_B T} 
\left({\vec c}-{\vec u}\right)^2 \right) :
\end{equation}
specifically;
\begin{eqnarray} \label{eq:MBeq0}
&& \int d^3 \vec c \, n({\vec c}) = \rho , \\  \label{eq:MBeq1}
&& \int d^3 \vec c \, n({\vec c}) c_\alpha 
= \rho u_\alpha , \\  \label{eq:MBeq2}
&& \int d^3 \vec c \, n({\vec c}) c_\alpha c_\beta
= \frac{\rho k_B T}{m_p} \delta_{\alpha\beta} + \rho u_\alpha u_\beta 
= \rho c_s^2 \delta_{\alpha\beta} + \rho u_\alpha u_\beta .
\end{eqnarray}
With these moments the Euler hydrodynamic equations 
(\cf\ Eqs. \ref{eq:navstokes1} and \ref{eq:navstokes2}),
\begin{eqnarray}
\label{eq:euler1}
&&\partial_t \rho + \partial_\alpha j_\alpha = 0 , \\
\label{eq:euler2}
&&\partial_t j_\alpha + 
\partial_\beta \left( \rho c_s^2 \delta_{\alpha \beta}
+ \rho u_\alpha u_\beta \right) = 0 ,
\end{eqnarray}
may be derived from the continuum version of the Chapman-Enskog
expansion \cite{cha60}. The viscous stress arises
from the non-equilibrium distribution
(\cf\ Sec. \ref {sec:fluct_lbe:chapman_enskog}).

An expansion of the Maxwell--Boltzmann equilibrium distribution
(Eq. \ref{eq:MBeq}) at low velocities suggests the following ansatz
 \cite{McN92,qian} for the discrete velocity equilibrium, 
\begin{equation} \label{eq:niequilansatz}
n_i^{eq} \left(\rho, \vec u \right) = 
a^{c_i} \rho \left( 1 + A \vec u \cdot \vec c_i
+ B ( \vec u \cdot \vec c_i )^2 + C u^2 \right) 
\end{equation}
with suitably adjusted coefficients $a^{c_i}$, $A$, $B$, and $C$.
The rationale for Eq. \ref{eq:niequilansatz} is that the
equilibrium momentum flux is quadratic in the flow
velocity $\vec u$ (Eq. \ref{eq:MBeq2}); it therefore makes sense
to construct a similar form for $n_i^{eq}$. A drawback of Eq.
\ref{eq:niequilansatz} is that $n_i^{eq}$ may be negative if $u$ becomes
sufficiently large. This can be avoided by more general equilibrium
distributions, which are equivalent to Eq. \ref{eq:niequilansatz}
up to order $u^2$ ~\cite{Wag98,Kar99,Bog03}.

The prefactors $a^{c_i} > 0$ are normalized such that
\begin{equation} \label{eq:w0}
\sum_i a^{c_i} = 1 ,
\end{equation}
which ensures that Eq. \ref{eq:equilmass} is satisfied in the
special case $\vec u = 0$.  The notation in Eq. \ref{eq:niequilansatz}
was chosen in order to indicate explicitly that the weights
depend only on the absolute value of the speed $c_i$, but not its
direction; this follows from the rotational symmetries of the LB model.
The coefficients $A$, $B$, $C$ are here independent
of $c_i$. There are other LB models, like
D3Q18 \cite{McN92}, where this condition is not imposed, and
$A$, $B$, and $C$ depend on $c_i$ as well; however, such models
are only hydrodynamically correct in the incompressible
limit \cite{Lad94}, and cannot be straightforwardly interpreted in terms
of statistical mechanics (see Sec. \ref{sec:fluct_lbe:statmech}).
We will not consider such models.

In a cubic lattice, symmetry dictates the following
relations for the low--order velocity moments of the weights,
\begin{eqnarray} \label{eq:w1}
&&\sum_i a^{c_i} c_{i \alpha}  =  0 , \\ \label{eq:w2}
&&\sum_i a^{c_i} c_{i \alpha} c_{i \beta} 
 =  C_2 \, \delta_{\alpha \beta} , \\ \label{eq:w3}
&&\sum_i a^{c_i} c_{i \alpha} c_{i \beta} c_{i \gamma}  =  0 , \\
\label{eq:w4}
&&\sum_i a^{c_i} c_{i \alpha} c_{i \beta} c_{i \gamma} c_{i \delta}
 =  C_4^\prime \, \delta_{\alpha \beta \gamma \delta}
+ C_4 \left( 
  \delta_{\alpha \beta}  \delta_{\gamma \delta}
+ \delta_{\alpha \gamma} \delta_{\beta \delta}
+ \delta_{\alpha \delta} \delta_{\beta \gamma}
\right) ,
\end{eqnarray}
where the values of the parameters $C_2$, $C_4$ and $C_4^\prime$
depend on the details of the choice of the coefficients $a^{c_i}$.
The tensor $\delta_{\alpha \beta \gamma \delta}$ is unity when $\alpha
= \beta = \gamma = \delta$ and zero otherwise. Rotational invariance of
the stress tensor requires that $C_4^\prime = 0$.

The results in Eqs. \ref{eq:w0}--\ref{eq:w4} allow us to calculate the
moments of Eq. \ref{eq:niequilansatz} up to second order.
Consistency with the mass density, momentum density and Euler
stress for a given $\rho$ and $\bu$, uniquely determines
the equilibrium distribution,
\begin{equation} \label{eq:original_equil_distrib}
n_i^{eq} \left( \rho, \vec u \right) =
a^{c_i} \rho \left( 1 + \frac{ \vec u \cdot \vec c_i }{c_s^2}
+ \frac{ \left( \vec u \cdot \vec c_i \right)^2 }{2 c_s^4}
- \frac{ u^2 }{2 c_s^2} \right) ,
\end{equation}
with the speed of sound $c_s^2 = C_2$, and the weights adjusted such
that $C_4^\prime = 0$ and $C_4 = C_2^2$. These two latter conditions,
together with the normalization condition Eq. \ref{eq:w0}, form a set
of three equations for the coefficients $a^{c_i}$.  Therefore at least
three speeds, or three shells of neighbors, are needed to satisfy the
constraints.  We consider the D3Q19 model, which incorporates the three
smallest speeds on
a simple--cubic lattice. Here one obtains $a^0 = 1/3$ for the
stationary particles, $a^1 = 1/18$ for the six nearest--neighbor
directions, and $a^{\sqrt{2}} = 1/36$ for the twelve next--nearest
neighbors. The speed of sound is then $c_s^2 = (1/3) (b / h)^2$.

We now turn back to the results of the previous subsection, since the
explicit form of $n_i^{eq}$ allows us to pursue the analysis further.
We first calculate the equilibrium third--order moment
(Eq. \ref{eq:definethirdmoment}) using Eq. \ref{eq:original_equil_distrib}:
\begin{equation} \label{eq:3rdmoment}
\Phi^{eq}_{\alpha \beta \gamma} = \rho c_s^2 \left(
u_\alpha \delta_{\beta \gamma} +
u_\beta  \delta_{\alpha \gamma} +
u_\gamma \delta_{\alpha \beta} \right) .
\end{equation}
In fact Eq. \ref{eq:3rdmoment}  is model independent to order $u^2$,
since only the linear term in $\bu \cdot \bc_i$ contributes.
To close the hydrodynamic equations for the mass and momentum densities
(Eqs. \ref{eq:chapenskresult1} and \ref{eq:chapenskresult2}) we need
expressions for the pre--collision and post--collision momentum fluxes,
$\pi_{\alpha\beta}^{\star neq}$ and $\pi_{\alpha\beta}^{neq}$.
From Eq. \ref{eq:eps1stress} we can obtain an expression for
$\pi_{\alpha\beta}^{\star neq} - \pi_{\alpha\beta}^{neq}$
in terms of the velocity gradient,
\begin{equation} \label{eq:newtonstress0}
\rho c_s^2 \left( \partial_{1\alpha} u_\beta 
                + \partial_{1\beta}  u_\alpha \right) 
= h^{-1} \left( \pi^{\star (1)}_{\alpha \beta} 
- \pi^{(1)}_{\alpha \beta} \right) ,
\end{equation}
where we have used Eqs. \ref{eq:eps1continuity} and \ref{eq:eps1momentum}
to rewrite the time derivative of $\pi_{\alpha \beta}^{(0)}$ in terms
of spatial derivatives of $\rho$ and $\bu$. In arriving at 
Eq. \ref{eq:newtonstress0}, we have neglected terms of order $u^3$,
consistent with the low Mach number limit we are considering.
Finally, Eq. \ref{eq:newtonstress0} can be rewritten in terms of
the unscaled variables,
\begin{equation} \label{eq:newtonstress}
\pi^{\star neq}_{\alpha \beta} - \pi^{neq}_{\alpha \beta} =
h \rho c_s^2 \left( \partial_\alpha u_\beta 
                + \partial_\beta  u_\alpha \right) .
\end{equation}
To obtain a further relation for the non-equilibrium momentum
fluxes, we must consider the collision operator in more detail.

\subsection{D3Q19 model II: Deterministic collision operator}
\label{sec:fluct_lbe:determ_coll}

In a deterministic model, the collision operator $\Delta_i$
is a unique function of the distribution $\bn$. Therefore,
we can obtain the Chapman--Enskog ordering of $\Delta_i$
via a Taylor expansion with respect to $\bn$:
\begin{eqnarray}
\nonumber
\Delta_i \left( \bn \right)
& = & \Delta_i \left( \bn^{(0)} + \varepsilon \bn^{(1)}
+ \varepsilon^2 \bn^{(2)} + \ldots \right) \\
\label{eq:deli1}
& = & \Delta_i \left( \bn^{(0)} \right)
+ \varepsilon \sum_j 
\left. \left( \frac{ \partial \Delta_i }{\partial n_j} \right)
\right\vert_{ \bn^{(0)} } n_j^{(1)}
+ O \left(\varepsilon^2\right) .
\end{eqnarray}
The analysis of Sec. \ref{sec:fluct_lbe:chapman_enskog} has
shown that $\Delta_i ( \bn^{(0)} ) = 0$ (Eq. \ref{eq:deli0}),
and that hydrodynamic behavior is determined by the order
$\varepsilon^1$ collision operator, 
\begin{equation} \label{eq:deli2}
\Delta_i^{(1)}
 = \sum_j \left.\left(\frac{\partial
 \Delta_i}{\partial n_j}\right)\right|_{ \bn^{(0)} } n_j^{(1)} .
\end{equation}
Although $\Delta_i^{(2)}$ appears at second order in the Chapman-Enskog
expansion (Eq. \ref{eq:eps2LBE}), it makes no contribution to the change
in mass and momentum densities (Eq. \ref{eq:delicons}). However,
$\Delta_i^{(1)}$ contributes to a first--order change in the viscous
stress (Eq. \ref{eq:pre2post}), which enters into the momentum equation
at second order (Eq. \ref{eq:eps2momentum}).
It is therefore reasonable to construct the collision operator
with the form
of $\Delta_i^{(1)}$:
\begin{equation}
\Delta_i = \sum_j {\mathcal L}_{ij} n_j^{neq},
\end{equation}
where ${\mathcal L}_{ij}$ is a matrix of constant coefficients.
Thus to lowest order in $\varepsilon$, the collision
process is a linear transformation between the non-equilibrium
distributions for each velocity:
\begin{equation} \label{eq:linear_collision_operator}
n_i^{\star neq} = \sum_j \left(\delta_{ij} +
{\mathcal L}_{ij}\right) n_j^{neq} .
\end{equation}
The simplest such collision operator is the lattice BGK
(Bhatnagar--Gross--Krook) model \cite{sauro}, ${\mathcal L}_{ij} = -
\delta_{ij} / \tau$, where the collisional relaxation time $\tau$ is
related to the viscosity. Here we will work within the more general
framework of the multi--relaxation time (MRT) model \cite{Dhu02}, for
which the lattice BGK model is a special case.

Polynomials in the dimensionless velocity vectors, ${\hat \bc}_i =
\bc_i /c$ ($c = b/h$), form a basis for a diagonal representation of
${\mathcal L}_{ij}$ \cite{Dhu02}, which allows for a more general and
stable LB model with the same level of computational complexity as the
BGK version \cite{Lal00}. Orthogonal basis vectors, $\be_{k}$, are
constructed from outer products of the vectors ${\hat \bc}_i$. For
example:
\begin{eqnarray}
&&e_{0i}  = 1 , \\
&&e_{1i}  = {\hat c}_{ix} , \\
&&e_{2i}  = {\hat c}_{iy} , \\
&&e_{3i}  = {\hat c}_{iz} .
\end{eqnarray}

\begin{table}[t] \caption{Basis vectors of the D3Q19 model.
Each row corresponds to a different
basis vector, with the actual polynomial in
${\hat c}_{i\alpha} =  c_{i\alpha} /c$ shown
in the second column.  The normalizing factor for each basis vector
is in the third column. The polynomials form an orthogonal set
when $q^{c_i} = a^{c_i}$ (Eq. \ref{eq:orthogonality}).}
\begin{center}
\begin{tabular}{r | c | c}
$k$ & $e_{ki}$ & $w_k$ \\
\hline
 0 & 1                                                    &  1  \\
 1 & ${\hat c}_{ix}$                                      & 1/3 \\
 2 & ${\hat c}_{iy}$                                      & 1/3 \\
 3 & ${\hat c}_{iz}$                                      & 1/3 \\
 4 & ${\hat c}_i^2-1$                                     & 2/3 \\
 5 & $3{\hat c}_{ix}^2-{\hat c}_i^2$                      & 4/3 \\
 6 & ${\hat c}_{iy}^2-{\hat c}_{iz}^2$                    & 4/9 \\
 7 & ${\hat c}_{ix} {\hat c}_{iy}$                        & 1/9 \\
 8 & ${\hat c}_{iy} {\hat c}_{iz}$                        & 1/9 \\
 9 & ${\hat c}_{iz} {\hat c}_{ix}$                        & 1/9 \\
10 & $(3{\hat c}_i^2-5){\hat c}_{ix}$                     & 2/3 \\
11 & $(3{\hat c}_i^2-5){\hat c}_{iy}$                     & 2/3 \\
12 & $(3{\hat c}_i^2-5){\hat c}_{iz}$                     & 2/3 \\
13 & $({\hat c}_{iy}^2-{\hat c}_{iz}^2){\hat c}_{ix}$     & 2/9 \\
14 & $({\hat c}_{iz}^2-{\hat c}_{ix}^2){\hat c}_{iy}$     & 2/9 \\
15 & $({\hat c}_{ix}^2-{\hat c}_{iy}^2){\hat c}_{iz}$     & 2/9 \\
16 & $3{\hat c}_i^4-6{\hat c}_i^2+1$                      &  2  \\
17 & $(2{\hat c}_i^2-3)(3{\hat c}_{ix}^2-{\hat c}_i^2)$   & 4/3 \\
18 & $(2{\hat c}_i^2-3)({\hat c}_{iy}^2-{\hat c}_{iz}^2)$ & 4/9 \\
\hline
\end{tabular}
\end{center}\label{tab:evectors_d3q19}
\end{table}

There are six quadratic polynomials, which are given in Table
\ref{tab:evectors_d3q19} as basis vectors $\be_4 - \be_9$.
A Gram--Schmidt procedure ensures that all the basis vectors
are mutually orthogonal with respect to a set of positive weights,
$q^{c_i} > 0$, 
\begin{equation} \label{eq:orthogonality}
\sum_i q^{c_i} e_{ki} e_{li} = w_k \delta_{kl} .
\end{equation}
The weights are restricted by the same symmetries as the coefficients
in the equilibrium distribution $a^{c_i}$, but are not necessarily the same;
in the D3Q19 model there are then three independent values of
$q^{c_i}$. The normalization factors, $w_k > 0$, are related to the
choice of basis vectors
\begin{equation} \label{eq:weights_unweighted}
w_k = \sum_i q^{c_i} e_{ki}^2 .
\end{equation}
Within the D3Q19 model, polynomials up to second order are complete,
but at third-order there is some deflation; for example, ${\hat
  c}_{ix}^3$ is equivalent to ${\hat c}_{ix}$. In fact, there are only
six independent third--order and three independent fourth--order
polynomials in the D3Q19 model. Beyond fourth order, all polynomials
deflate to lower orders, so the basis vectors in Table
\ref{tab:evectors_d3q19} form a complete set for the D3Q19 model.

The basis vectors can be used to construct a complete set of moments
of the LB distribution,
\begin{equation} \label{forward}
m_k = \sum_i e_{ki} n_i ,
\end{equation}
which allows for a diagonal representation of the collision operator
\cite{Dhu92,Dhu02}, as will be made clear later.
Hydrodynamic variables are related to the moments up to
quadratic order in ${\hat \bc}_i$
(\cf\ Table \ref{tab:evectors_d3q19}):
\begin{eqnarray} \label{eq:hmodes_beg}
&&\rho = m_0 , \\ 
&&j_x  = m_1 c, \\
&&j_y  = m_2 c, \\
&&j_z  = m_3 c, \\
&&\pi_{xx}  = (m_0+m_4+m_5) c^2/3, \\
&&\pi_{yy}  = (2m_0+2m_4-m_5+3m_6) c^2/6, \\
&&\pi_{zz}  = (2m_0+2m_4-m_5-3m_6) c^2/6, \\
&&\pi_{xy}  =  m_7 c^2, \\
&&\pi_{yz}  =  m_8 c^2, \\ \label{eq:hmodes_end}
&&\pi_{zx}  =  m_9 c^2 .
\end{eqnarray}
There are additional degrees of freedom in the D3Q19 model beyond
those required for the conserved variables and stresses (Eqs.
\ref{eq:hmodes_beg}--\ref{eq:hmodes_end}). These ``kinetic'' or ``ghost''
\cite{Adh05} moments do not play a role in the large-scale dynamics
\cite{DSL}, but they are important for proper thermalization
\cite{Adh05} and near boundaries \cite{Gin03}.

The basis vectors in Table \ref{tab:evectors_d3q19} are complete but
not unique. Besides trivial variations in the Gram--Schmidt
orthogonalization, there is a substantive difference that depends on
the choice of the weighting factors $q^{c_i}$: these factors determine
both the result of the orthogonalization procedure, as well as the
back transformation from moments $m_k$ to populations $n_i$. This
is most easily seen from the observation that Eq. \ref{eq:orthogonality}
can be rewritten as the standard orthonormality relation \cite{DSL}
\begin{equation} \label{eq:standard_orthogonality_forward}
\sum_i \hat{e}_{ki} \hat{e}_{li} = \delta_{kl} ,
\end{equation}
where we have introduced the orthonormal basis vectors
\begin{equation}\label{eq:evectors_orthonormal}
\hat{e}_{ki} = \sqrt{\frac{q^{c_i}}{w_k}} e_{ki} .
\end{equation}
Equation \ref{eq:standard_orthogonality_forward} implies the
backward relation
\begin{equation} \label{eq:standard_orthogonality_backward}
\sum_k \hat{e}_{ki} \hat{e}_{kj} = \delta_{ij} ,
\end{equation}
or, in terms of unnormalized basis vectors,
\begin{equation} \label{eq:orthogonality_backward}
\sum_k w_k^{-1} e_{ki} e_{kj} = \frac{1}{q^{c_i}} \delta_{ij} .
\end{equation}
In the normalized basis the transformations between distribution and
moments are
\begin{eqnarray} \label{eq:orthonormal_trafo_1}
&&{\hat m}_k = \frac{m_k}{\sqrt{w_k}}  =  \sum_i \hat{e}_{ki} 
\frac{n_i}{\sqrt{q^{c_i}}} =  \sum_i \hat{e}_{ki} 
{\hat n}_i, \\ \label{eq:orthonormal_trafo_2}
&&{\hat n}_i = \frac{n_i}{\sqrt{q^{c_i}}}  =  \sum_k \hat{e}_{ki}
\frac{m_k}{\sqrt{w_k}} =  \sum_k \hat{e}_{ki} {\hat m}_k;
\end{eqnarray}
we will make use of these relations in Sec. \ref{sec:fluct_lbe:noise}.
The analog of Eq.~\ref{eq:linear_collision_operator} for the
normalized basis is
\begin{equation} \label{eq:linear_collision_operator-norm}
{\hat n}_i^{\star neq} = \sum_j \left( \delta_{ij} +
{\hat \mathcal L}_{ij} \right) {\hat n}_j^{neq} ,
\end{equation}
with
\begin{equation} 
{\hat \mathcal L}_{ij} = \sqrt{\frac{q^{c_j}}{q^{c_i}}}
\mathcal L_{ij} .
\end{equation}
In terms of unnormalized basis vectors the back transformation is given
by
\begin{equation} \label{eq:back_transformation}
n_i = q^{c_i} \sum_k w_k^{-1} e_{ki} m_k .
\end{equation}

The most obvious choice is to set $q^{c_i} = 1$ \cite{Dhu92,Dhu02},
but then the basis vectors of the kinetic modes, $\be_{10} - \be_{18}$,
are not orthogonal to the equilibrium distribution, and the
moments $m_{10} - m_{18}$ have both equilibrium and non-equilibrium
contributions \cite{Dhu92,Dhu02}. The statistical mechanical connection
is more straightforward if the weights $q^{c_i}$ are matched to the weights
in the equilibrium distribution, setting $q^{c_i} = a^{c_i}$; this
eliminates the projection of $n_i^{eq}$ on the kinetic moments. The
weighted orthogonality relation defines a different but
equivalent set of basis vectors to those given in Refs.
\cite{Dhu92,Dhu02}, and these are the ones given in Table
\ref{tab:evectors_d3q19}. A comparison of the two sets of basis
vectors can be found in Ref. \cite{Chu07}.

The basis vectors can be used to construct a collision operator that
automatically satisfies all the lattice symmetries,
\begin{equation}\label{eq:lco_eigen}
\hat {\cal L}_{ij} = \sum_k \lambda_k {\hat e}_{ki}  {\hat e}_{kj} ,
\end{equation}
which is a symmetric matrix, while ${\cal L}_{ij}$, in general, is
not symmetric. The orthogonality of the basis
vectors ensures that each moment relaxes independently under the
action of the linearized collision operator,
\begin{equation}\label{eq:lco_update}
{\hat m}_k^{\star neq} = \gamma_k {\hat m}_k^{neq},
\end{equation}
where $\gamma_k = 1 + \lambda_k$.
For the conserved modes $k = 0, \ldots, 3$ the value $\gamma_k$ is
immaterial, since $m_k^{\star neq} = m_k^{neq} = 0$.  For the other
modes, $k > 3$, linear stability requires that
\begin{equation}
\left\vert \gamma_k \right\vert \le 1;
\end{equation}
\ie\ the effect of collisions must be to cause the nonequilibrium
distribution to decrease rather than increase. The eigenvalues, $\gamma_k$,
may be positive or negative, with $\gamma_k < 0$ corresponding to
``over--relaxation''.

The number of independent eigenvalues is limited
by symmetry. There are at most six independent $\gamma_k$'s in the
D3Q19 model, corresponding to a bulk viscous mode with eigenvalue
$\gamma_v$, five symmetry--related shear modes, which must have the same
eigenvalue, $\gamma_s$, and nine kinetic modes, broken down into
symmetry--related groups: $e_{10} - e_{12}$, $e_{13} - e_{15}$,
$e_{16}$, $e_{17} - e_{18}$.
The eigenvalues $\gamma_s$ and $\gamma_v$ can be related to the
shear and bulk viscosities by decomposing the stress tensor
into traceless--symmetric (shear) and trace (bulk) components,
\begin{equation}
\pi_{\alpha \beta} = \overline \pi_{\alpha \beta} + \frac{1}{3}
\pi_{\gamma \gamma} \delta_{\alpha \beta} ;
\end{equation}
the overbar is used to denote a traceless tensor.
Equation \ref{eq:lco_update} implies the
following relations between pre-- and post--collisional stresses:
\begin{eqnarray}
\label{eq:stressrelax1}
&&\overline \pi_{\alpha \beta}^{\star neq}
 =  \gamma_s \overline \pi_{\alpha \beta}^{neq} , \\ 
\label{eq:stressrelax2}
&&\pi_{\alpha \alpha}^{\star neq}
 =  \gamma_v \pi_{\alpha \alpha}^{neq} .
\end{eqnarray}
Additional relations between the pre-- and post--collision stresses
have already been provided (Eq. \ref{eq:newtonstress}):
\begin{eqnarray} \label{eq:newtonstress1}
&&\overline{\pi}^{\star neq}_{\alpha \beta} -
\overline{\pi}^{neq}_{\alpha \beta} =
h \rho c_s^2 \left( \overline{\partial_\alpha u}_\beta 
                  + \overline{\partial_\beta  u}_\alpha \right) , \\
\label{eq:newtonstress2}
&&\pi^{\star neq}_{\alpha \alpha} - \pi^{neq}_{\alpha \alpha} =
2h \rho c_s^2 \partial_\alpha u_\alpha .
\end{eqnarray}
Equations \ref{eq:stressrelax1}--\ref{eq:newtonstress2} can
be solved to relate the pre-- and post--collision
stresses to the velocity gradient:
\begin{eqnarray} \label{eq:pisolve1}
&&\overline{\pi}^{neq}_{\alpha \beta} =
	- \frac{h \rho c_s^2}{1 - \gamma_s}
	  \left( \overline{\partial_\alpha u}_\beta 
    + \overline{\partial_\beta  u}_\alpha \right) , \\ \label{eq:pisolve2}
&&\overline{\pi}^{\star neq}_{\alpha \beta} =
	- \frac{h \rho c_s^2 \gamma_s}{1 - \gamma_s}
	  \left( \overline{\partial_\alpha u}_\beta 
    + \overline{\partial_\beta  u}_\alpha \right) , \\ \label{eq:pisolve3}
&&\pi^{neq}_{\alpha \alpha} =
	- \frac{2h \rho c_s^2}{1 - \gamma_v}
\partial_\alpha u_\alpha , \\ \label{eq:pisolve4}
&&\pi^{\star neq}_{\alpha \alpha} =
	- \frac{2h \rho c_s^2 \gamma_v}{1 - \gamma_v}
\partial_\alpha u_\alpha .
\end{eqnarray}
From Eq. \ref{eq:chapenskresult2} we then find the usual Newtonian form
for the viscous stress, and can identify the shear and bulk viscosities:
\begin{eqnarray} \label{eq:shearviscosity}
&&\eta   =  \frac{h \rho c_s^2}{2} \frac{1 + \gamma_s}{1 - \gamma_s} , \\
\label{eq:bulkviscosity}
&&\eta_v  =  \frac{h \rho c_s^2}{3} \frac{1 + \gamma_v}{1 - \gamma_v} .
\end{eqnarray}

Lattice symmetry dictates that there are at most four independent
eigenvalues of the kinetic modes:
(see Table \ref{tab:evectors_d3q19}):
$\gamma_{3a}$ (modes $10-12$), $\gamma_{3b}$ (modes $13-15$),
$\gamma_{4a}$ (mode $16$), and $\gamma_{4b}$ (modes $17-18$).
In a number of implementations of the MRT model
\cite{McN92,McN93,Lad94a,Lad01} the kinetic eigenvalues are set to zero,
so that these modes are projected out by the collision operator, although
they reoccur at the next time step.
Recently, it has been shown that the kinetic eigenvalues can be tuned
to improve the accuracy of the boundary conditions at solid surfaces
\cite{Gin03}. A useful simplification is to use only two independent
relaxation rates, with
$\gamma_v = \gamma_s = \gamma_{4a} = \gamma_{4b} = \gamma_e$
and $\gamma_{3a} = \gamma_{3b} = \gamma_o$. The optimal boundary
conditions are obtained with specific relations between $\gamma_e$ and
$\gamma_o$ \cite{Gin03,Chu07}.

\subsection{D3Q19 model III: Thermal noise}
\label{sec:fluct_lbe:noise}

In the fluctuating LB model \cite{Lad94,Lad01}, thermal noise is
included by adding a stochastic contribution, $\Delta_i^\prime$,
to the collision operator:
\begin{equation} \label{eq:coll_operator_plus_noise}
\Delta_i = \sum_j {\cal L}_{ij} n_j^{neq} + \Delta_i^\prime .
\end{equation}
The collision operator must still conserve mass and momentum exactly,
\begin{equation}
\sum_i \Delta_i^\prime = \sum_i \Delta_i^\prime \vec c_i = 0 ,
\end{equation}
while the statistical properties of $\Delta_i^\prime$ 
include a vanishing mean, $\left< \Delta_i^\prime \right> = 0$, and
a nontrivial covariance matrix,
$\left< \Delta_i^\prime \Delta_j^\prime \right>$, that gives
the correct fluctuations at the hydrodynamic level
(see Eqs. \ref{eq:navstokes2} and \ref{eq:macrostressfluct}):
\begin{equation} \label{eq:site_macrostressfluct}
\left< \sigma^f_{\alpha \beta} \sigma^f_{\gamma \delta} \right> = 
\frac{2 k_B T}{b^3 h} \eta_{\alpha \beta \gamma \delta}.
\end{equation}
The stochastic collision operator is assumed to be local in space
and time, so that there are no correlations between the
noise at different lattice sites or at different times.
The delta functions in space and time have been replaced
by $b^{-3}$ and $h^{-1}$, respectively, so that the double
integral of Eq. \ref{eq:macrostressfluct} with respect to
$\br^\prime$ and $t^\prime$, over a small space--time region
of size $b^3 h$, matches the corresponding
integral of Eq. \ref{eq:site_macrostressfluct}.

Splitting the tensor into the trace, $\sigma^f_{\alpha \alpha}$,
and traceless, $\bar{\sigma}^f_{\alpha \beta}$, parts gives
the equivalent relations
\begin{eqnarray} \label{eq:hydrofluc1}
&&\left< \bar{\sigma}^f_{\alpha \beta} \bar{\sigma}^f_{\gamma \delta} \right>
 = \frac{2 k_B T \eta} {b^3 h}
\left[ \delta_{\alpha \gamma} \delta_{\beta \delta}
+ \delta_{\alpha \delta} \delta_{\beta \gamma}
- \frac{2}{3} \delta_{\alpha \beta} \delta_{\gamma \delta} \right] , \\
\label{eq:hydrofluc2}
&&\left< \sigma^f_{\alpha \alpha} \sigma^f_{\beta \beta} \right>
 = \frac{18 k_B T \eta_v} {b^3 h} , \\
&&\left< \bar{\sigma}^f_{\alpha \beta} \sigma^f_{\gamma \gamma} \right>
 =  0 .
\end{eqnarray}
Although temperature does not appear directly in the D3Q19 LB model,
we can determine the appropriate fluctuation level through the
equation of state for an isothermal ideal gas of particles of mass
$m_p$, $k_B T = m_p c_s^2 = \mu b^3 c_s^2$ \cite{DSL}.
Taking into account the results for
$\eta$ and $\eta_v$ (Eqs. \ref{eq:shearviscosity} and
\ref{eq:bulkviscosity}), we can write the desired correlations in
terms of the LB variables:
\begin{eqnarray}
&&\frac{\left< \bar{\sigma}^f_{\alpha \beta} 
               \bar{\sigma}^f_{\gamma \delta} \right>}
{\mu \rho c_s^4} = \frac{1 + \gamma_s}{1 - \gamma_s}
\left[ \delta_{\alpha \gamma} \delta_{\beta \delta}
+ \delta_{\alpha \delta} \delta_{\beta \gamma}
- \frac{2}{3} \delta_{\alpha \beta} \delta_{\gamma \delta} \right] , \\
&&\frac{\left< \sigma^f_{\alpha \alpha} 
               \sigma^f_{\beta \beta} \right>}
{\mu \rho c_s^4} = 
6 \frac{1 + \gamma_v}{1 - \gamma_v} , \\
&&\left< \bar{\sigma}^f_{\alpha \beta}
              \sigma^f_{\gamma \gamma} \right>
 =  0 .
\end{eqnarray}

The stress fluctuations $\sigma^f_{\alpha \beta}$ are \emph{different}
from the random stresses $\sigma^r_{\alpha \beta}$ that arise in the
LB algorithm itself,
\begin{equation}
\sigma^r_{\alpha \beta} = \sum_i \Delta_i^\prime c_{i \alpha} c_{i \beta} .
\end{equation}
The reason is that $\sigma^f_{\alpha \beta}$ pertains to fluctuations
on the $t_1$ time scale, which interact with the hydrodynamic flow field,
while $\sigma^r_{\alpha \beta}$ represents added noise on the lattice
time scale, $h$.
We use the Chapman--Enskog procedure to work backwards from the known
fluctuations in $\sigma^f_{\alpha \beta}$ to determine the
covariance matrix for $\sigma^r_{\alpha \beta}$ \cite{DSL}.
The stress update rule including random noise $\sigma^r_{\alpha\beta}$ is
(\cf\ Eqs. \ref{eq:stressrelax1} and \ref{eq:stressrelax2}):
\begin{eqnarray}
\label{eq:stressrelax_andnoise1}
&&\bar{\pi}_{\alpha \beta}^{\star neq}
 =  \gamma_s \bar{\pi}_{\alpha \beta}^{neq} + \bar{\sigma}^r_{\alpha \beta} ,
\\
\label{eq:stressrelax_andnoise2}
&&\pi_{\alpha \alpha}^{\star neq}
 =  \gamma_v \pi_{\alpha \alpha}^{neq} + \sigma^r_{\alpha \alpha} .
\end{eqnarray}
Eqs. \ref{eq:newtonstress1} and \ref{eq:newtonstress2}
remain valid and, together with Eqs. \ref{eq:stressrelax_andnoise1}
and \ref{eq:stressrelax_andnoise2}, can be solved for the pre-- and
post--collisional stresses, $\pi^{neq}_{\alpha \beta}$ and
$\pi^{\star neq}_{\alpha \beta}$, as before:
\begin{eqnarray}
&&\bar{\pi}^{neq}_{\alpha \beta} =
	- \frac{h \rho c_s^2}{1 - \gamma_s}
	  \left( \overline{\partial_\alpha u}_\beta 
    + \overline{\partial_\beta  u}_\alpha \right)
	+ \frac{1}{1 - \gamma_s} {\overline \sigma^r_{\alpha \beta}} , \\
&&\bar{\pi}^{\star neq}_{\alpha \beta} =
	- \frac{h \rho c_s^2 \gamma_s}{1 - \gamma_s}
	  \left( \overline{\partial_\alpha u}_\beta 
    + \overline{\partial_\beta  u}_\alpha \right)
	+ \frac{1}{1 - \gamma_s} {\overline \sigma^r_{\alpha \beta}} , \\
&&\pi^{neq}_{\alpha \alpha} =
	- \frac{2h \rho c_s^2}{1 - \gamma_v}
		\partial_\alpha u_\alpha 
	+ \frac{1}{1 - \gamma_v} \sigma^r_{\alpha \alpha}, \\
&&\pi^{\star neq}_{\alpha \alpha} =
	- \frac{2h \rho c_s^2 \gamma_v}{1 - \gamma_v}
		\partial_\alpha u_\alpha 
	+ \frac{1}{1 - \gamma_v} \sigma^r_{\alpha \alpha} .
\end{eqnarray}
Comparing Eq. \ref{eq:chapenskresult2} with Eq. \ref{eq:navstokes2} we
can read off the relations between the hydrodynamic fluctuations
and the random noise,
\begin{eqnarray}
&&\bar \sigma^f_{\alpha \beta}  =  
 - \frac{1}{1 - \gamma_s} \bar \sigma^r_{\alpha \beta} , \\
&&\sigma^f_{\alpha \alpha}  =  
 - \frac{1}{1 - \gamma_v} \sigma^r_{\alpha \alpha} .
\end{eqnarray}
Therefore, the random noise inserted at the microscopic (LB) level
must have the following covariances:
\begin{eqnarray} \label{eq:Rstresscorr1}
&&\frac{\left< \bar{\sigma}^r_{\alpha \beta} 
\bar{\sigma}^r_{\gamma \delta} \right>}
{\mu \rho c_s^4}
 =  \left( 1 - \gamma_s^2 \right)
\left[ \delta_{\alpha \gamma} \delta_{\beta \delta}
+ \delta_{\alpha \delta} \delta_{\beta \gamma}
- \frac{2}{3} \delta_{\alpha \beta} \delta_{\gamma \delta} \right] , \\
\label{eq:Rstresscorr2}
&&\frac{\left< \sigma^r_{\alpha \alpha} 
\sigma^r_{\beta \beta} \right>}
{\mu \rho c_s^4}
 = 
6\left( 1 - \gamma_v^2 \right) , \\
&&\label{eq:Rstresscorr3}
\left< \bar{\sigma}^r_{\alpha \beta} \sigma^r_{\gamma \gamma} \right>
 =  0 .
\end{eqnarray}

The random stress has a typical amplitude of $\sqrt{\mu \rho} c_s^2$
and is obtained from the second--order moment of the fluctuations
in $n_i^{neq}$. Therefore a typical fluctuation in the population
density is of order $\sqrt{\mu \rho}$. Combining this scaling with
Eqs. \ref{eq:orthonormal_trafo_1} and \ref{eq:orthonormal_trafo_2},
suggests dimensionless variables
\begin{eqnarray}
\label{eq:normalized_ni}
&&\hat{n}_i  =  \frac{n_i}{\sqrt{a^{c_i} \mu \rho}} , \\
\label{eq:normalized_mk}
&&\hat{m}_k  =  \frac{m_k}{\sqrt{w_k \mu \rho}} ,
\end{eqnarray}
which transform using the symmetric basis vectors defined in
Eq. \ref{eq:evectors_orthonormal},
\begin{eqnarray}
\label{eq:orthonormal_transformation_1}
&&\hat{m}_k  =  \sum_i \hat{e}_{ki} \hat{n}_i , \\
\label{eq:orthonormal_transformation_2}
&&\hat{n}_i  =  \sum_k \hat{e}_{ki} \hat{m}_k .
\end{eqnarray}
The stochastic collision operator can then be implemented independently
for each mode $m_k$,
\begin{equation} \label{eq:stochupdaterule1}
\hat{m}_k^{\star neq} = \gamma_k \hat{m}_k^{neq} + \varphi_k r_k ,
\end{equation}
where $r_k$ are independent Gaussian random variables with zero mean
and unit variance. The dimensionless constants $\varphi_k$ are
determined by expressing the random stresses $\sigma^r_{\alpha \beta}$
in terms of the $r_k$ and $\varphi_k$ and then calculating
the covariance matrix \cite{DSL}. For example,
$\sigma^{r}_{xy} = \sqrt{\mu \rho} c_s^2 \varphi_7 r_7$, while
$\sigma^{r}_{\alpha\alpha} = \sqrt{6 \mu \rho} c_s^2 \varphi_4 r_4$.
Comparison with
Eqs. \ref{eq:Rstresscorr1} -- \ref{eq:Rstresscorr3} shows that
the correct stress correlations are obtained for
\begin{equation} \label{eq:detailedbalance1}
\varphi_k = \left( 1 - \gamma_k^2 \right)^{1/2} .
\end{equation}

At the hydrodynamic scale, only fluctuations
in stress contribute to the time evolution of the momentum density
(Eq. \ref{eq:navstokes2}) so in principle it is sufficient to add random
fluctuations to the modes $m_4, \ldots, m_9$ only:
In the original derivation of the fluctuating LB equation
\cite{Lad94,Lad01}, the kinetic modes were projected out entirely,
\ie\ $\gamma_k = \varphi_k = 0$ for $k = 10, 11, \ldots, 18$.
More recently, Adhikari \etal\ \cite{Adh05} have argued
that the kinetic modes should be thermalized as well. They extended
Eq. \ref{eq:stochupdaterule1} to the kinetic modes
($k = 10, \ldots, 18$), with $\gamma_k = 0$
(as in Refs. \cite{Lad94,Lad01}) but with $\varphi_k = 1$, which then
satisfies Eq. \ref{eq:detailedbalance1}. It was demonstrated
numerically that this leads to more accurate fluctuations at short length
scales, but the theoretical justification remained somewhat
obscure. From the discussion so far, we can see that both procedures
give the same random stresses $\sigma^r_{\alpha \beta}$,
and hence are not different from the point of view of fluctuating
hydrodynamics. This has been clarified recently
\cite{DSL}, by analyzing the LB model in terms of
statistical mechanics. A purely microscopic approach was taken,
in which the stochastic collisions were viewed as a Monte Carlo
\cite{landbind} process. Knowledge of the probability distribution of
the LB variables $\bn$ then makes it possible
to check whether or not a given collision rule satisfies the 
condition of detailed balance. It can be shown \cite{DSL} that the kinetic
modes must be thermalized in order to satisfy detailed balance, in
agreement with the procedure proposed in Ref. \cite{Adh05}.
The theory will be outlined in Sec. \ref{sec:fluct_lbe:statmech}.

\subsection{Statistical mechanics of lattice-Boltzmann models}
\label{sec:fluct_lbe:statmech}

The starting point of the statistical mechanical development in
Ref. \cite{DSL} is the notion of a generalized lattice gas. We define
$\nu_i(\br,t)$ in Eq. \ref{eq:GLG} as the \emph{number} of particles
with velocity $\vec c_i$ at site $\vec r$ at time $t$. In contrast with
the standard LB model, $\nu_i$ is a (positive) integer;
in contrast with lattice-gas models, $\nu_i \gg 1$.
The state at a particular lattice site, $\bnu(\br,t)$, is modified
by the collision process, subject to the constraints of mass and
momentum conservation; the post--collision state, $\bnu^\star(\br,t)$, 
is then propagated to the neighboring sites (Eq. \ref{eq:GLG}).

Although a deterministic GLG collision operator would be difficult
to construct, we can nevertheless determine the distribution in a
homogeneous equilibrium state from the conservation laws alone. 
First we note that there is an entropy associated with each $\nu_i$,
\begin{equation} \label{eq:boltzmann_entropy_function}
S_i = - \left( \nu_i \ln \nu_i - \nu_i - \nu_i \ln {\bar \nu}_i 
+ {\bar \nu}_i \right) ,
\end{equation}
where ${\bar \nu}_i$ is the mean value of $\nu_i$ in the homogeneous
state. Each velocity direction $i$ at each lattice point has a
degeneracy $\exp(S_i)$, which can be derived from a Bernoulli process.
Particles are selected for the velocity direction $i$ with probability
$p_0$, with $p_0$ chosen so that on average a total of
$\bar \nu = N_r p_0$ particles will be selected from a reservoir of
$N_r$ particles.
Then the probability to select exactly $\nu$ particles
is given by the binomial distribution,
\begin{equation} \label{eq:binomial_distribution}
\nonumber
p(\nu) =
\frac{N_r !}{\nu ! \left(N_r - \nu \right)!} 
\left( \frac{\bar \nu}{N_r} \right)^\nu 
\left( 1 - \frac{\bar \nu}{N_r} \right)^{N_r - \nu} .
\end{equation}
Equation \ref{eq:boltzmann_entropy_function} results from calculating
$\ln p(\nu)$ in the limit of $N_r \to \infty$, at fixed $\bar \nu$.
Under the usual
assumption that in the equilibrium state the populations corresponding
to different lattice sites and different directions are uncorrelated,
the entropy per lattice site is $S(\bnu) = \sum_i S_i$.

The populations at a given lattice site are sampled from a probability
distribution proportional to $\exp \left[ S \left( \bnu \right) \right]$, 
but subject to the
constraints of fixed mass and momentum density, which characterize the
homogeneous state:
\begin{equation} \label{eq:Pdist}
P \left( \bnu \right) \propto
\exp \left[ S \left( \bnu \right) \right]
\delta \left( \mu \sum_i \nu_i - \rho \right)
\delta \left( \mu \sum_i \nu_i \vec c_i - \vec j \right) .
\end{equation}
Consistency with the formalism developed in the previous sections requires
\begin{equation}
\mu \bar \nu_i = \rho a^{c_i} .
\end{equation}
The equilibrium or mean populations for a given $\rho$ and $\bj$
are found by maximizing $P$ or, more conveniently, by maximizing 
$S$ and taking into account the conservation laws by
Lagrange multipliers:
\begin{eqnarray}
&&\frac{\partial S}{\partial \nu_i} + \lambda_\rho 
+ \vec \lambda_{\vec j} \cdot \vec c_i  =  0 ,
\label{eq:Smax}\\
&&\mu \sum_i \nu_i - \rho  =  0 , 
\label{eq:constraint_mass} \\
&&\mu \sum_i \nu_i \vec c_i - \vec j  =  0 . 
\label{eq:constraint_mom}
\end{eqnarray}
The exact solution is
\begin{equation} \label{eq:equildist_lagrange}
\nu_i^{eq} = {\bar \nu}_i \exp \left( \lambda_\rho + 
\vec \lambda_{\vec j} \cdot \vec c_i \right) ,
\end{equation}
where the Lagrange multipliers, $\lambda_\rho$ and $\vec \lambda_{\vec
j}$, are found from the constraint equations \ref{eq:constraint_mass}
and \ref{eq:constraint_mom}. Solving these equations in terms of a
power series in $u$, and disregarding terms of order $O(u^3)$, one
finds the standard equilibrium distribution given in
Eq. \ref{eq:original_equil_distrib}. This approach has been previously
proposed within the framework of the ``entropic lattice-Boltzmann''
method \cite{Kar99,Bog03}, which however, focuses exclusively on the
deterministic LB model.

Within the statistical--mechanical framework we have developed
for the LB model, the population densities $n_i$ fluctuate around
mean values determined by the hydrodynamic flow fields. Thus
the non-equilibrium distribution is sampled from $P( \bn^{neq} )$
which is Gaussian distributed about
the equilibrium \cite{DSL},
\begin{equation} \label{eq:Pdistneq}
P \left( \bn^{neq} \right)
\propto 
\exp \left( -  \sum_i \frac{ \left( n_i^{neq} \right)^2 }
                           { 2 \mu n_i^{eq} } \right)
\delta \left( \sum_i n_i^{neq} \right)
\delta \left( \sum_i \vec c_i \, n_i^{neq} \right) .
\end{equation}
The variance of the fluctuations is controlled by the mass density
$\mu$, associated with an LB particle.  A small number of particles
gives rise to large fluctuations and \emph{vice versa}. For simplicity
we will ignore the effects of flow on the variance of the
distribution, replacing $n_i^{eq}$ by its $u = 0$ value. This can be
justified at the macroscopic level by the Chapman-Enskog expansion
\cite{DSL}. Rewriting Eq. \ref{eq:Pdistneq} in terms of normalized
variables $\hat{n}_i$ (see Eq. \ref{eq:normalized_ni}), and
transforming to the normalized modes (see Eqs. \ref{eq:normalized_mk}
and \ref{eq:orthonormal_transformation_2}) eliminates the explicit
constraints,
\begin{equation}
P \left( \hat{\bm}^{neq} \right)
\propto \exp \left( - \frac{1}{2} \sum_{k > 3}
\hat{m}_k^{neq \, 2} \right) .
\end{equation}
Fluctuations only arise in the non--conserved modes, while the
conserved modes have no non-equilibrium contribution, \ie\
$m_k^{neq} = 0$ for $k \le 3$.

We now reinterpret the update rule Eq. \ref{eq:stochupdaterule1},
\begin{equation} 
\hat{m}_k^{\star neq} = \gamma_k \hat{m}_k^{neq} + \varphi_k r_k ,
\end{equation}
as a Monte Carlo move. The transition probability is then
identical to the probability of generating
the random variable $r_k$,
\begin{equation}
\omega \left( \hat{m}_k^{neq} \to \hat{m}_k^{\star neq} \right)
= \left( 2 \pi \varphi_k^2 \right)^{-1/2}
\exp \left[ - \frac{ \left(  \hat{m}_k^{\star neq} - \gamma_k
\hat{m}_k^{neq} \right)^2 }{2 \varphi_k^2} \right] ,
\end{equation}
and for the reverse transition the same formula holds, with the pre--
and post--collisional populations exchanged.
The condition of detailed balance \cite{landbind},
\begin{equation} \label{eq:detailedbalance2}
\frac{ 
\omega \left( \hat{m}_k^{neq} \to \hat{m}_k^{\star neq} \right)
}
{
\omega \left( \hat{m}_k^{\star neq} \to \hat{m}_k^{neq} \right)
}
=
\frac{ \exp \left( - (\hat{m}_k^{\star neq})^2 / 2 \right) }
     { \exp \left( - (\hat{m}^{neq})^2 / 2 \right) } ,
\end{equation}
then holds if and only if
\begin{equation} \label{eq:detailedbalance3}
\varphi_k = \left( 1 - \gamma_k^2 \right)^{1/2} ,
\end{equation}
as before (Eq. \ref{eq:detailedbalance1}).
The important point is that this relation, which in the previous
subsection was only proved for the stress modes, can now be shown to
hold for \emph{all} non--conserved modes. It is a necessary condition
for consistent sampling of the thermal fluctuations, not just
on the macroscopic hydrodynamic level (for which the stress modes alone
are sufficient), but also on the microscopic LB level itself.
Although assigning $\gamma_k = 0$ (and $\varphi_k = 1$) to all kinetic
modes is obvious and straightforward \cite{Adh05}, the present
analysis shows that this is not necessary. Other values of $\gamma_k$
and $\varphi_k$ are possible as well, so long as they satisfy
Eq. \ref{eq:detailedbalance3}, and specific values may be
desirable for a more accurate treatment of boundary conditions
\cite{Gin03,Chu07}.

\subsection{External forces}
\label{sec:fluct_lbe:forces}

An external force density $\vec f(\vec r, t)$ can be introduced
into the LB algorithm by an additional
collision operator $\Delta_i^{\prime \prime}$,
\begin{equation} \label{eq:coll_operator_plus_noise_plus_force}
\Delta_i = \sum_j {\cal L}_{ij} \left( n_j - n_j^{eq} \right) 
+ \Delta_i^{\prime \prime} .
\end{equation}
For simplicity, we only consider the deterministic case; the analysis
of the fluctuating part ($\Delta_i^\prime$) remains the same.
Application of the new collision operator should leave the mass
density unchanged, but increase the momentum density by $h \vec f$.
This implies the following conditions on the moments of
$\Delta_i^{\prime \prime}$:
\begin{eqnarray}
\label{eq:moment_zero_external_force}
&&\sum_i \Delta_i^{\prime \prime}  =  0 , \\
\label{eq:moment_one_external_force}
&&\sum_i \Delta_i^{\prime \prime} \vec c_i  =  h \vec f .
\end{eqnarray}
Consequently the definition of the fluid velocity is no longer unique:
one can legitimately choose any value for $\vec u$ between $\rho^{-1}
\left(\sum_i n_i \vec c_i \right)$ and $\rho^{-1} \left( \sum_i n_i
\vec c_i + h \vec f \right)$ (\ie\ between the pre--collisional and
post--collisional states). However, numerical \cite{Lad94} and
theoretical \cite{Gin94,Lad01,Guo02} analysis shows that the optimum value
is just the arithmetic mean of the pre-- and post--collisional
velocities. We \emph{define} the momentum density as
\begin{equation} \label{eq:special_choice_momentum_density}
\vec j = \sum_i n_i \vec c_i + \frac{h}{2} \vec f ,
\end{equation}
and the corresponding flow velocity as $\vec u = \vec j / \rho$.
Consistency with Eq. \ref{eq:equilmom} requires that we use
this value for $\vec u$ to calculate $n_i^{eq}$
(Eq. \ref{eq:original_equil_distrib}):
\begin{eqnarray}
&&\sum_i n_i^{eq} \vec c_i  =  \vec j , \\
&&\sum_i n_i^{neq} \vec c_i  =  - \frac{h}{2} \vec f .
\end{eqnarray}
In Ref. \cite{Lad01} the usual moment condition $\sum_i n_i^{neq} \vec
c_i = 0$ was maintained. In comparison with the present approach this
makes a small error of order $f^2$ to the distribution, which leads to
spurious terms in the Chapman-Enskog analysis. In contrast the present
approach leads to a clean result, entirely equivalent to the
force-free case. This may be of consequence when there are strongly
inhomogeneous forces, such as are considered in
Sec. \ref{sec:coupling}. However, it should also be noted that these
differences vanish in the low Reynolds number limit.

Since $\Delta_i^{\prime \prime}( \bn^{eq} ) \ne 0$,
the Chapman--Enskog expansion of $\Delta_i^{\prime \prime}$
starts at order $\varepsilon^1$ (\cf\ Eq. \ref{eq:eps0LBE}):
\begin{equation}
\Delta_i^{\prime \prime} = 
\varepsilon \Delta_i^{\prime \prime (1)} +
\varepsilon^2 \Delta_i^{\prime \prime (2)} + \ldots .
\end{equation}
Following the procedure of Sec. \ref{sec:fluct_lbe:chapman_enskog},
we take moments of Eqs. \ref{eq:eps1LBE} and \ref{eq:eps2LBEb}
and obtain similar equations for the mass and momentum density
(\cf\ Eqs. \ref{eq:chapenskresult1} and \ref{eq:chapenskresult2}):
\begin{eqnarray}
\label{eq:chapenskresult1plusf}
&&\partial_t \rho + \partial_\alpha j_\alpha  =  0 , \\
\label{eq:chapenskresult2plusf}
&&\partial_t j_\alpha + 
\partial_\beta \pi^{eq}_{\alpha \beta}
+ \frac{1}{2} \partial_\beta \left( 
\pi_{\alpha \beta}^{\star neq} + \pi^{neq}_{\alpha \beta}
\right)  =  f_\alpha .
\end{eqnarray}
However, the second moment leads to a force-dependent contribution to
the non-equilibrium momentum flux, which can be derived as before,
beginning with Eq. \ref{eq:eps1stress} and substituting the equilibrium
expressions for $\pi_{\alpha \beta}^{eq}$
(Eq. \ref{eq:equileulerstress}) and $\Phi_{\alpha \beta \gamma}^{eq}$
(Eq. \ref{eq:3rdmoment}). The time derivative of the momentum flux now
generates terms involving $\bu \vec {f} + \vec {f} \bu$ from the
momentum conservation equation on the $t_1$ time scale (\cf\
Eq. \ref{eq:newtonstress0}):
\begin{equation} \label{newtonstressplusf}
\pi^{\star neq}_{\alpha \beta} - \pi^{neq}_{\alpha \beta} =
h \rho c_s^2 \left( \partial_\alpha u_\beta 
                + \partial_\beta  u_\alpha \right) +
h \left( u_\alpha f_\beta + u_\beta f_\alpha \right) .
\end{equation}

Spurious terms proportional to $\bu \vec f$ can be eliminated
from Eq. \ref{eq:chapenskresult2plusf}, by including the second
moment of $\Delta_i^{\prime \prime}$,
\begin{equation}
\Sigma_{\alpha \beta} = \frac{1}{h}\sum_i \Delta_i^{\prime \prime}
c_{i \alpha} c_{i \beta} ,
\end{equation}
so that the stress update is now (\cf\ Eq. \ref{eq:stressrelax1}
and \ref{eq:stressrelax2})
\begin{equation} \label{eq:stress_updaterule_with_force}
\pi^{\star neq}_{\alpha \beta} = 
\gamma_s \bar \pi^{neq}_{\alpha \beta} +
\frac{1}{3} \gamma_v \pi^{neq}_{\gamma \gamma} \delta_{\alpha \beta} +
h \Sigma_{\alpha \beta} .
\end{equation}
Equations \ref{newtonstressplusf} and
\ref{eq:stress_updaterule_with_force}
form a linear system for $\pi^{neq}_{\alpha \beta}$ and
$\pi^{\star neq}_{\alpha \beta}$. Solving these equations as before
(Eqs. \ref{eq:pisolve1} -- \ref{eq:pisolve4}), and inserting
the result into Eq. \ref{eq:chapenskresult2plusf}, we
obtain a Newtonian stress with unchanged
values for the viscosities by choosing $\Sigma_{\alpha \beta}$
such that
\begin{equation} \label{eq:moment_two_external_force}
\Sigma_{\alpha \beta} = \frac{1}{2} \left( 1 + \gamma_s \right)
\left[ u_\alpha f_\beta + u_\beta f_\alpha
- \frac{2}{3} u_\gamma f_\gamma \delta_{\alpha \beta} \right]
+ \frac{1}{3} \left( 1 + \gamma_v \right) u_\gamma f_\gamma
\delta_{\alpha \beta} .
\end{equation}
The moment conditions expressed by Eqs. 
\ref{eq:moment_zero_external_force},
\ref{eq:moment_one_external_force} and
\ref{eq:moment_two_external_force}
are uniquely satisfied by the choice
\begin{equation}
\Delta_i^{\prime \prime} = a^{c_i} \left[
\frac{h}{c_s^2} f_\alpha c_{i \alpha}
+ \frac{h}{2 c_s^4} \Sigma_{\alpha \beta}
\left( c_{i \alpha} c_{i \beta} - c_s^2 \delta_{\alpha \beta}
\right) \right] ,
\end{equation}
where $\Delta_i^{\prime \prime}$ only affects the modes $m_1,
\ldots, m_9$. This result has been derived previously~\cite{Guo02}
within the context of the LBGK model; here we have presented the
derivation in the more general MRT framework.

%% file: coupling.tex
The fundamental algorithmic problem in soft matter simulations is
the coupling between the solid and fluid phases.
A key attraction of LB methods is the simplicity with which
geometrically complex boundaries can be incorporated. The first
correct implementation of a moving boundary condition was described in
the proceedings of a workshop on lattice-gas cellular
automata \cite{Lad89c}; a more accessible source is Ref. \cite{Lad90a}.
The idea was to modify the bounce-back rule for stationary surfaces such
that the steady-state distribution was consistent with the local
surface velocity. By constructing the boundary-node interactions along
the individual links, the viscous stress remains unchanged.  Subsequently,
we showed numerically that this algorithm gives accurate
hydrodynamic interactions between spherical particles suspended
in a lattice-gas fluid \cite{Lad90a}. Nevertheless, it quickly became
clear that the fluctuating LB model was a more useful computational tool,
for the reasons outlined in Sec. \ref{sec:flbe}. The LG algorithm
for the moving boundary condition carries over in a simple and direct
way to the LB method \cite{Lad94}.  A number of improvements to the
bounce-back boundary condition have been proposed over the years, and
we will summarize some of the more practical approaches
in Sec. \ref{sec:bounceback}.

More recently an entirely different approach has been proposed
\cite{ahlrichs:98,Ahl99} in which particles couple to the fluid
through a frictional drag. This method has the advantage of greatly
reducing the number of LB grid points in the simulation, at the cost
of a representation that is only correct in the far field. The method
has been applied to polymers \cite{Ahl99,ahlrichs:01,fyta} and to
suspended solid particles \cite{vladimir,vladimir2,vladimir3,%
  chatterji2005cmd,chatterji2007eph}. In the latter case the surface
is described by a number of sources distributed over the surface of
the particle. The distributed forces resemble the Immersed Boundary
(IB) methods \cite{Pes02}, which are common in finite-difference and
finite-element simulations; this connection has only been recognized
recently \cite{nash07}. We will summarize these developments and add
some new ideas and interpretation of the force coupling methods. In a
related development \cite{Fen03,Shi05}, conventional immersed boundary
methods are being used in conjunction with an LB fluid. However, the
coupling in this case is implicit, solving for the velocity of the
interface through a force balance, which corresponds to the high
friction limit of Refs. \cite{vladimir,vladimir2,vladimir3,%
  chatterji2005cmd,chatterji2007eph}. Here, we will only consider
inertial coupling, since the theory for thermal fluctuations has not
been worked out for the implicit schemes.

\subsection{Boundary conditions}\label{sec:bc}

\begin{figure}
\centering
\includegraphics[width=\smallwidth]{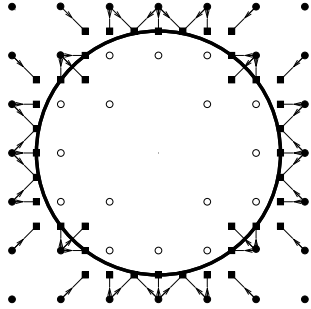}
\caption{Location of boundary nodes for a curved surface.
The velocities along links cutting the boundary
surface are indicated by arrows. The locations of the boundary
nodes are shown by solid squares, and the fluid nodes by solid
circles. The open circles indicate nodes in the solid adjacent to fluid
nodes.} \label{fig:bc}
\end{figure}

To simulate the hydrodynamic interactions between solid particles in
suspension, the lattice-Boltzmann model must be modified to
incorporate the boundary conditions imposed on the fluid by the solid
particles. The basic methodology is illustrated in Fig. \ref{fig:bc}.
The solid particles are defined by a boundary surface, which can be of
any size or shape; in Fig. \ref{fig:bc} it is a circle. When placed on
the lattice, the boundary surface cuts some of the links between
lattice nodes.  The fluid particles moving along these links interact
with the solid surface at boundary nodes placed halfway along the
links.  Thus a discrete representation of the particle surface is
obtained, which becomes more and more precise as the particle gets
larger.

In early work, the lattice nodes on either side of the boundary
surface were treated in an identical fashion \cite{Lad90a,Lad94}, so
that fluid filled the whole volume of space, both inside and outside
the solid particles. Although the fluid motion inside the particle
closely follows that of a rigid solid body \cite{Lad94a}, at short
times the inertial lag of the fluid is noticeable, and the
contribution of the interior fluid to the particle force and torque
reduces the stability of the particle velocity update. Today, most
simulations exclude interior fluid, although the implementation is
more difficult when the particles move.  The moving boundary condition
\cite{Lad94} without interior fluid \cite{Aid98} is then implemented
as follows. We take the set of fluid nodes $\br$ just outside the
particle surface, and for each node all the velocities $\bc_b$ such
that $\br + \bc_b h$ lies inside the particle surface. An example of a
set of boundary node velocities is shown by the arrows in
Fig. \ref{fig:bc}. Each of the corresponding population densities is
then updated according to a simple rule which takes into account the
motion of the particle surface \cite{Lad94};
\begin{equation}\label{eq:MBB}
n_{b^\prime}(\br, t+h) = 
n_b^\ast(\br, t) - \frac{2a^{c_b} \rho \bu_b \cdot \bc_b}{c_s^2},
\end{equation}
where $n_b^\ast(\br, t)$ is the post-collision distribution at
$(\br, t)$ in the direction $\bc_b$, and $\bc_{b^\prime} = -
\bc_b$. The local velocity of the particle surface,
\begin{equation}\label{NodeVel} 
\bu_b = \bU + {\bf \Omega \times} (\br_b - {\bf R}),
\end{equation}
is determined by the particle velocity $\bU$, angular velocity
$\bf \Omega$, and center of mass $\bf R$; $\br_b = \br +
\hh \bc_b$ is the location of the boundary node.

\begin{figure}
\centering
\includegraphics[width=\fullwidth]{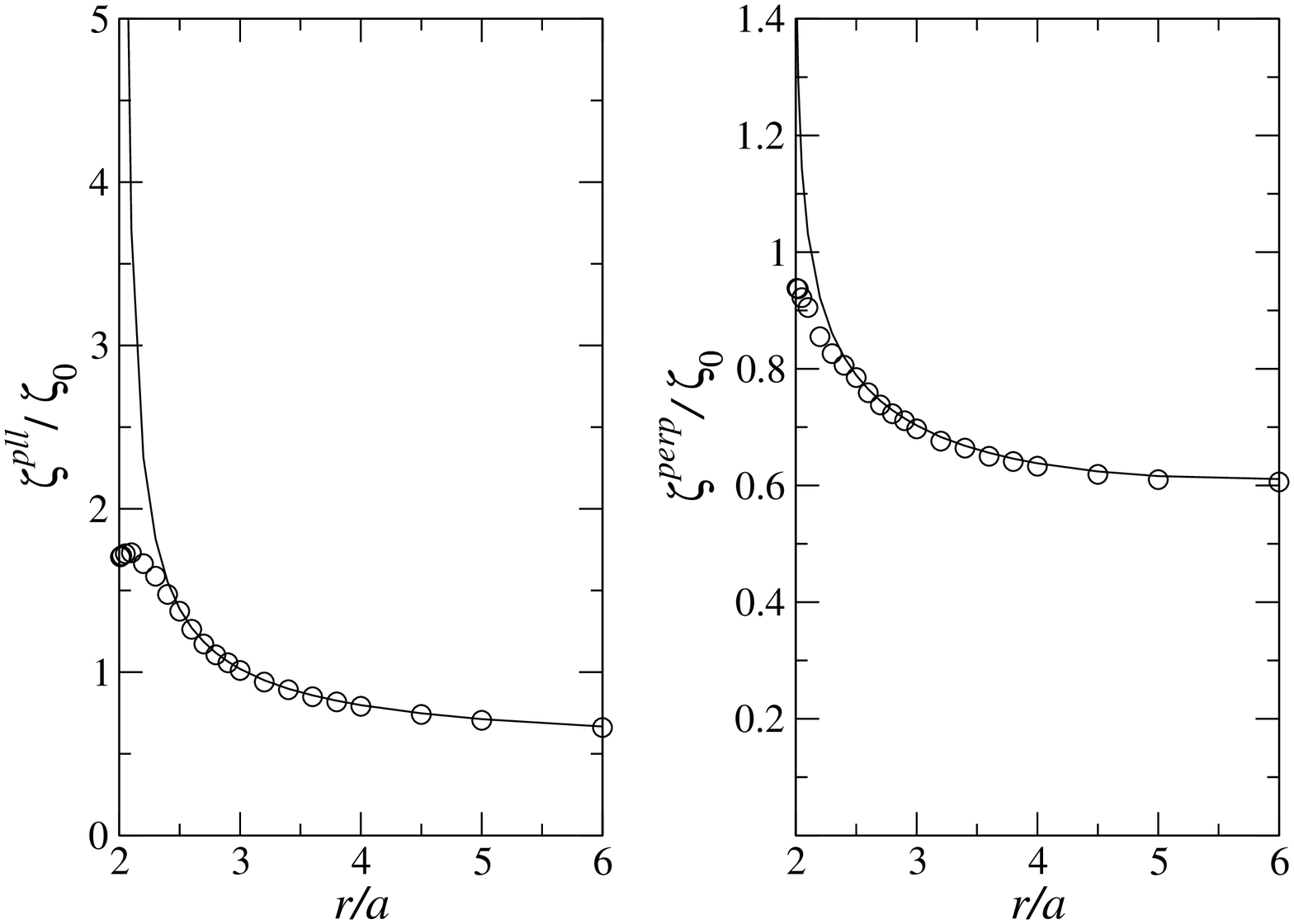}
\caption{Hydrodynamic interactions from LB simulations with
particles of radius $a = 2.5 b$. The solid symbols are the LB
friction coefficients, $\zeta^{pll}$ and $\zeta^{perp}$, for the
relative motion of two spheres along the line of centers (left) and
perpendicular to the line of centers (right).  Results are compared with
essentially exact results from a multipole code \cite{Cic00} in the
same geometry (solid lines).} \label{fig:hi-nolub}
\end{figure}

As a result of the boundary node updates, momentum is exchanged
locally between the fluid and the solid particle, but the combined
momentum of solid and fluid is conserved. The forces exerted at
the boundary nodes can be calculated from the momentum transferred
in Eq. \ref{eq:MBB}, and the particle forces and torques are then
obtained by summing over all the
boundary nodes associated with a particular particle. It can be
shown analytically that the force on a planar wall in a linear
shear flow is exact \cite{Lad94}, and several numerical examples
of lattice-Boltzmann simulations of hydrodynamic interactions are
given in Ref. \cite{Lad94a}. Figure \ref{fig:hi-nolub} illustrates
the accuracy that can be achieved with the MRT collision operator
described in Sec. \ref{sec:fluct_lbe:determ_coll}. Even with
small particles, only $5b$ in diameter, the hydrodynamic interactions are
within 1\% of a precise numerical solution \cite{Cic00}, down to
separations between the particle surfaces $s = r - 2a \sim b$, 
corresponding to $s \sim 0.4a$, where $a$ is the sphere radius.
Periodic boundaries with a unit cell size $L = 12a$ were used,
with the pair inclined at $30^\circ$ to a symmetry axis; other
geometries give a very similar level of agreement. We emphasize that
there are no adjustable parameters in these comparisons. In particular,
in contrast to previous work \cite{Lad94a,Ngu02}, there is no need to
calibrate the particle radius; the correct particle size arises
automatically when the eigenvalues of the kinetic modes of the MRT model
have the appropriate dependence on the shear viscosity \cite{Chu07}.

To understand the physics of the moving boundary condition, one
can imagine an ensemble of particles, moving at constant speed
$\bc_b$, impinging on a massive wall oriented perpendicular to the
particle motion. The wall itself is  moving with velocity $\bu_b
\ll \bc_b$. The velocity of the particles after collision with the
wall is $-\bc_b + 2\bu_b$ and the force exerted on the wall is
proportional to $\bc_b - \bu_b$.  Since the velocities in the
lattice-Boltzmann model are discrete, the desired boundary
condition cannot be implemented directly, but we can instead
modify the density of returning particles so that the momentum
transferred to the wall is the same as in the continuous velocity
case. It can be seen that this implementation of the no-slip
boundary condition leads to a small mass transfer across a moving
solid-fluid interface. This is physically correct and arises from
the discrete motion of the solid surface.  Thus during a time step
$h$ the fluid is flowing continuously, while the solid particle
is fixed in space. If the fluid cannot flow across the surface
there will be large artificial pressure gradients, arising from
the compression and expansion of fluid near the surface.  For a
uniformly moving particle, it is straightforward to show that the
mass transfer across the surface in a time step $h$
(Eq. \ref{eq:MBB}) is exactly recovered when the particle moves to
its new position. For example, each fluid node adjacent to a
planar wall has 5 links intersecting the wall. If the wall is
advancing into the fluid with a velocity $\bU$, then the mass flux
across the interface (from Eq. \ref{eq:MBB}) is $\rho \bU$. Apart
from small compressibility effects, this is exactly the rate at
which fluid mass is absorbed by the moving wall. For sliding
motion, Eq. \ref{eq:MBB} correctly predicts no net mass transfer
across the interface.

\subsection{Particle motion}

An explicit update of the particle velocity
\begin{equation}\label{eq:explicit}
\bU(t + h) = \bU(t) + \frac{h}{m} \bF(t)
\end{equation}
has been found to be unstable \cite{Lad94a} unless the particle
radius is large or the particle mass density is much higher than
the surrounding fluid. In previous work \cite{Lad94a} the
instability was reduced, but not eliminated, by averaging the
forces and torques over two successive time steps. Subsequently,
an implicit update of the particle velocity was
proposed \cite{Low95} as a means of ensuring stability. 
A generalized version of that idea, which can be adapted
to situations where two particles are in near contact, was developed in
Ref. \cite{Ngu02}. Here we sketch an elaboration of this idea, which is
consistent with a Trotter decomposition of the Liouville
operator
\cite{ricci_ciccotti,bussiparrinello,thalmann,DeF06,Ser06}. 
We will only consider the update of the position and linear
velocity explicitly; the extension to rotational motion is
straightforward \cite{Ngu02}.

The equations of motion for the suspended particles are written as
\begin{eqnarray}\label{eq:EOM-R}
{\dot \bR}_i & = & \bU_i, \\ \label{eq:EOM-U}
m{\dot \bU}_i & = &
\bF_i^h(\bR_i,\bU_i) + \bF_i^c(\bR^N),
\end{eqnarray}
where we have separated the forces into a hydrodynamic component
$\bF_i^h$, which depends on the particle position and velocity,
and a conservative force $\bF_i^c$, which depends on the positions
of all particles. The hydrodynamic force depends on the fluid degrees
of freedom as well, but these remain unchanged during
the particle update and need not be considered 
as dynamical variables here.

A second-order Trotter decomposition 
\cite{ricci_ciccotti,bussiparrinello,thalmann,DeF06,Ser06}
breaks the
update of a single time step into three independent components: a
half time step update of the positions at constant velocity,
a full time step update of the velocities with fixed positions,
and a further half time step update of the positions using the new
velocities:
\begin{eqnarray}\label{eq:EOMD-1}
&&\bR_i(t+\hh) = \bR_i(t) + \frac{h}{2} \bU_i(t), \\ \label{eq:EOMD-2}
&&{\dot \bU}_i = \frac{1}{m}\left[\bF_i^h(\bR_i(t+\hh),\bU_i) +
  \bF_i^c(\bR^N(t+\hh)) \right], \\ \label{eq:EOMD-3}
&&\bR_i(t+h) = \bR_i(t+\hh) + \frac{h}{2} \bU_i(t+h).
\end{eqnarray}
In the absence of velocity-dependent forces this is just the Verlet scheme,
but the solid-fluid boundary conditions (Eq. \ref{eq:MBB}) introduce
a hydrodynamic force that depends linearly on the particle velocity
\cite{Low95,Ngu02},
\begin{equation}\label{eq:force}
\bF_i^h(\bR_i,\bU_i) = \bF_0^h(\bR_i) -
\bzeta(\bR_i) \cdot \bU_i .
\end{equation}
The velocity independent force is calculated at the half-time step
\begin{eqnarray}\label{eq:force0}
\bF_0^h(\bR_i(t + \hh)) = \frac{b^3}{h} \sum_{b} 2
n^{\ast}_b (\br, t) \bc_b ,
\end{eqnarray}
where the sum is over all the boundary nodes, $b$, describing the
particle surface and $\bc_b$ points towards the particle center.
The location of the boundary nodes is determined by the particle
coordinates $\bR_i(t + \hh)$, which should be evaluated at the half time
step as indicated. The post-collision populations, $n_b^\star$, are
calculated at time $t$ but arrive at the boundary nodes at the
half time step also. The components of the matrix
\begin{equation}\label{eq:zeta}
\bzeta(\bR_i(t + \hh)) = \frac{2 \rho b^3}{c_s^2 h} \sum_{b}
a^{c_b} \bc_b \bc_b 
\end{equation}
are high-frequency friction coefficients, which describe the
instantaneous force on a particle in response to a sudden change
in velocity. Complete expressions, including rotation,
are given in Ref. \cite{Ngu02}.

The LB fluid and the solid particles are coupled by an
instantaneous momentum transfer at the half time step, which is
therefore presumed to be conservative: 
\begin{eqnarray}\label{eq:EOMU-1}
&&\bU_i(t+\hh) = \bU_i(t) + \frac{h}{2m} \bF_i^c(\bR^N(t+\hh)), \\
 \label{eq:EOMU-2}
&&\bU_i^{\star}(t+\hh) = \bU_i(t+\hh) +
      \frac{h}{m}\bF_i^h(\bR_i(t+\hh),{\tilde \bU}_i(t+\hh)), \\
\label{eq:EOMU-3}
&&\bU_i(t+h) = \bU_i^\star(t+\hh) +
      \frac{h}{2m} \bF_i^c(\bR^N(t+\hh)).
\end{eqnarray}
However it is not entirely clear what velocity should be used
in Eq. \ref{eq:EOMU-2}: among the possibilities discussed in
Ref. \cite{Ngu02} are an explicit update
${\tilde \bU}_i(t+\hh) = \bU_i(t+\hh)$, an implicit update
${\tilde \bU}_i(t+\hh) = \bU_i^\star(t+\hh)$, and a semi-implicit update
${\tilde \bU}_i(t+\hh) = [\bU_i(t+\hh) + \bU_i^\star(t+\hh)]/2$.
It has been pointed out \cite{Ust05} that, even for a Langevin equation
with constant friction, there are deviations in the temperature
for finite values of $h$. However the semi-implicit scheme satisfies the
FDT exactly for constant friction.
Here we will consider a different model for the velocity,
assuming that the hydrodynamic force is distributed over the time step.
For simplicity we consider a single component of the velocity,
\begin{equation} \label{eq:EOMU}
m{\dot U} = -\zeta U + F_0^h + F^c,
\end{equation}
where $\zeta$, $F_0^h$ and $F^c$ are constant in this context.
The solution of Eq. \ref{eq:EOMU} over a time interval $h$ is
\begin{equation} \label{eq:EOMUsol}
U(t+h) = U(t)\exp(-\alpha) +
\frac{F_0^h + F^c}{\zeta}\left[1 - \exp(-\alpha)\right],
\end{equation}
where $\alpha = \zeta h/m$ is the dimensionless time step.
Equation \ref{eq:EOMUsol} is stable for all values of $\alpha$, 
satisfies the FDT exactly, and, when there is no conservative force,
encompasses previous algorithms as limiting cases. Both explicit
\cite{Lad94} and implicit \cite{Low95,Ngu02} schemes are consistent with an
expansion of Eq. \ref{eq:EOMUsol} to linear order in $\alpha$,
while the semi-implicit method \cite{Ngu02}
can be derived from a second order expansion in $\alpha$. The steady-state
velocity $U(t+h) = U(t)$ satisfies the force balance $F^h+F^c = 0$
exactly.  This new result may lead to more
accurate integration of the particle positions and velocities
in the large $\alpha$ limit.

To complete the update, the velocity ${\tilde \bU}_i(t+\hh)$ is
needed to calculate the momentum transfer to the fluid (Eq. \ref{eq:MBB}).
An explicit update \cite{Lad94} can be done in a single pass since
${\tilde \bU}_i(t+\hh) = \bU_i(t)$ is already known, but
an implicit or semi-implicit update 
requires two passes through the boundary nodes. The first pass is
used to calculate $\bF_0^h$ so that Eq. 
\ref{eq:EOMU-2} can be solved
for ${\tilde \bU}_i(t+\hh)$ \cite{Ngu02}. This velocity is used
to update the population densities in a second sweep through the
boundary nodes. In the present case we calculate ${\tilde \bU}_i(t+\hh)$
by enforcing consistency between the sequential update 
Eqs. \ref{eq:EOMU-1}--\ref{eq:EOMU-3} and Eq. \ref{eq:EOMUsol}:
\begin{equation}
\alpha {\tilde U}_i(t+\hh) = U_i(t)\left[1 - \exp(-\alpha)\right] +
\left(F_0^h + F^c\right)
\left(\frac{\alpha}{\zeta} - \frac{1 - \exp(-\alpha)}{\zeta}\right).
\end{equation}
This ensures overall momentum conservation as before.

When there are short-range conservative forces between the particles,
the LB time step is frequently too large for accurate integration
of the interparticle forces. The LB time step can be divided into an
integer number of substeps, but the question then arises as to how
to best incorporate the hydrodynamic forces, since $F^h$ should,
in principle, be calculated at $t+\hh$.
One possibility is to use the fact that $\bzeta$ varies slowly with
particle position and accept the small error associated with using
$R(t)$ rather than $R(t+\hh)$. Or this solution could be used as
a predictor step for calculating $R(t+\hh)$, which could
then be followed by one or more corrector cycles with increasingly
more accurate calculations of $\bzeta(t+\hh)$. The corrector
cycles should not involve a significant overhead since the boundary
nodes would be largely the same from one cycle to the next, and
the time-consuming lookup of LB population densities could be avoided.

Although the momentum exchange between fluid and solid
occurs instantaneously at the half time step, in calculating
${\tilde \bU}_i(t+\hh)$ we have made the assumption that the
hydrodynamic force is distributed over the time step.
We actually attempted to derive an update for the velocity assuming that the
hydrodynamic force acts over a very small fraction of the time step, but
this has not lead to a sensible result as yet.
It is not entirely clear if the assumption that the hydrodynamic force
acts over the whole time step is valid, and does not,
for example, produce an artificial dissipation. To resolve this
question will require a detailed analysis of the fully coupled system,
along the lines given in Sec. \ref{sec:forcecoupling} for the
simpler case of frictional coupling.  A similar analysis for
solid-fluid boundary conditions is an open area for further research.

\subsection{Surfaces near contact}

When two particle surfaces come within one grid spacing, fluid nodes
are excluded from regions between the solid surfaces, leading to a
loss of mass conservation. This happens because boundary updates at
each link cause mass transfer across the solid-fluid interface, which
is necessary to accommodate the discrete motion of the particle
surface (see Sec. \ref{sec:bc}). The total mass transfer in or out of
an isolated particle is
\begin{equation}
\Delta M = -\frac{2 h^3 \rho}{c_s^2} \left[\bU \cdot \sum_b
a^{c_b} \bc_b \right] = 0,
\end{equation}
regardless of the particle's size or shape.

Although the sum $\sum_{b} a^{c_b} \bc_b$
is zero for any closed surface \cite{Ngu02}, when two
particles are close to contact some of the boundary nodes are
missing and the surfaces are no longer closed. In this case
$\Delta M \not= 0$ and mass conservation is no longer ensured. Two
particles that remain in close proximity never reach a steady
state, no matter how slowly they move, since fluid is constantly
being added or removed, depending on the particle positions and
velocities. If the two particles move as a rigid body mass
conservation is restored, but in general this is not the case.
The accumulation or loss of mass occurs slowly, and in many
dynamical simulations it fluctuates with changing particle
configuration but shows no long-term drift. However, we typically enforce
mass conservation, particle-by-particle, by redistributing the
excess mass among the boundary nodes \cite{Ngu02}. An alternative idea
is to ensure that there is always at least one fluid node in
the gap between the particle surfaces. In dense suspensions it would
be quite inaccurate to insert an artificial excluded volume around the
particles, but a more promising idea is to cut back the particle
surfaces along planes perpendicular to the line of centers
\cite{Gha95,Din03}.
It remains to be seen if the hydrodynamic interactions retain the 
level of accuracy shown in Fig. \ref{fig:hi-nolub}.

When two particles are in near contact, the fluid flow in the gap
cannot be resolved. For particle sizes that are typically used in
multiparticle simulations ($a < 5 b$), the lubrication breakdown
in the calculation of the hydrodynamic interaction occurs at gaps
of the order of $0.1a$. However, in some flows, notably the shearing of
a dense suspension, qualitatively important physics occurs at
smaller separations, typically down to $0.01a$. Here we outline a
method to implement lubrication corrections into a
lattice-Boltzmann simulation.

\begin{figure}
\centering
\includegraphics[width=\fullwidth]{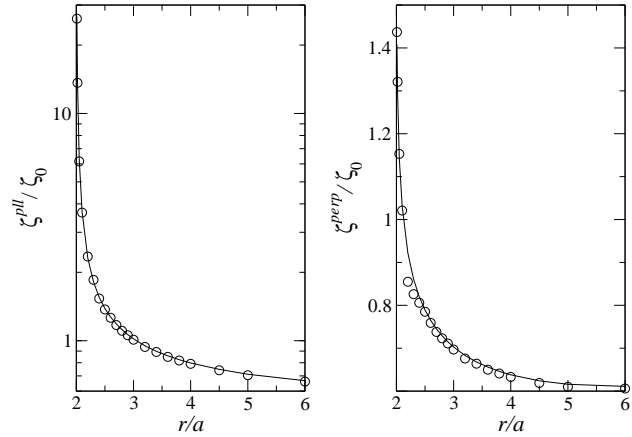}
\caption{Hydrodynamic interactions including lubrication, with
particles of radius $a = 2.5 b$. The solid symbols are the LB
friction coefficients, $\zeta^{pll}$ and $\zeta^{perp}$, for the
relative motion of two spheres along the line of centers (left) and
perpendicular to the line of centers (right). Results are compared with
essentially exact results from a multipole code \cite{Cic00} in the
same geometry (solid lines).} \label{fig:hi-lub}
\end{figure}

For particles close to contact, the lubrication force, torque, and
stresslet can be calculated from a sum of pairwise-additive
contributions \cite{Bra88b}, and if we consider only singular
terms, they can be calculated from the particle velocities
alone \cite{Cla89}.  In lattice-Boltzmann
simulations \cite{Lad01,Lad97b} the calculated forces follow
the Stokes flow results down to a fixed separation, approximately equal to
the grid spacing $b$, and remain roughly constant thereafter
(see Fig. \ref{fig:hi-nolub}). The simplest lubrication correction is
to take the difference between the lubrication force at a gap $s$ and the
force at some cut off distance $s_c$; \ie\
\begin{eqnarray}
&&\bF^l = - 6 \pi \eta \frac{a_1^2 a_2^2}{(a_1+a_2)^2}
\left(\frac{1}{s} - \frac{1}{s_c} \right)
\bU_{12} \cdot \hat{\bf R}_{12} \hat{\bf R}_{12} ~~~ s < s_c \\
&&\bF^l = 0 ~~~ s > s_c,
\end{eqnarray}
where $\bU_{12} = \bU_1 - \bU_2$, $s = |{\bf R_{\rm 12}}| - a_1 -
a_2$ is the gap between the two surfaces, and the unit vector
$\hat{\bf R}_{\rm 12} = \bf R_{\rm 12} / | R_{\rm 12} |$.
Numerical tests of this procedure for the older 10-moment LB model
are reported in Ref. \cite{Ngu02}. Results for the MRT model are shown
in Fig. \ref{fig:hi-lub}, using a cutoff distance $s_c = 1.1b$ for the
parallel component and $s_c = 0.7b$ for the perpendicular component.
Even this simple form for the correction gives an accurate description of
the lubrication regime, with the largest deviations occurring near the patch
points. A more accurate correction can be obtained by calibrating
each distance separately as in Stokesian dynamics and related methods
\cite{Bra88,Cic00}.

\subsection{Improvements to the bounce-back boundary condition}
\label{sec:bounceback}

The bounce-back boundary condition remains the most popular choice for
simulations of suspensions, because of its robustness and simplicity.
The results in Figs. \ref{fig:hi-nolub} and \ref{fig:hi-lub} show
that accurate hydrodynamic interactions, within 1--2\%, can be
achieved with quite small particles, particularly when combined with
the MRT model. The reason that bounce-back
works so well, despite being only first-order accurate, is
that the errors in the momentum transfer tend to cancel when averaged
over a random sampling of boundary node positions \cite{Chu07}.
In fact bounce-back can sometimes be more accurate than interpolation,
where the errors, though locally smaller, do not cancel.

\begin{figure}
\centering
\includegraphics[width=\fullwidth]{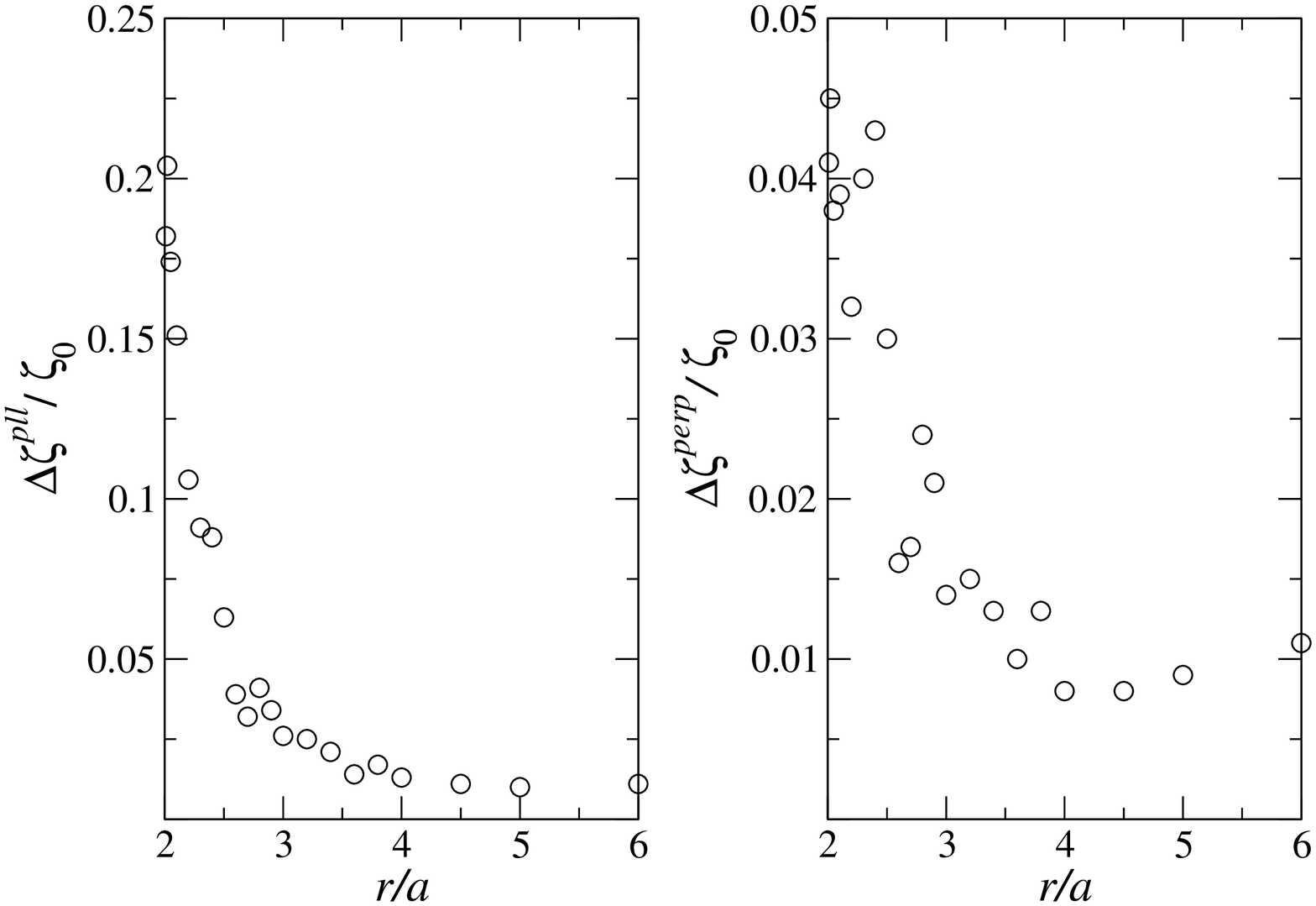}
\caption{Variation in friction coefficients with grid location for
particles of radius $a = 2.5 b$. The solid symbols are the variance in
the LB friction coefficients, $\Delta\zeta^{pll}$ and $\Delta\zeta^{perp}$
for the relative motion of two spheres along the line
of centers (left) and perpendicular to the line of centers (right).
Results are a single standard deviation in the friction coefficients,
calculated from 100 independent positions with respect to the grid.}
\label{fig:hi-var}
\end{figure}

The most important deficiency of the bounce-back algorithm is the
dependence of the force on the position of the nodes with respect to
the grid. The results in Figs. \ref{fig:hi-nolub} and \ref{fig:hi-lub}
are averages over 100 independent configurations, in which the
relative positions of the particles are the same but the pair is displaced
randomly with respect to the underlying lattice. However, the variance
in the friction for randomly sampled grid locations is small, typically
of the order of 1\%, as can be seen in Fig. \ref{fig:hi-var}. Nevertheless
there is a much larger fluctuation in the force around the particle
surface, which is particularly problematic if the particles are
deformable \cite{Bux05,Ale06}.  Thus while the bounce-back method
is quite accurate on average, locally the errors can be large.
A detailed analytical and numerical
critique of the bounce-back algorithm can be found in Ref. \cite{Jun05b},
together with an analysis of several of the modifications mentioned below.
The most practical higher-order boundary conditions are adapted from
the link bounce-back algorithm outlined in Sec. \ref{sec:bc}.

More sophisticated boundary conditions have been developed using
finite-volume methods \cite{Che98a,Che98b} and interpolation
\cite{Bou01,Fil98,Mei99}.  A simple, physically motivated
interpolation scheme has been proposed \cite{Bou01,Lal03}, which both
improves the accuracy of the bounce-back rule and is unconditionally
stable for all boundary positions; the scheme has both linear and
quadratic versions. A more general framework for this class of
interpolation schemes has been extensively analyzed in a comprehensive
and seminal paper \cite{Gin03}; the Multi-Reflection Rule proposed in
Ref. \cite{Gin03} is the most accurate boundary condition yet
discovered for lattice-Boltzmann methods. However interpolation
requires additional fluid nodes in the gap between adjacent particle
surfaces. The bounce-back rule requires only one grid point between
the surfaces but linear interpolation requires at least two grid
points, while quadratic interpolation and multi-reflection require
three. Recently, it was proposed that only the equilibrium
distribution needs to be interpolated \cite{Chu07}. Although this is
more complex to implement than linear interpolation, it has the
advantage that the velocity distribution at the boundary surface may
be used to provide an additional interpolation point. In this way the
span of fluid nodes can be reduced to that of the bounce-back rule,
while obtaining second-order accuracy in the flow field. In
conjunction with an appropriate choice of collision operator
\cite{Gin03}, the location of the hydrodynamic boundary remains
independent of fluid viscosity, unlike the linear and quadratic
interpolations \cite{Bou01}. For viscous fluids, where $\gamma_s > 0$,
the equilibrium interpolation rule is more accurate than either linear
or quadratic interpolation \cite{Chu07}.

\subsection{Force coupling}
\label{sec:forcecoupling}

The force-coupling algorithm \cite{ahlrichs:98,Ahl99} starts from a
system of mass points which are coupled \emph{dissipatively} to the
hydrodynamic continuum. The particles are specified by positions
$\vec r_i$, momenta $\vec p_i$, masses $m_i$, and phenomenological
friction coefficients $\Gamma_i$. They interact via a potential $V
\left( \vec r^N \right)$, giving rise to conservative forces
$\vec F_i^c = - \partial V / \partial \vec r_i$. The fluid exerts a
drag force on each particle based on the difference between the
particle velocity and the fluid velocity $\bu_i = \bu(\br_i)$,
\begin{equation}\label{eq:fdrag}
\bF_i^d = -\Gamma_i \left(\frac{\bp_i}{m_i} - \bu_i\right).
\end{equation}
Momentum conservation requires that an equal and opposite force be
applied to the fluid. Both discrete and continuous degrees of freedom
are subject to Langevin noise in order to balance the frictional and
viscous losses, and thereby keep the temperature constant. The
algorithm can be applied to any Navier--Stokes solver, not just to
LB models. For this reason, we will discuss the coupling within a
(continuum) Navier--Stokes framework, with a general equation of state
$p(\rho)$. We use the abbreviations $\eta_{\alpha \beta \gamma
\delta}$ for the viscosity tensor (Eq. \ref{eq:macrostressviscous2}),
and
\begin{equation}
\pi^{E}_{\alpha \beta} = p \delta_{\alpha \beta} + \rho u_\alpha u_\beta
\end{equation}
for the inviscid momentum flux or Euler stress
(Eq. \ref{eq:equileulerstress}).
Since the fluid equations are solved on a grid, whereas the particles
move continuously, it will be necessary to \emph{interpolate} the flow
field from nearby lattice sites to the particle positions \cite{Ahl99}.

The addition of a point force into the continuum fluid equations
introduces a singularity into the flow field, which causes both
mathematical and numerical difficulties. On the other hand, the flow
field around a finite-sized particle can be generated
by a distributed force located entirely inside the
particle \cite{Max01,Lom02}. This flow field is everywhere finite,
and the force density appearing in Eq. \ref{eq:NS} can be written as
\begin{equation}\label{eq:ffluid}
{\bf f}(\br) = - \sum_i \bF_i^d \Delta(\br,\br_i),
\end{equation}
where $\Delta(\br)$ is a weight function with compact support and
normalization
\begin{equation}\label{eq:Dnorm}
\int d^3\br \Delta(\br,\br_i) = 1.
\end{equation}
Compact support limits the set of nodes $\br$ to those in the vicinity
of $\br_i$ and ensures that the interactions remain local. Away from
solid boundaries, translational invariance requires that
\begin{equation}\label{eq:Dinvar}
\Delta(\br,\br_i) = \Delta(\br - \br_i).
\end{equation}

The function $\Delta(\br,\br_i)$ plays a dual role, both interpolating
the fluid velocity field to the particle position,
\begin{equation} \label{eq:ufluid}
{\bf u}(\br_i) = \int d^3\br\, \Delta(\br,\br_i) \bu(\br),
\end{equation}
and then redistributing the reactive force to the fluid, according to
Eq. \ref{eq:ffluid}. Within the context of polymer
simulations, $\Delta$ has been regarded as an interpolating function
for point forces, but it can equally well be regarded as a model for a
specific distributed force, contained within an envelope described by
$\Delta(\br - \br_i)$. The flow fields from a point force and a
distributed force are similar at large distances from the source, but
the distributed source has the advantage that the near field also
corresponds to a physical system, namely finite-size particles. We
will adopt the distributed source interpretation both here and in
Sec. \ref{sec:volumeforce}.

The Langevin equations of motion for the coupled fluid--particle system
are:
\begin{eqnarray}
&&\frac{d}{dt} \vec r_i = \frac{1}{m_i} \vec p_i , \\
&&\frac{d}{dt} \vec p_i = \vec F_i^c + \bF_i^d + \bF_i^f , \\
&&\partial_t \rho + \partial_\alpha j_\alpha = 0 , \\
&&\partial_t j_\alpha + \partial_\beta \pi^{E}_{\alpha \beta} =
\partial_\beta \eta_{\alpha \beta \gamma \delta} 
\partial_\gamma u_\delta + f^h_\alpha + 
\partial_\beta \sigma^f_{\alpha \beta},
\end{eqnarray}
where the force density applied to the fluid includes both 
dissipative and random forces,
\begin{equation}\label{eq:ftotal}
{\bf f}^h(\br) = -\sum_i \left(\bF_i^d + \bF_i^f\right) \Delta(\br,\br_i).
\end{equation}
The Langevin noises for the particles and fluid,
$\vec F_i^f$ and $\sigma^f_{\alpha \beta}$, satisfy the usual moment
conditions:
\begin{eqnarray}
&&\left< F^f_{i \alpha} \right> = 0, \\
&&\left< \sigma^f_{\alpha \beta} \right> = 0 , \\
&&\left< F^f_{i \alpha} \left( t \right)
       F^f_{j \beta} \left( t^\prime \right) \right> =
2 k_B T \Gamma_i \delta_{ij} \delta_{\alpha \beta}
\delta \left( t - t^\prime \right) , \\
&&\left< \sigma^f_{\alpha \beta} \left( \vec r, t \right)
\sigma^f_{\gamma \delta} \left(\vec r^\prime, t^\prime \right) \right> =
2 k_B T \eta_{\alpha \beta \gamma \delta}
\delta \left( \vec r - \vec r^\prime \right)
\delta \left( t - t^\prime \right) .
\end{eqnarray}
By construction, this coupling is local, and conserves both the total
mass
\begin{equation}
M = \sum_i m_i + \int d^3 \vec r \rho
\end{equation}
and the total momentum
\begin{equation}
\vec P = \sum_i \vec p_i + \int d^3 \vec r \rho \vec u .
\end{equation}
Galilean invariance is ensured by using velocity differences in the
coupling between particles and fluid (Eq. \ref{eq:fdrag}). A finer
point is that the interpolation uses $\vec u$ (and not $\vec j$),
so that the velocity field enters strictly linearly. We will prove that
the fluctuation--dissipation theorem (FDT) holds for this
coupled system, proceeding in three steps that successively
take more terms into account.

Let us look first at the conservative system
where the particles and fluid are completely decoupled:
\begin{eqnarray}
&&\frac{d}{dt} \vec r_i = \frac{1}{m_i} \vec p_i , \\
&&\frac{d}{dt} \vec p_i = \vec F_i^c , \\
&&\partial_t \rho + \partial_\alpha j_\alpha = 0 , \\
&&\partial_t j_\alpha + \partial_\beta \pi^{E}_{\alpha \beta} = 0 .
\end{eqnarray}
The dynamics of the particles \emph{and} the Euler fluid can be described
within the framework of Hamiltonian mechanics \cite{zakharov:97}.
The Hamiltonians for the particles
\begin{equation}
{\cal H}_p = \sum_i \frac{\vec p_i^2}{2 m_i} + V,
\end{equation}
and fluid,
\begin{equation}
{\cal H}_f = \int d^3 \vec r \left( \frac{1}{2} \rho \vec u^2
+ \epsilon \left( \rho \right) \right),
\end{equation}
are conserved quantities, with $\epsilon (\rho)$ the
internal energy density of the fluid.

As a second step, we consider a system where particles and fluid are
still decoupled, but are subject to dissipation and noise:
\begin{eqnarray}
&&\frac{d}{dt} \vec r_i = \frac{1}{m_i} \vec p_i , \\
&&\frac{d}{dt} \vec p_i = \vec F_i^c
             - \frac{\Gamma_i}{m_i} \vec p_i + \vec F_i^f , \\
&&\partial_t \rho + \partial_\alpha j_\alpha = 0 , \\
&&\partial_t j_\alpha + \partial_\beta \pi^{E}_{\alpha \beta} =
\partial_\beta \eta_{\alpha \beta \gamma \delta} 
\partial_\gamma u_\delta + \partial_\beta \sigma^f_{\alpha \beta} .
\end{eqnarray}
These Langevin equations are known to satisfy the
FDT \cite{risken,chandra,bdlangevin,Lan59,foxuhlenbeck}. We 
briefly sketch the formalism used for the proof, since this will be
needed for the final step in which we consider the fully coupled
system.

Instead of describing the stochastic dynamics via a Langevin equation,
we use the Fokker--Planck equation, which is the
evolution equation for the probability density in phase space. For an
$N$-particle system,
\begin{equation}
\partial_t 
P \left( \vec r^N, \vec p^N \right)
= \left( {\cal L}_1 + {\cal L}_2 + {\cal L}_3 \right)
P \left( \vec r^N, \vec p^N \right) ,
\end{equation}
where $\br^N, \bp^N$ denote the positions and momenta of all $N$
particles. The three operators ${\cal L}_1$, ${\cal L}_2$, and ${\cal L}_3$
describe the Hamiltonian, frictional, and stochastic part of the
dynamics; they can be found via the Kramers--Moyal
expansion \cite{risken,bdlangevin}:
\begin{eqnarray}
&&{\cal L}_1 = - \sum_i \left( 
\frac{\partial}{\partial \vec r_i} \cdot \frac{\vec p_i}{m_i} +
\frac{\partial}{\partial \vec p_i} \cdot \vec F_i^c
\right) , \\
&&{\cal L}_2 = \sum_i \frac{\Gamma_i}{m_i}
\frac{\partial}{\partial \vec p_i} \cdot \vec p_i , \\
&&{\cal L}_3 = k_B T \sum_i \Gamma_i 
\frac{\partial^2}{\partial \vec p_i^2} .
\end{eqnarray}
The FDT holds if the Boltzmann factor,
$\exp \left(-{\cal H}_p/k_B T \right)$,
is a stationary solution of the Fokker--Planck equation. Using
$\beta = (k_B T)^{-1}$ to define the inverse temperature, we have
\begin{equation}
{\cal L}_1 \exp \left( - \beta {\cal H}_p \right) = 0
\end{equation}
as a direct consequence of energy conservation in
Hamiltonian systems. Furthermore, the relation
\begin{equation}
\left( {\cal L}_2 + {\cal L}_3 \right) 
\exp \left( - \beta {\cal H}_p \right) = 0
\end{equation}
can be shown by direct differentiation.

For the fluid system, the phase space comprises all possible configurations
of the fields $\rho (\vec r)$, $\vec j (\vec r)$, which we denote as
$\left[ \rho \right], \left[ \vec j \right]$. The Fokker-Planck equation
for the fluid degrees of freedom can be written as
\begin{equation}
\partial_t 
P \left( \left[ \rho \right], \left[ \vec j \right] \right)
= \left( {\cal L}_4 + {\cal L}_5 + {\cal L}_6 \right)
P \left( \left[ \rho \right], \left[ \vec j \right] \right) ,
\end{equation}
where ${\cal L}_4$, ${\cal L}_5$, and ${\cal L}_6$ describe the
Hamiltonian, viscous, and stochastic components,
\begin{eqnarray}
&&{\cal L}_4 =
\int d^3 \vec r \left(
\frac{\delta}{\delta \rho} \partial_\alpha j_\alpha +
\frac{\delta}{\delta j_\alpha} \partial_\beta \pi^{E}_{\alpha \beta}
\right) ,
\\
&&{\cal L}_5 =
- \eta_{\alpha \beta \gamma \delta} 
\int d^3 \vec r \frac{\delta}{\delta j_\alpha}
\partial_\beta \partial_\gamma u_\delta ,
\\
&&{\cal L}_6 = k_B T \eta_{\alpha \beta \gamma \delta}
\int d^3 \vec r \int d^3 \vec r^\prime
\frac{\delta}{\delta j_\alpha (\vec r)}
\frac{\delta}{\delta j_\gamma (\vec r^\prime) } \left[
\frac{\partial}{\partial r_\beta}
\frac{\partial}{\partial r_\delta^\prime}
\delta \left( \vec r - \vec r^\prime \right) \right] ,
\end{eqnarray}
and $\delta \ldots / \delta \ldots$ represents
a functional derivative \cite{functional_derivative}.
Replacing $\partial / \partial r_\delta^\prime$ in the last equation
with $- \partial / \partial r_\delta$ enables
integration over $\vec r^\prime$:
\begin{equation}
{\cal L}_6 =  - k_B T \eta_{\alpha \beta \gamma \delta}
\int d^3 \vec r \frac{\delta}{\delta j_\alpha}
\partial_\beta \partial_\gamma
\frac{\delta}{\delta j_\delta} ,
\end{equation}
where we have exploited the symmetry of the viscosity tensor with
respect to the indexes $\gamma$ and $\delta$.
Functional differentiation of the Boltzmann factor with respect to
$\bj$,
\begin{equation}\label{eq:fdiff}
\frac{\delta}{\delta j_\delta} \exp \left( - \beta {\cal H}_f \right)
= - \beta u_\delta \exp \left( - \beta {\cal H}_f \right) ,
\end{equation}
then shows that
\begin{equation}
\left( {\cal L}_5 + {\cal L}_6 \right)
\exp \left( - \beta {\cal H}_f \right) = 0 .
\end{equation}
Finally, the relation
\begin{equation}
{\cal L}_4 \exp \left( - \beta {\cal H}_f \right) = 0
\end{equation}
follows from energy conservation in Hamiltonian dynamics.

We now turn to the coupled system, with
Hamiltonian ${\cal H} = {\cal H}_p + {\cal H}_f$. The
Fokker--Planck equation in the full phase space reads
\begin{equation}
\partial_t 
P \left( \vec r^N, \vec p^N,
\left[ \rho \right], \left[ \vec j \right] \right)
= \left( \sum_{i = 1}^{10} {\cal L}_i \right)
P \left( \vec r^N, \vec p^N,
\left[ \rho \right], \left[ \vec j \right] \right) ,
\end{equation}
with the operators ${\cal L}_7-{\cal L}_{10}$ to describe the coupling
in the equations of motion:
\begin{eqnarray}
&&{\cal L}_7 = - \sum_i \Gamma_i \frac{\partial}{\partial p_{i \alpha}}
             u_{i \alpha} , \\
&&{\cal L}_8  =  - \sum_i \Gamma_i \int d^3 \vec r
\Delta \left( \vec r, \vec r_i \right)
\frac{\delta}{\delta j_\alpha (\vec r)}
\left( \frac{1}{m_i} p_{i \alpha} - u_{i \alpha} \right) , \\
&&{\cal L}_9 = k_B T \sum_i \Gamma_i
\int d^3 \vec r \Delta \left( \vec r, \vec r_i \right)
\frac{\delta}{\delta j_\alpha (\vec r)}
\int d^3 \vec r^\prime \Delta \left( \vec r^\prime, \vec r_i \right)
\frac{\delta}{\delta j_\alpha (\vec r^\prime)} , \\
&&{\cal L}_{10} =
 - 2 k_B T \sum_i \Gamma_i
\frac{\partial}{\partial p_{i \alpha}}
\int d^3 \vec r \Delta \left( \vec r, \vec r_i \right)
\frac{\delta}{\delta j_\alpha (\vec r)} .
\end{eqnarray}
The coupling of the fluid velocity to the particles is
described by ${\cal L}_7$, while ${\cal L}_8$ describes the
drag on the fluid. The stochastic contributions
include fluid--fluid correlations via ${\cal L}_9$,
and fluid--particle cross correlations via ${\cal L}_{10}$.

It should be noted that $\vec u_i$, being the result of the
interpolation, depends on the fields $\left[\rho\right]$ and
$\left[\vec j\right]$, so that $\delta u_i / \delta j_{\alpha} (\vec
r)$ is nonzero. Hence, the corresponding operators in ${\cal L}_8$ do
not commute. Explicit functional differentiation shows that
Eq. \ref{eq:fdiff} holds in an analogous way for the interpolated
velocity,
\begin{equation}
\int d^3 \vec r \Delta \left( \vec r, \vec r_i \right)
\frac{\delta}{\delta j_\alpha (\vec r)}
\exp \left( - \beta {\cal H} \right) =
- \beta u_{i \alpha} 
\exp \left( - \beta {\cal H} \right) .
\end{equation}
Explicit calculations, as outlined for the uncoupled system,
show that
\begin{equation}
\left( {\cal L}_7 + {\cal L}_8 + {\cal L}_9 + {\cal L}_{10} \right)
\exp \left( - \beta {\cal H} \right) = 0 ,
\end{equation}
which implies the exact FDT for the fully coupled system,
\begin{equation}
\left( \sum_{i = 1}^{10} {\cal L}_i \right)
\exp \left( - \beta {\cal H} \right) = 0 .
\end{equation}
An important consequence of this result is that a consistent simulation
needs to thermalize \emph{both} fluid and particle degrees of freedom;
any other choice will violate the FDT.

The Langevin integrator for the particles is constructed in much
the same way as the velocity Verlet algorithm for Molecular Dynamics.
Although a Langevin analog to the Verlet algorithm has been known for
some time \cite{vangunsteren_berendsen}, straightforward derivations
have become available only recently, by applying
operator--splitting techniques that were previously limited to
Hamiltonian systems \cite{mclachlan}.  We employ a
second--order integrator \cite{ricci_ciccotti,bussiparrinello,thalmann},
which reduces to the velocity
Verlet scheme in the limit of vanishing friction. Higher--order
schemes are known \cite{forbert_chin}, but
they are considerably more complicated. Specifically, we
approximately integrate the equations
\begin{eqnarray} \label{eq:posode}
&&\frac{d}{dt} \vec r_i = \frac{1}{m_i} \vec p_i , \\ \label{eq:momode}
&&\frac{d}{dt} \vec p_i = \vec F_i^c + \vec F_i^d + \vec F_i^f ,
\end{eqnarray}
assuming that the fluid velocity $\vec u_i$ is constant over a time
step $h$. This corresponds to the Fokker--Planck equation for the particles
\begin{equation}
\partial_t
P \left( \vec r^N, \vec p^N, t \right)
= \left( {\cal L}_r + {\cal L}_p \right)
P \left( \vec r^N, \vec p^N, t \right)
\end{equation}
with
\begin{eqnarray}
&&{\cal L}_r = - \sum_i
\frac{\partial}{\partial \vec r_i} \cdot \frac{\vec p_i}{m_i} , \\
&&{\cal L}_p = - \sum_i
\frac{\partial}{\partial \vec p_i} \cdot \vec F_i^c
+ \sum_i \Gamma_i
\frac{\partial}{\partial \vec p_i} \cdot \left(
\frac{\vec p_i}{m_i} - \vec u_i \right)
+ k_B T \sum_i \Gamma_i
\frac{\partial^2}{\partial \vec p_i^2} .
\end{eqnarray}
The formal solution
\begin{equation}
P \left( \vec r^N, \vec p^N, h \right)
=
\exp \left[
\left( {\cal L}_r + {\cal L}_p \right)
h \right]
P \left( \vec r^N, \vec p^N, 0 \right)
\end{equation}
is approximated by a second-order Trotter decomposition,
\begin{equation} \label{eq:pTrotter}
\exp \left[
\left( {\cal L}_r + {\cal L}_p \right)
h \right] 
=
\exp \left( {\cal L}_r h / 2 \right)
\exp \left( {\cal L}_p h  \right)
\exp \left( {\cal L}_r h / 2 \right) + O(h^3) .
\end{equation}
The operator splitting implies the following algorithm: a half time step
update of the coordinates with \emph{constant momenta},
\begin{equation}
\vec r_i (t + h/2) = \vec r_i (t) + \frac{h}{2} \frac{\vec p_i (t)}{m_i} ,
\end{equation}
followed by a full time step momentum update,
with \emph{constant coordinates},
\begin{equation} \label{eq:langevin_momentum_update}
\vec p_i (t + h) = C_i^{(1)}(h) \vec p_i (t)
+ C_i^{(2)}(h) \left[ \vec F_i^c + \Gamma_i \vec u_i \right]
+ C_i^{(3)}(h) \vec \theta_i  ,
\end{equation}
and finally another half time step coordinate update,
with \emph{constant momenta},
\begin{equation}
\vec r_i (t + h) = \vec r_i (t + h/2) + \frac{h}{2} 
\frac{\vec p_i (t + h)}{m_i} .
\end{equation}
The coefficients in Eq. \ref{eq:langevin_momentum_update} are
\begin{eqnarray}
&& C_i^{(1)}(h) = \exp \left( - \frac{\Gamma_i}{m_i} h \right) , \\
&& C_i^{(2)}(h) = \frac{m_i}{\Gamma_i} \left[ 1 -
              \exp \left( - \frac{\Gamma_i}{m_i} h \right) \right] , \\
&& C_i^{(3)}(h) = \sqrt{ m_i k_B T \left[ 1
- \exp \left( - 2 \frac{\Gamma_i}{m_i} h \right) \right] } ,
\end{eqnarray}
where $\theta_{i \alpha}$ are Gaussian random variables with zero mean
and unit variance. Equation \ref{eq:langevin_momentum_update} is the
\emph{exact} solution of the momentum update, since
Eq. \ref{eq:momode} is a linear Langevin equation describing Brownian
motion in a harmonic potential \cite{chandra}. Thus the only
source of error in integrating the particle motion is
derived from the Trotter decomposition itself,
Eq. \ref{eq:pTrotter}. Nevertheless, it is important to limit the
range of random numbers, to ensure that very large steps do not
occasionally occur. It is therefore both desirable and more efficient
to use distributions of random variates with finite range, which
reproduce the Gaussian moments up to a certain order; in the present
case, $O(h^2)$ accuracy requires that moments up to the fourth
cumulant are correct. One possible choice is
\begin{equation}
P(\theta) = \frac{2}{3} \delta \left( \theta \right)
        + \frac{1}{6} \delta \left( \theta - \sqrt{3} \right)
        + \frac{1}{6} \delta \left( \theta + \sqrt{3} \right) .
\end{equation}

After the update of the particle momenta, the fluid force density,
$\vec f^h(\br)$ (Eq. \ref{eq:ftotal}), is distributed to the
surrounding lattice sites. After one or more particle updates, the
fluid variables are updated for a single LB time step, with the
external forces being taken into account in the collision
operator. This scheme is probably only first order accurate overall.
It remains a challenge for the future to develop a unified framework
to describe the fully coupled system and analyze its convergence
properties; the algorithm could then perhaps be improved in a
systematic fashion.

The input friction coefficient is \emph{not} the same
as the long--time friction coefficient, which is measured by the
ratio of the particle velocity to the applied force
\cite{Ahl99}.
Consider an isolated particle with ``bare'' (or input) friction
coefficient $\Gamma$ dragged through the fluid by a constant
force $\vec F$, resulting in a steady particle velocity $\vec U$.
The force balance requires that
\begin{equation} \label{eq:dragexperiment_particle_equation}
{\vec F} = -{\vec F}^d = \Gamma (\vec U - \vec u_0) ,
\end{equation}
where $\vec u_0$ is the fluid velocity at the particle center,
$\vec r_0$,
\begin{equation}\label{eq:u0}
\vec u_0 = \int d^3 \vec r \Delta(\br, \vec r_0) \vec u(\vec r) .
\end{equation}
In the absence of thermal fluctuations, the deterministic fluid
velocity field can be calculated in the Stokes flow approximation
using the Green's function appropriate to the boundary
conditions \cite{Hap86},
\begin{equation} \label{eq:u}
\vec u (\vec r) = \int d^3 \vec r^\prime
\tensor T (\vec r,\vec r^\prime) \cdot {\vec F}
\Delta (\vec r^\prime, \vec r_0) .
\end{equation}
In an unbounded fluid, the Green's function reduces to the
Oseen tensor \cite{Hap86},
$\tensor T(\br,\br^\prime)=\tensor O(\br-\br^\prime)$, with
\begin{equation}
\tensor O(\br) = \frac{1}{8 \pi \eta r}
\left(\tensor 1 + \frac{\br\br}{r^2} \right),
\end{equation}
but Green's functions are also known for periodic boundary conditions
\cite{Has59} and planar boundaries \cite{Bla71, Lir76} as well.
Combining Eqs. \ref{eq:dragexperiment_particle_equation}, \ref{eq:u0}
and \ref{eq:u},
\begin{equation} \label{eq:combined_u_dragexperiment}
\vec u_0 = \tensor{T}^{av} \cdot \vec F =
\Gamma \tensor{T}^{av} \cdot \left( \vec U - \vec u_0 \right),
\end{equation}
where $\tensor{T}^{av}$ is the Green's function averaged over
the particle envelope,
\begin{equation} \label{eq:Tbar}
\tensor{T}^{av} (\br_0) = \int d^3 \vec r \int d^3 \vec r^\prime
\Delta (\vec r, \vec r_0) \Delta (\vec r^\prime, \vec r_0)
\tensor T (\vec r , \vec r^\prime) .
\end{equation}

In a system with translational invariance, $\tensor{T}^{av}$ is
independent of $\br_0$ and proportional to the unit tensor by
symmetry. Then from Eq. \ref{eq:combined_u_dragexperiment}
\begin{equation}
\vec F = \frac{\Gamma}{1 + \mu_\infty\Gamma} \vec U,
\end{equation}
where $\mu_\infty = T^{av}_{\alpha \alpha}/3$ accounts for the
renormalization of $\Gamma$. If the size of the particle $a$ is
associated with the range of interaction of $\Delta$, then dimensional
analysis of Eq. \ref{eq:Tbar} suggests that $\mu_\infty \sim (\eta
a)^{-1}$. The effective friction coefficient, defined via $\vec F =
\Gamma_{eff} \vec U$, is therefore diminished by the flow field
induced by the applied force,
\begin{equation} \label{eq:effsize}
\frac{1}{\Gamma_{eff}} = \frac{1}{\Gamma} + \mu_\infty .
\end{equation}
This relation has been verified numerically by extensive computer
experiments \cite{Ahl99,Ust05}. In the strong coupling limit,
$\Gamma \rightarrow \infty$, the effective friction saturates to the
limiting value $\Gamma_{eff} = \mu_\infty^{-1}$. This suggests assigning
an effective radius to the interpolating function,
\begin{equation} \label{eq:aeff0}
\frac{1}{a} = 6 \pi \eta \mu_\infty.
\end{equation}
However, for smaller values of the input friction, the effective particle
size is given by
\begin{equation} \label{eq:aeff}
\frac{1}{a} = 6 \pi \eta \left( \frac{1}{\Gamma} + \mu_\infty \right) =
\frac{1}{a_0} + \frac{1}{gb},
\end{equation}
where $a_0 = \Gamma/6 \pi \eta$ is the input particle radius and
$gb = (6 \pi \eta \mu_\infty)^{-1}$ depends on the 
interpolating function.  The interesting physical parameter is
$\Gamma_{eff}$, which describes the long--time behavior of the
coupled system. Thus, to approach the continuum limit, one should keep
$\Gamma_{eff}$ constant as the lattice spacing is decreased,
and change the bare coupling $\Gamma$ as necessary.

It is not yet known how $\tensor{T}^{av} (\br_0)$ behaves in the
vicinity of a solid boundary, when translational invariance is broken.
The compact support of the weighting
function $\Delta$ suggests that $\tensor{T}^{av}$ is a local correction
and therefore largely independent of macroscopic boundary conditions.
Numerical simulations with periodic boundary conditions
(Sec. \ref{sec:volumeforce}) show that $g$ is
independent of system size and fluid viscosity. The weak
system-size dependence reported in Ref. \cite{Ust05} is
entirely accounted for by the difference between the periodic
Green's function \cite{Has59} and the Oseen tensor. Thus in a periodic
unit cell of length $L$, Eq. \ref{eq:aeff} requires a correction
of order $1/L$ \cite{Lad90b},
\begin{equation} \label{eq:aeffcorr}
\frac{1}{gb} = \frac{1}{a} + \frac{2.84}{L} - \frac{1}{a_0} .
\end{equation}

\subsection{Interpolating functions}
\label{sec:volumeforce}

In this section we consider translationally invariant interpolating
functions, $\Delta(\br, \br_0)$ $=$ $\Delta(\br - \br_0)$ in more
detail. For a discrete lattice, the interpolation procedure (\cf\
Eq. \ref{eq:ufluid}) reads
\begin{equation} \label{eq:ufluid_discrete}
{\bf u}(\br_0) = \sum_\br \Delta(\br - \br_0) \bu(\br),
\end{equation}
where $\br_0$ is the position of the particle, and $\br$ denotes the
lattice sites. The normalization condition
\begin{equation}\label{eq:DnormDiscrete}
\sum_\br \Delta(\br - \br_0) = 1
\end{equation}
must hold for all particle positions, $\br_0$, in order for the
conservation laws to be satisfied exactly. We will assume the analysis
of Sec. \ref{sec:forcecoupling} carries over to the discrete system
with no more than second-order discretization errors. Proof of this
assumption remains for future work; here we numerically compare
various choices of $\Delta$.

Previous work, incorporating force coupling into spectral codes, used
an isotropic Gaussian distribution for $\Delta(r)$ \cite{Max01}, but
this is not commensurate with cubic lattice symmetry. Thus we take
$\Delta$ as a product of one-dimensional functions \cite{Pes02}
\begin{equation}\label{eq:Delta}
\Delta(x,y,z) = \phi\left(\frac{x}{b}\right)
        \phi\left(\frac{y}{b}\right) \phi\left(\frac{z}{b}\right) .
\end{equation}
We first consider the two-point linear interpolating polynomial
\begin{equation}\label{eq:2point}
\phi_2(u) = \left\{
\begin{array}{lcl}
1 - \mod{u} & \hspace{3em} & \mod{u} \le 1 , \\ 
0           & \hspace{3em} & \mod{u} \ge 1 ,
\end{array} \right.
\end{equation}
which satisfies the following moment conditions for all real-valued
$u$ and integer $j$:
\begin{eqnarray}\label{eq:C1}
\sum_j \phi(u-j) &=& 1 , \\ \label{eq:C2}
\sum_j j \phi(u-j) &=& u .
\end{eqnarray}
It exactly conserves momentum and angular momentum of the particle
and fluid. However, $\phi_2$ violates the condition
\begin{equation}\label{eq:C3}
\sum_j \phi^2 (u-j) = C,
\end{equation}
where $C$ is a constant, independent of $u$. The importance of this
condition is explained in Ref. \cite{Pes02}.

The conditions in Eqs. \ref{eq:C1}-\ref{eq:C3} can be satisfied by a
three-point interpolation function,
\begin{equation}\label{eq:3point}
\phi_3(u) = \left\{
\begin{array}{lcl}
\frac{1}{3} \left(1 + \sqrt{1-3u^2}\right)
&\hspace{3em}&
0 \le \mod{u} \le \frac{1}{2}\\
\frac{1}{6} \left(5-3\mod{u} - \sqrt{-2+6\mod{u}-3u^2}\right)
&\hspace{3em}&
\frac{1}{2} \le \mod{u} \le \frac{3}{2}\\
0
&\hspace{3em}&
\frac{3}{2} \le \mod{u}.
\end{array} \right.
\end{equation}
An important property that emerges from $\phi_3$ is that the first
derivative, $\phi_3^\prime(u)$, is continuous throughout the whole domain
of $u$. This ensures that the velocity field varies smoothly across
the grid, with a continuous spatial derivative $\nabla \bu$. By
contrast, linear interpolation leads to a continuous velocity but
discontinuous derivatives.

In order to test the various interpolation schemes we have determined
the settling velocity of a single particle in a periodic unit cell.
A small force was applied to the particle and a compensating pressure
gradient (or uniform force density) was added to the fluid, so that the
net force on the system was zero. The steady-state particle velocity was
determined, without allowing the particle to move on the grid
\cite{Lad94a}. This procedure is valid in Stokes flow, where an
arbitrarily small velocity may be assumed, and gives a clean result for
the variation in settling velocity with grid position.
We used Eq. \ref{eq:aeffcorr} to convert the measured mobility to a
single parameter $g$, which does not depend on system size ($L$) or
fluid viscosity ($\eta$).  The mobility of the particle is reduced
by the periodic images \cite{Has59}, but the correction in
Eq. \ref{eq:aeffcorr} accounts for the effects of the periodic boundaries
quantitatively \cite{Lad90a,Lad94a,Lad00}. In simulations of polymer
solutions an average of $g$ over all grid positions is used.

\begin{figure}
\centering
\includegraphics[clip,width=\fullwidth]{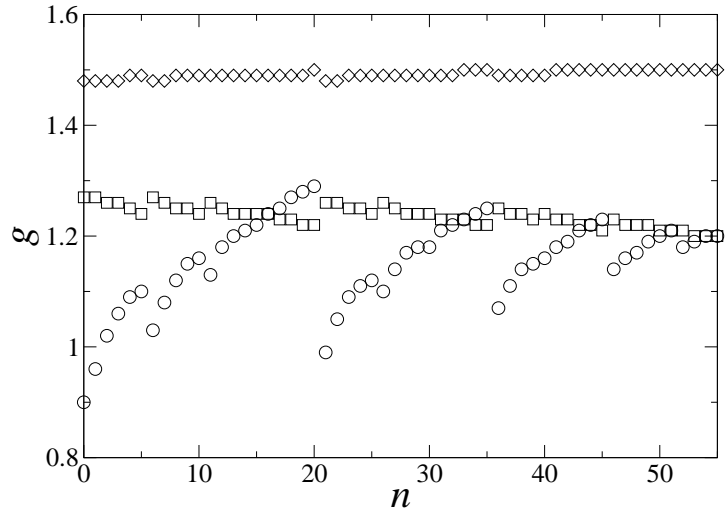}
\caption{Variation in settling velocity with grid location. The effective
hydrodynamic radius was determined from $U = F/6 \pi \eta a$ and
converted to $g$ using Eq. \ref{eq:aeffcorr}; $g$ was found to be
independent of $\eta$ and $L$, as expected. Results are shown at 56
different grid positions (labeled by the index $n$), systematically
varying the coordinates in steps of $0.1b$. Particles with input
radius $a_0 = b$ were placed at
coordinates $(ib/10,jb/10,kb/10)$, with $0 \le i \le j \le k \le 5$.
Results are shown for the two-point (circles), three-point (squares) and
four-point (diamonds) interpolation schemes.}
\label{fig:one}
\end{figure}

The smoother velocity field derived from three-point interpolation
means that the particle velocity is less
dependent on the underlying grid than with linear interpolation, as
can be seen in Fig. \ref{fig:one}. Here we show the variation in
effective particle size, as determined by the parameter $g$
(Eq. \ref{eq:aeffcorr}), for linear (two point) interpolation
(Eq. \ref{eq:2point}), three-point interpolation (Eq.
\ref{eq:3point}), and the four-point interpolation,
\begin{equation}\label{eq:4point}
\phi_4(u) = \left\{
\begin{array}{lcl}
\frac{1}{8} \left(3-2\mod{u} + \sqrt{1+4\mod{u}-4u^2}\right)
&\hspace{3em}&0 \le \mod{u} \le 1\\
\frac{1}{8} \left(5-2\mod{u} - \sqrt{-7+12\mod{u}-4u^2}\right)
&\hspace{3em}&1 \le \mod{u} \le 2\\
0
&\hspace{3em}& 2 \le \mod{u},
\end{array} \right.
\end{equation}
that is commonly used in immersed boundary methods \cite{Pes02}. As
was recently noticed \cite{nash07}, the four-point interpolation leads
to a much smaller variation in effective friction than linear
interpolation. The parameter $g$ varies with grid position by up to
20\% in the case of linear interpolation, but by less
than 1\% with four-point interpolation.  On the other hand, linear
interpolation requires an envelope volume of 8 grid points, while the
four-point scheme requires 64 grid points. Away from the
strong-coupling limit, the grid dependence of the settling velocity is
reduced since there is a non-negligible input mobility apart from the
lattice contribution.

Four-point interpolation is only necessary when using
centered-difference approximations to the velocity and pressure fields
\cite{Pes02}, a situation that does not arise in LB simulations.  We
see that the three-point scheme is also much smoother than linear
interpolation, with about a 3\% variation in $g$. It is not as smooth as
the four-point interpolation, but requires only 27 grid points.
Furthermore, the smaller span of nodes means that the boundary surface
is more tightly localized, and in fact the hydrodynamic interactions
obtained with three-point interpolations are just as accurate as those
obtained with four-point interpolation, as shown below.

\begin{figure}
\centering
\includegraphics[width=\fullwidth]{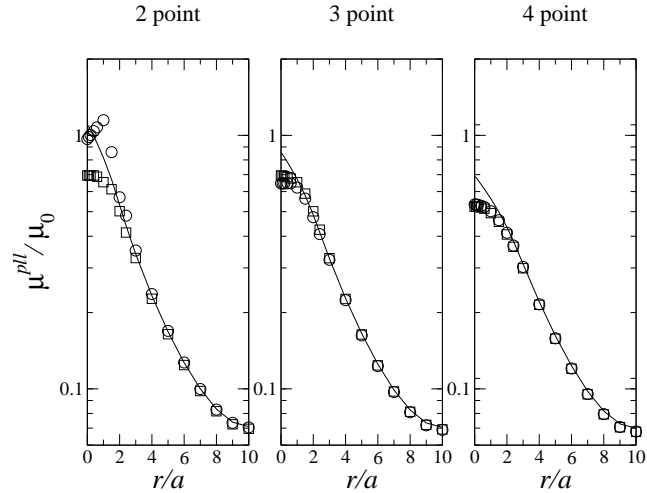}
\caption{Hydrodynamic interactions between a pair of spherical
particles using the force-coupling method. The normalized mobility
$\mu_{12}^{pll}/\mu_0$ is plotted at various separations $r$.
We fit the effective hydrodynamic radius of each
interpolating function to numerical solutions of the Stokes equations
with a \emph{uniform} force density on the sphere. These results correspond
to a Rotne-Prager description of the hydrodynamic mobility and
do not include lubrication. We show results at two different grid
locations, $(0,0,0)$ (circles) and $(b/2,b/2,b/2)$ (squares), for
two-point (left), three-point (center) and four-point (right)
interpolation schemes.}
\label{fig:two}
\end{figure}

An important test of the force coupling scheme is its ability to
represent the hydrodynamic interactions between two spherical
particles. As an example of the accuracy of the different
interpolation schemes, in Fig. \ref{fig:two} we show the hydrodynamic
interactions between two spheres moving along the line of centers. A
small force is applied to sphere 1, in the direction of the vector
between 1 and 2, and the velocity of sphere 2 is determined. From this
we can calculate the hydrodynamic mobility $\mu_{12}^{pll}$. The
results are normalized by the mobility of the isolated sphere
$\mu_0 = (6 \pi \eta a)^{-1}$. Results were obtained for the two-point,
three-point, and 4-point interpolation schemes, using a source
particle placed on a grid point $(0,0,0)$ in one instance and in the
center of the voxel $(b/2,b/2,b/2)$ in the other. The simulations were
carried out in a periodic unit cell, with 20 grid points in each
direction. Results are compared with a spectral solution of the
Stokes equations in a periodic geometry, assuming the force density on
the particle surface is constant. For an isolated pair of particles
this level of approximation includes the Oseen interaction and the
Faxen correction; it corresponds to the Rotne-Prager (RP) interaction
\cite{Rot69} used in most Brownian dynamics simulations of
hydrodynamically interacting particles:
\begin{equation}\label{eq:hi-rp}
\bU_2 = \left[\tensor{T}(\bR_{12}) + \frac{a^2}{12 \pi \eta R_{12}^5}
\left(R_{12}^2\tensor{1} - 3 \bR_{12}\bR_{12}\right)\right] \cdot \bF_1.
\end{equation}
The periodic RP tensor can be calculated by Ewald summation
\cite{beenakker:86} or by direct summation of Fourier components,
which converges if used in conjunction with finite-volume sources
\cite{Maz82,Lad88}.

The results for the two-point interpolation show a significant
dependence on the exact grid position when the particles are close to
each other, $r < 4a$, but the results for the higher-order schemes are
essentially independent of the grid. The three-point integration
scheme is of comparable accuracy to the four-point scheme but requires
less than half the number of grid points. It would seem to be the best
choice for applications even though the particle motion is not quite
as smooth. When the particles are widely separated, the simulations
match almost perfectly with both the Oseen and RP solutions for the
same periodic geometry; the typical errors are of the order of 0.1\%
of the Stokes velocity.

When the spheres are closer together, $r < 3a$,
then the simulated hydrodynamic interactions match the
Rotne-Prager interaction rather better than the Oseen
interaction. This confirms that the weight function does make the
particles behave as volume sources, rather than points. The best fit
between simulation results and Stokes flow is obtained for an
effective particle radius that is roughly $0.33 w$, where $w$ is the
range of the weight function. So for two-point interpolation ($w=2b$)
the effective size is about $0.7 b$, for three point interpolation
($w=3b$) it is about $1.0b$ and for four-point interpolation ($w=4b$)
it is about $1.3b$. The actual values used in Fig. \ref{fig:two} are
$0.8b$, $1.0b$ and $1.2b$ respectively. The optimal hydrodynamic
radius is quite large, $a \sim b$, corresponding to strong
coupling, $a_0 \sim 5b$. It remains an open question whether it is
practical or desirable to run the fluctuating simulations with such
large input friction.

%% file: apps.tex
In this section we will discuss applications of the LB method to
simulations of soft matter. We will briefly summarize some of our published
work in this area, with the aim of indicating the breadth of possible
applications of the method. Results will be summarized from simulations
that cover a wide range of experimental length and time scales--from
nanometers to millimeters and from nanoseconds to seconds.

\subsection{Short-time diffusion of colloids}
The first application of the fluctuating LB model was to the short-time
diffusion of hard-sphere colloids. At the time, a new experimental
technique--Diffusing Wave Spectroscopy (DWS)~\cite{Wei89}--enabled the
study of the dynamics of colloids on time scales of a few nanoseconds.
In general, the diffusion of a colloidal particle is a Markov process,
but at such short times the developing hydrodynamic flow field gives
rise to additional long-range correlations, 
analogous to the ``long-time tails'' in molecular dynamics~\cite{Ald70}.
Although the existence of long-time tails had been established
theoretically~\cite{Ern70,Dor75} and by molecular dynamics
simulations~\cite{Ald70}, these experiments marked the first
direct observation of correlated hydrodynamic fluctuations.
Brownian and Stokesian dynamics both neglect long-range dynamic
correlations, using instead the Stokes-flow approximation, which is
typically only valid on time scales longer then one microsecond.

\begin{figure}
\centering
\includegraphics[width=\fullwidth]{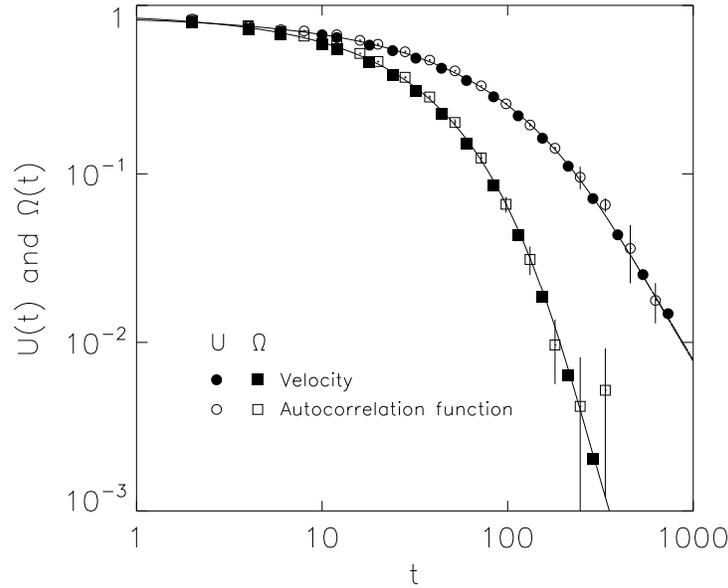}
\caption{Decay of translational ($U$) and rotational ($\Omega$) velocity
correlations of a suspended sphere.
The time-dependent velocities of the sphere are shown as solid symbols;
the relaxation of the corresponding velocity autocorrelation functions
are shown as open symbols (with statistical error bars).  A sufficiently
large fluid volume was used so that the periodic boundary conditions
had no effect on the numerical results for times up to $t = 1000$ in
lattice units ($h = b = 1$).  The
solid lines are theoretical results, obtained by an inverse
Laplace transform of the frequency dependent friction
coefficients~\cite{Hau73} of a sphere of appropriate size ($a = 2.6$)
and mass ($\rho_s/\rho = 12$); the kinematic viscosity of the pure
fluid $\eta_{kin} = 1/6$.}
\label{fig:ltt}
\end{figure}

Long-time tails occur naturally in the dynamics of lattice-gas models of
colloidal suspensions~\cite{Hoe91} and even of the lattice gases
themselves~\cite{Hoe91a}. It might be supposed that such correlations
would be absent in a Boltzmann-level model, due to the Stosszahlansatz
closure assumption. The fluctuating LB model described in
Sec.~\ref{sec:flbe} does not have any long-time tails in the
stress autocorrelation functions, but mode coupling between the diffusion
of fluid momentum and the diffusion of the colloidal particle does lead to
an algebraic decay of the velocity correlation function of a suspended
sphere \cite{Lad93a}. In these simulations the particle-fluid coupling
was implemented via the link-bounce-back (BB)
algorithm~\cite{Lad90a,Lad94} described in Sec.~\ref{sec:bc}.
Figure~\ref{fig:ltt} shows the decay of translational and
rotational velocity from two different types of computer experiment.
In one case an initial velocity is imposed, which decays away
due to viscous dissipation, and in the other the particle is set in motion
by stress fluctuations in the fluid. Figure~\ref{fig:ltt} shows that,
within statistical errors, the normalized velocity correlation
functions are identical to the steady decay of the translational
and rotational velocities of the sphere; thus
our simulations satisfy the fluctuation-dissipation theorem.  Moreover
the simulations agree almost perfectly with theoretical results derived
from the frequency-dependent friction coefficients~\cite{Hau73},
even though there are no adjustable parameters in these comparisons;
thus we see that the fluctuating lattice-Boltzmann equation can account
for the hydrodynamic memory effects that lead to long-time
tails~\cite{Ald70}.

\begin{figure}
\centering
\includegraphics[width=\fullwidth]{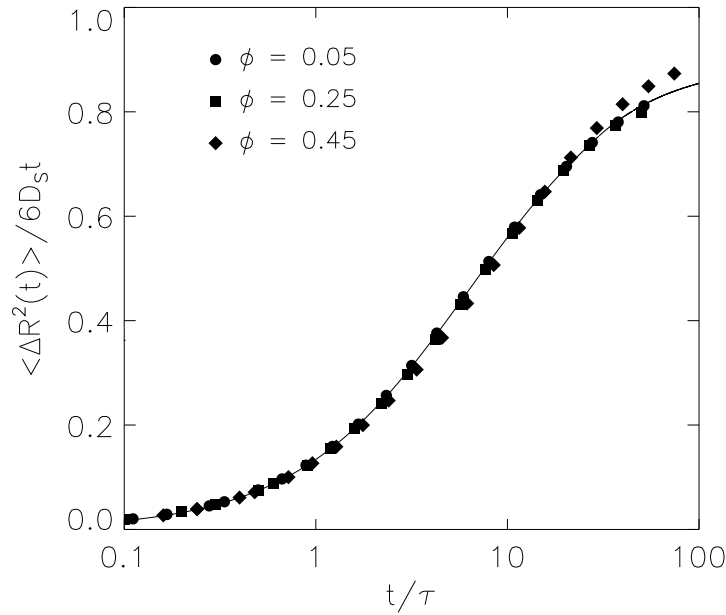}
\caption{Scaled mean-square displacement
$\left< \Delta R^2 (t) \right>/6D_s t$ at short times, {\it vs.}
reduced time $t/\tau$.  Simulation results for 128 spheres
(solid symbols) are shown at packing fractions $\phi$
of 5\%, 25\% and 45\%; the solid line is the isolated sphere result.
The suspension viscosity at these packing fractions is $1.14 \eta_0$,
$2.17 \eta_0$, and $5.6 \eta_0$ respectively, where $\eta_0$ is the
viscosity of the pure fluid.}
\label{fig:msqd}
\end{figure}

The simulations were also used to measure self diffusion in dense colloidal
suspensions, up to a solids volume fraction of 45\%. The simulation data,
shown in Fig.~\ref{fig:msqd}, exhibits the same scaling with amplitude
and time found in the DWS experiments~\cite{Zhu92}.  In
Fig.~\ref{fig:msqd} the amplitude of the mean-square displacement has been
normalized by its limiting value $6 D_s t$, where
$D_s$ is the short-time self-diffusion coefficient. $D_s(\phi)$ is a
monotonically decreasing function of concentration, because neighboring
particles increasingly restrict the hydrodynamic flow field generated by
the diffusing particle.  Although the colloidal particles are freely
moving in the fluid, the no-slip boundary condition induces stresslets and
higher force multipoles on the particle surfaces; $D_s(\phi)$
is one average measure of these hydrodynamic interactions. The normalized
mean-square displacement has a single relaxation time, so that when the
time axis is scaled by $\tau$ all the data collapse onto a single curve,
which is the same as for an isolated sphere. The relaxation time $\tau$
is the viscous diffusion time $\rho a^2/\eta$, of a single particle in a
fluid of viscosity $\eta$, where $\eta(\phi)$ is numerically similar to
the high-frequency viscosity of the suspension~\cite{Lad90b}.  This
observation, which is in line with experimental
measurements~\cite{Zhu92,Kao93}, suggests that the short-time diffusion
is essentially mean field like.

\subsection{Dynamic scaling in polymer solutions}
\label{sec:apps:polymer_scaling}

\begin{figure}
\centering
\includegraphics[width=\fullwidth]{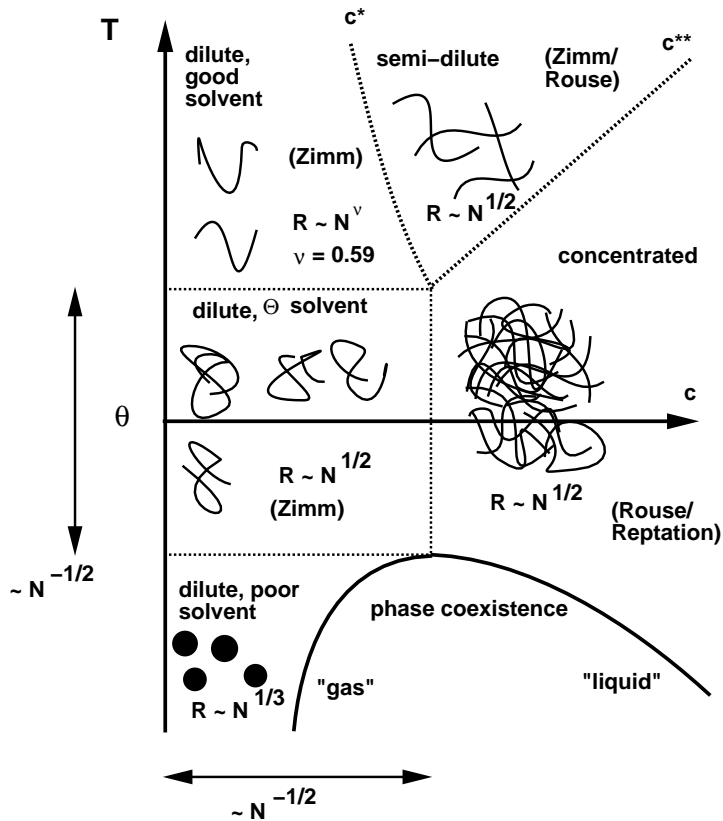}
\caption{Phase diagram of a polymer solution,
  in the $c-T$ plane, where $c$ is the monomer concentration and
  $T$ is the temperature
  (parameterizing solvent quality). The static properties
  are characterized by scaling laws which describe
  the dependence of the chain size $R$ (gyration
  radius or end--to--end--distance for example) on the degree of
  polymerization $N$. In the dilute limit ($c \to 0$) the
  so--called theta transition occurs, where at $T = \Theta$,
  single isolated chains collapse from a swollen random coil
  to a compact globule. For finite chain length $N$, this
  transition is ``smeared out'' over a temperature region
  $\Delta T \propto N^{-1/2}$, in which the chain conformations
  are Gaussian. Below $\Theta$, there is phase coexistence between a
  ``gas'' of globules and a ``liquid'' of strongly interpenetrating
  Gaussian chains. The corresponding critical point
  occurs at a very low concentration, $c_{c} \propto N^{-1/2}$,
  and in the vicinity of $\Theta$, $\Theta - T_{c} \propto N^{-1/2}$.
  The crossover region, which connects the regime of swollen
  isolated coils with that of the concentrated (Gaussian) solution
  at high temperatures, is called the semidilute regime. The
  dynamics is characterized by the Zimm model in the dilute
  limit where hydrodynamic interactions are important, and
  by the Rouse model for dense systems where they are screened.
  For very dense systems or sufficiently long chains,
  where curvilinear motion dominates, the Rouse model
  must be replaced by the reptation model (or the crossover
  behavior between these two cases). The Rouse and Zimm models
  are briefly described in the text.
}
\label{fig:solutionphasdiag}
\end{figure}

The classical theory \cite{doi:86} of the equilibrium dynamics of
polymer chains in solution is the Zimm model \cite{zimm}, which
considers a single flexible chain in a good solvent, such that its
conformations are given by a random coil with excluded volume segments:
\begin{equation}
R \sim b N^\nu .
\end{equation}
Here, $R$ is the size of the coil, measured in terms of the
gyration radius or the end--to--end distance, while $N$ denotes the
number of monomers in the chain or the degree of polymerization, and
$b$ is the monomer size. Long--range interactions like
electrostatics, or effects of poor solvent quality, are not
considered.  Furthermore, the solution is considered to be dilute, such
that the chains do not overlap and a single--chain picture is
sufficient. In other words, the standard Zimm model applies in the
upper left corner of the generic phase diagram given in Fig.
\ref{fig:solutionphasdiag}. Here the exponent $\nu$ takes the value
$\nu \approx 0.59$ in three dimensions, because the excluded volume
interaction leads to swelling of the chain when compared to an
ideal random coil ($\nu = 1/2$). A polymer with excluded-volume is thus
a self--similar random fractal with dimension $1/\nu$.

The Zimm model is based on the Rouse model \cite{doi:86,rouse},
but includes long-range hydrodynamic interactions between
the segments. Both models predict self--similarity, not
only with respect to space, but also with respect to time. Therefore
the dynamics is conveniently described in terms of an
exponent $z$, connecting the chain relaxation time $\tau_R$ with
the size of the coil $R$:
\begin{equation}
\tau_R \propto R^z .
\end{equation}
The internal degrees of freedom completely re--organize on a time
scale $\tau_R$, leading to statistically independent conformations.
This is also the time that the chain needs to diffuse through a
distance equal to its own size:
\begin{equation}
D_{cm} \tau_R \sim R^2 ,
\end{equation}
where $D_{cm}$ is the translational diffusion coefficient, describing
the center--of--mass motion. These two relations can be combined to
determine the scaling of $D_{cm}$ with chain size:
\begin{equation}
D_{cm} \propto R^{2 - z} .
\end{equation}
In the \emph{Zimm} model, the hydrodynamic interactions result in
strongly correlated motions, such that the coil as a
whole behaves like a Stokes sphere, $D_{cm} \propto
R^{-1}$, or
\begin{equation}
z = 3.
\end{equation}
The \emph{Rouse} model neglects hydrodynamic interactions and the
monomer friction coefficients
add up to give a total friction coefficient that is
linearly proportional to $N$. Since $D_{cm} \propto
N^{-1} \propto R^{-1/\nu}$,
\begin{equation}
z = 2 + \frac{1}{\nu} ,
\end{equation}
corresponding to slower dynamics. 

Self--similarity implies that the relaxation of the internal degrees
of freedom also scales with the exponent $z$ on time scales $\tau_b
\ll t \ll \tau_R$, where $\tau_b$ is the relaxation time on the monomer
scale $b$. In the space--time window $b \ll l \ll R$, $\tau_b \ll t
\ll \tau_R$, there is a scaling of the mean square displacement of a monomer,
\begin{equation}
\left< \Delta r^2 \right> \propto t^{2 / z},
\end{equation}
while the dynamic structure factor of a single chain of $N$ monomers
\begin{equation}
S(k,t) =
\frac{1}{N} \left< \sum_{ij=1}^N \exp \left( i \vec k \cdot 
( \vec r_i(t) - \vec r_j (0) ) \right) \right>
\end{equation}
scales as
\begin{equation}
S(k,t) = k^{-1/\nu} f \left( k^2 t^{2/z} \right) .
\end{equation}

The Zimm model applies to dilute solutions, and, therefore, to the
dynamics of a single solvated chain.  It has become a benchmark
system, used to test the validity of mesoscopic simulation methods. A
single chain, modeled by bead--spring interactions, coupled to a
surrounding solvent to account for hydrodynamic interactions, has been
successfully simulated via (i) Molecular Dynamics
\cite{pierleoni:92,smith:92,duenweg:93}, (ii) Dissipative Particle
Dynamics \cite{schlijper:95,spenley:00}, Multi--Particle Collision
Dynamics \cite{malevanets_yeomans:00,mussawisade:05}, and by
Lattice Boltzmann \cite{Ahl99}, applying the dissipative
coupling described in Sec. \ref{sec:forcecoupling}. These studies
are nowadays all sufficiently accurate to be able to clearly
distinguish between Rouse and Zimm scaling. However, a more
demanding goal is to verify not only the exponent, but also
the prefactor of the dynamic scaling law.

The Kirkwood approximation to the diffusion constant,
\begin{equation}
D^{(K)} = \frac{k_B T}{\Gamma N} + \frac{k_B T}{6 \pi \eta}
\left< \frac{1}{R_H} \right> ,
\end{equation}
($\Gamma$ is the monomer friction coefficient) can be calculated
from a conformational average of the hydrodynamic radius $R_H$,
\begin{equation}\label{eq:R_H}
\left\langle \frac{1}{R_H} \right\rangle = 
\frac{1}{N^2} \sum_{i \ne j}
\left\langle \frac{1}{r_{ij}} \right\rangle .
\end{equation}
Highly accurate results for a single chain in a structureless solvent
have been obtained by Monte Carlo methods \cite{Lad92,duenweg:02}.
However, a naive comparison will fail badly. The
expression for $R_H$ (Eq. \ref{eq:R_H}) assumes an infinite system, but
in a simulation the system is confined to a periodic unit
cell, which is typically not substantially larger than the size
of the coil. The effects of periodic boundaries can be accounted for
quantitatively,
by replacing the Oseen tensor with an Ewald sum that includes the
hydrodynamic interactions with the periodic images \cite{beenakker:86}.
The consequences of this have been worked out in detail for polymers
\cite{duenweg:93} and colloids \cite{Lad90b}.
The main result is that the hydrodynamic radius must be
replaced by a system--size dependent effective hydrodynamic radius;
the leading--order correction is proportional to $R/L$, where $L$ is
the linear dimension of the periodic simulation cell. Interestingly,
\emph{internal} modes, such as Rouse modes \cite{doi:86}, where the motion
of the center of mass has been subtracted, have a much
weaker finite--size effect, which scales as $L^{-3}$, corresponding
to a dipolar hydrodynamic interaction with the periodic
images \cite{Ahl99}. Taking the finite--size effects into
account, the predictions of the Zimm model are nicely confirmed.

\begin{figure}
\centering
\includegraphics[width=7cm,angle=270]{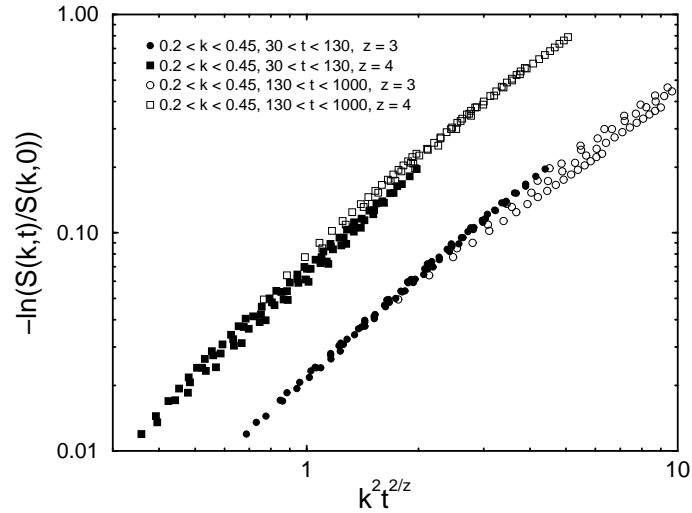}
\caption{Scaling of the single--chain dynamic structure
  factor data, showing both Rouse and Zimm scaling
  \protect\cite{ahlrichs:01}. The wave number $k$ has
  been restricted such that only length scales above
  the blob size are probed ($k \xi < 1$), while the
  size of the polymer chain as a whole does not yet
  matter ($k R_G > 1$). The data are labeled according
  to the time regimes; solid symbols refer to the
  short--time regime below the blob relaxation time,
  $t < \tau_\xi$, while open symbols are for later
  times $t > \tau_\xi$. The upper curve is for Rouse
  scaling ($z=4$) and the lower curve for Zimm
  scaling ($z=3$). One sees that Zimm scaling works
  better in the short--time regime, while Rouse
  scaling holds for later times.
}
\label{fig:patrick}
\end{figure}

However, the Zimm model is no longer valid as soon as the chains start
to overlap. Here a double screening mechanism sets in: (i) Screening of
excluded volume interactions (Flory screening). In a dense melt, the
chain conformations are not those of a self--avoiding walk, but rather
those of a random walk ($\nu = 1/2$). Essentially, this is an entropic
packing effect: A swollen coil would take too much configuration space
from the surrounding chains. This effect can be understood
in terms of a self--consistent mean--field theory, which is expected to
work well for dense systems where density fluctuations are suppressed
\cite{grosberg}. (ii) Screening of hydrodynamic interactions. In
dense melts, the dynamics is not Zimm--like, but rather Rouse--like,
or governed by reptation \cite{doi:86}. Reptation occurs for long
chains in dense systems, where topological constraints enforce an
essentially curvilinear motion. We will not be concerned with these
latter effects, but rather with the mechanism which leads to the
suppression of hydrodynamic interactions. Based upon the results of
computer simulations \cite{ahlrichs:01}, we were able to develop a
simple picture, which essentially confirmed the previous work by de
Gennes \cite{degennes:76}, and completed it. The basic mechanism is
chain--chain collisions. A monomer encountering another chain will
deform it elastically, inducing a stress along the polymer backbone
instead of propagating the signal into the
surroundings. Since the chain arrangements are random, the fluid momentum
is also randomized, such that momentum correlations
(or hydrodynamic interactions) are destroyed.

The crossover region between dilute and dense systems is called the
semidilute regime. A semidilute solution is characterized by strongly
overlapping chains which are however so long and so dilute that the
monomer concentration can still be considered as vanishingly small.
Apart from $b$ and $R$, there is now a third important length scale,
the ``blob size'' $\xi$ with $b \ll \xi \ll R$. Essentially, $\xi$
is the length scale on which interactions with the surrounding
chains become important; this length scale controls the crossover from
dilute to dense behavior. The chain conformations are characterized by
$\nu = 0.59$ on length scales much smaller than $\xi$, while on length
scales substantially above $\xi$ the exponent is $\nu = 1/2$.
The challenge for computer simulations is that both
behaviors need to be resolved simultaneously, which is only possible for $N
> 10^3$. Roughly $30$ monomers are needed to resolve
the random fractal structure within the blob, while another $30$ blobs
per chain are needed to observe the random walk regime. Furthermore, a
many--chain system should be run without self--overlaps, and this
leads to the conclusion \cite{ahlrichs:01} that the smallest system to
simulate semidilute dynamics contains roughly $5 \times 10^4$ monomers
and $5 \times 10^5$ LB lattice sites.

The picture which emerges from these simulations \cite{ahlrichs:01}
can be summarized as follows. Initially, the dynamics is Zimm--like,
even for length scales beyond
the blob size. The reason is that hydrodynamic signals can spread
easily throughout the system, and just drag the chains with them. This
continues until chain--chain collisions start to play a role. The
relevant time scale is the blob relaxation time $\tau_\xi
\propto \xi^3$, \ie\ the time a blob needs to move its own size. From
then on, the screening mechanism described above
becomes important, and the dynamics is Rouse--like. This is
only observable on length scales beyond the blob size, since
on smaller scales all dynamic correlations have decayed already.
This is nicely borne out by single--chain dynamic structure factor
data (see Fig. \ref{fig:patrick}), and explains the previous
observation of ``incomplete screening'' \cite{richter:84}
in a straightforward and natural way.

\subsection{Polymer migration in confined geometries}

Flexible polymers in a pressure-driven flow field migrate towards
the center of the channel, because of hydrodynamic interactions.
The local shear rate stretches the polymer and the resulting
tension in the chain generates an additional flow field around the
polymer. This flow field becomes asymmetric near a no-slip
boundary and results in a net drift towards the center of the
channel \cite{Fan05,Jen04,Ma05}. Recent simulations
\cite{Ma05,Ust06} show that hydrodynamic lift is the dominant
migration mechanism in pressure-driven flow, rather than spatial
gradients in shear rate. Recently, we used numerical simulations to
investigate a flexible polymer driven by a combination of fluid flow
and external body force~\cite{Ust07}, but
ignoring the complications arising from counterion screening in
electrophoretic flows. We used the fluctuating LB model
(Sec.~\ref{sec:flbe}) in conjunction with the point-force coupling
scheme described in Sec.~\ref{sec:forcecoupling}.

\begin{figure}
\centering
\includegraphics[width=\fullwidth]{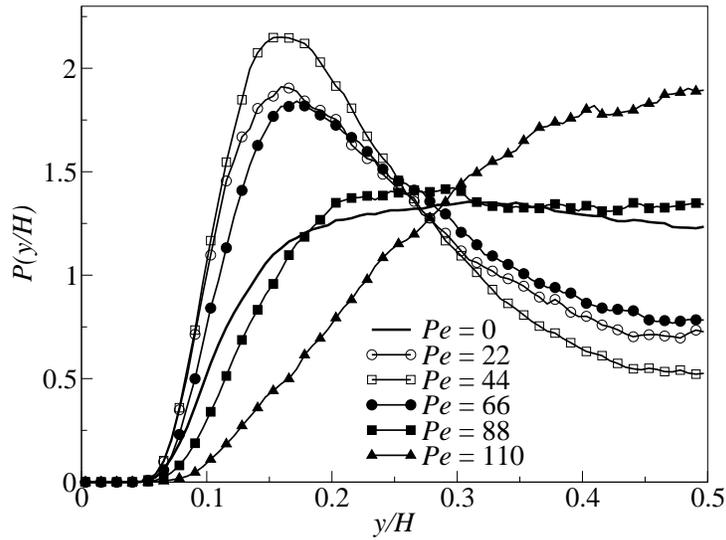}
\caption{Center of mass distributions for countercurrent
application of an external body force and pressure-driven flow
in a channel of width $H$.
The solid curve shows the level of migration under the
pressure-driven flow only. The flow Peclet number in all cases is
$Pe_f=12.5$. The boundary is at $y/H=0$ and the center of the
channel is at $y/H=0.5$; $H = 8 R_g$.} \label{fig:polymigration}
\end{figure}

We were surprised to find that the polymer migrates towards
the channel center under the action of a body-force alone, while
in combination with a pressure-driven flow the polymer can move
either towards the channel wall or towards the channel center.
The external field and pressure gradient
result in two different Peclet numbers: $Pe=\overline{U} R_g /D$
and $Pe_f=\overline{\gamma}{R_g}^2/D$. Here $\overline{U}$ is the
average polymer velocity with respect to the fluid,
and $\overline{\gamma}$ is
the average shear rate. The interplay between force and flow can
lead to a wide variety of steady-state distributions of the polymer center
of mass across the channel \cite{Ust07}. For example, 
in a countercurrent application of the two fields, the polymer
tends to orient in different quadrants depending on the relative
magnitude of the two driving forces. The polymer then drifts either
towards the walls or towards the center depending on its mean orientation.
The results in Fig.~\ref{fig:polymigration} show migration towards
the boundaries when the external force is small ($Pe < 30$), but
increasing the force eventually reverses the orientation of the
polymer and the polymer again migrates towards the center ($Pe >
100$).

The simulations mimic recent experimental observations of the migration
of DNA in combined electric and pressure-driven flow
fields~\cite{Zhe02,Zhe03}. The similarities between these results suggest
that hydrodynamic interactions in polyelectrolyte solutions are
only partially screened. In fact, within the Debye-H\"{u}ckel
approximation, there is a residual dipolar flow field \cite{Lon01}.
Although this flow is weak in comparison to the electrophoretic
velocity, its dipolar orientation enables it to drive a transverse
migration of the polymer. A recent kinetic theory calculation~\cite{But07}
supports these qualitative observations.

\subsection{Sedimentation}

The previous examples have focused on sub-micron sized particles,
colloids and polymers, where Brownian motion is an essential component
of the dynamics. For particles larger than a few microns, Brownian
motion is negligible under normal laboratory conditions and a
suspension of such particles can be simulated using the deterministic
version of the LB equation (see Sec.~\ref{sec:flbe}). There is an
interesting regime of particle sizes, from 1--100 micron depending on
solvent, where Brownian motion is negligible, yet inertial effects are
still unimportant. This corresponds to the region of low Reynolds
number ($Re = Ud/\eta_{kin}$) but high Peclet number ($Pe =
Ud/D$). Here $U$ is the characteristic particle velocity, $d$ is the
diameter, and $D$ is the
particle diffusion coefficient. Because of the large difference in
time scale between diffusion of momentum and particle diffusion, it is
quite feasible for $Pe$ to be $6-10$ orders of
magnitude larger than $Re$. We have carried out a number of
simulations in this regime, with the aim of elucidating the role of
suspension microstructure in controlling the amplitude of the velocity
fluctuations as the suspension settles.

In a sedimenting suspension, spatial and temporal variations in particle
concentration drive large fluctuations in the particle velocities, of the
same order of magnitude as the mean settling velocity. For particles larger
than a few microns, this hydrodynamic diffusion dominates the thermal
Brownian motion, and in the absence of inertia ($Re \ll 1$),
the particle velocities are determined entirely by the instantaneous
particle positions.  If the particles
are randomly distributed, then the velocity fluctuations will
diverge with increasing container size~\cite{Caf85}, although the
density fluctuations may eventually drain out of the system by
convection~\cite{Hin88}. However, experimental measurements
indicate that the velocity fluctuations converge to a finite value
as the container dimensions are increased~\cite{Nic95,Seg97}, but
the mechanism by which the velocity fluctuations saturate is not
yet clear. Some time ago, Koch and Shaqfeh suggested that the
distribution of
pairs of particles could be modified by shearing forces induced by
the motion of a third particle, and that these changes in
microstructure could in turn lead to a screening of the long-range
hydrodynamic interactions driving the velocity
fluctuations~\cite{Koc91}. However, detailed numerical simulations
found no evidence of the predicted microstructural
changes~\cite{Lad96}. Instead the velocity fluctuations in
homogeneous suspensions (with periodic boundary conditions) were
found to diverge with increasing cell size. More recently, it has been
proposed that long wavelength density fluctuations can be
suppressed by a convection diffusion mechanism~\cite{Ton98,Lev98},
but a bulk screening mechanism cannot be reconciled with the
results of computer simulations~\cite{Lad96,Lad97b}.
Alternatively, it has been suggested that the vertical walls of
the container may modify, although not eliminate, the divergence
of the velocity fluctuations~\cite{Bre99}.  Most recently, it has
been shown~\cite{Luk00,Muc03} that a small vertical density gradient can
damp out diverging velocity fluctuations.

\begin{figure}
\centering
\includegraphics[width=\fullwidth]{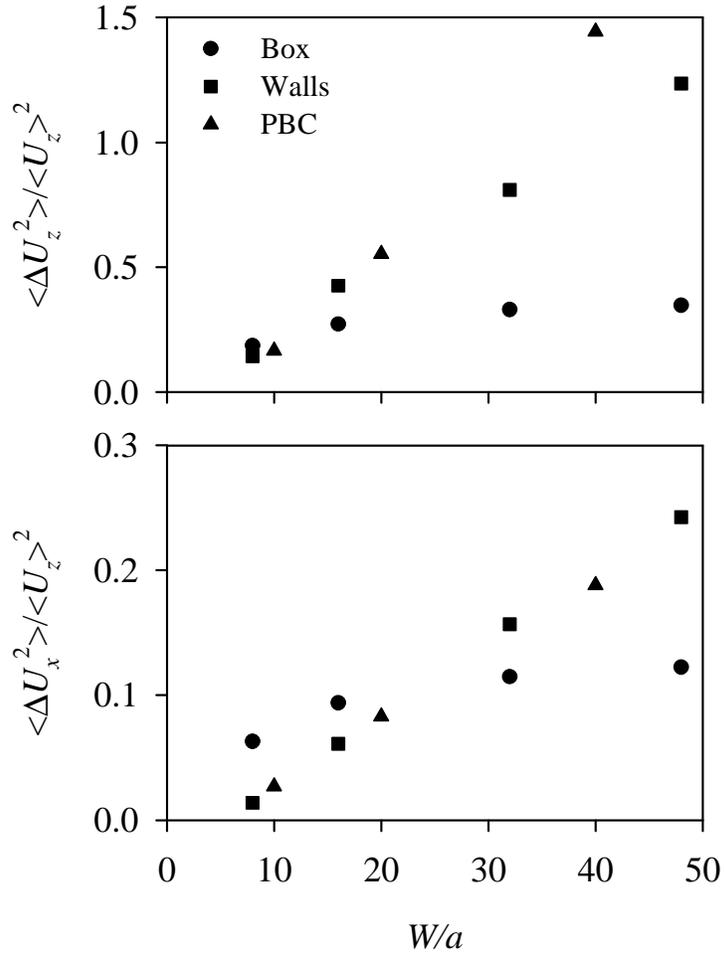}
\caption{Relative velocity fluctuations $\left<\Delta
U_\alpha^2\right>/\left<U_z\right>^2$ as a function of container
width.  The vertical ($\left<\Delta U_z^2\right>$) and horizontal
($\left<\Delta U_x^2\right>$) fluctuations are shown for three
different boundary conditions: a cell bounded in all three
directions by no-slip walls (Box), a cell bounded by vertical
walls (Walls), and a cell that is periodic in all three directions
(PBC)~\cite{Lad96,Lad97b}. The statistical errors are comparable
to the size of the symbols.} \label{fig:velfluct}
\end{figure}

Lattice-Boltzmann simulations were used to test these theoretical ideas,
comparing the behavior of the velocity fluctuations in three different
geometries \cite{Lad02}.  We found striking differences in the
level of velocity
fluctuations, depending on the macroscopic boundary conditions.
In a geometry similar to those used in laboratory
experiments~\cite{Nic95,Seg97}, namely a rigid container bounded
in all three directions, we found that the calculated
velocity fluctuations saturate with increasing container
dimensions, as observed experimentally, but contrary to earlier
simulations with periodic boundary conditions~\cite{Lad96,Lad97b}.
The main result is illustrated in Fig.~\ref{fig:velfluct},
and suggests that the velocity fluctuations in a bounded container are
independent of container width for sufficiently large containers.
On the other hand, in vertically homogeneous suspensions velocity
fluctuations are proportional to the container width, regardless
of the boundary conditions in the horizontal plane.  The
significance of this result is that it establishes 
that velocity fluctuations in a sedimenting suspension depend
on the macroscopic boundary conditions and that laboratory
measurements~\cite{Nic95,Seg97,Gua01} are not necessarily characteristic
of a uniform suspension, as had been supposed. Instead, the simulations
show that vertical variations in particle concentration are
responsible for suppressing the velocity fluctuations, which
otherwise diverge with increasing container size, in agreement
with theory~\cite{Caf85} and earlier simulations~\cite{Lad96,Lad97b}.
The upper and lower boundaries apparently act as sinks for the fluctuation
energy~\cite{Hin88}, while in homogeneous suspensions velocity fluctuations
remain proportional to the system size~\cite{Caf85}.

\subsection{Inertial migration in pressure-driven flow}
At still larger particle sizes, typically in excess of 100 microns,
the inertia of the fluid can no longer be ignored. For suspended
particles in a gravitational field, the Reynolds number grows in
proportion to the cube of the particle size. Inertia breaks the
symmetry inherent in Stokes flow and leads to new phenomena, and in
particular the possibility of lateral migration of particles. A
particle in a shear flow experiences a transverse force at non-zero
$Re$, with a direction that depends on the velocity of the particle
with respect to the fluid velocity at its center. Thus if the particle
is moving slightly faster than the fluid it moves crosswise to the
flow in the direction of lower fluid velocity and
vice-versa~\cite{Saf65}; if it is moving with the local stream
velocity then it does not migrate in the lateral direction at all. Now
in Poiseuille flow, a spherical particle moves faster than the surrounding
fluid because of the Faxen force, proportional to the curvature in
the fluid velocity field. Thus particles tend to migrate towards the
channel walls~\cite{Ho74}. However, near the wall the particle is
slowed down by the additional drag with the wall and so eventually
migrates the other way. At small Reynolds numbers ($Re < 100$), these
forces balance when the particle is at a radial position of roughly
$0.6 R$, where $R$ is the radius of the pipe.  In a cylindrical pipe,
a uniformly distributed suspension of particles rearranges to form a
stable ring located at approximately $0.6 R$~\cite{Seg62}. Theoretical
calculations for small particles in plane Poiseuille flow give similar
equilibrium positions to those observed experimentally
\cite{Sch89,Asm99}.  The profile of the lateral force across the
channel shows only one equilibrium position, which shifts closer to
the boundary wall as the Reynolds number increases. Our interest in
this problem was sparked by two recent experimental observations:
first that particles tend to align near the walls to make linear
chains of more or less equally-spaced particles \cite{Seg62,Mat04a},
and second that at high Reynolds numbers ($Re \sim 1000$) an
additional inner ring of particles was observed when the ratio of
particle diameter $d$ to cylinder diameter $D$ was of the order of
1:10 \cite{Mat04}. Large particles introduce an additional Reynolds
number, $Re_p = Re\cdot(d/D)^2$, which may not be small, as assumed
theoretically \cite{Sch89,Asm99}. We used the LB method to
investigate inertial migration of neutrally buoyant particles in the
range of Reynolds numbers from 100 to 1000~\cite{Chu06}. Individual
particles in a channel with a square cross section migrate to one of
a small number of equilibrium positions in the cross-sectional plane,
located near an edge or at the center of a face; we could not identify
any stable postions for single particles near the center of the channel.

\begin{figure}
\centering \mbox{
        \includegraphics[width=\halfwidth]{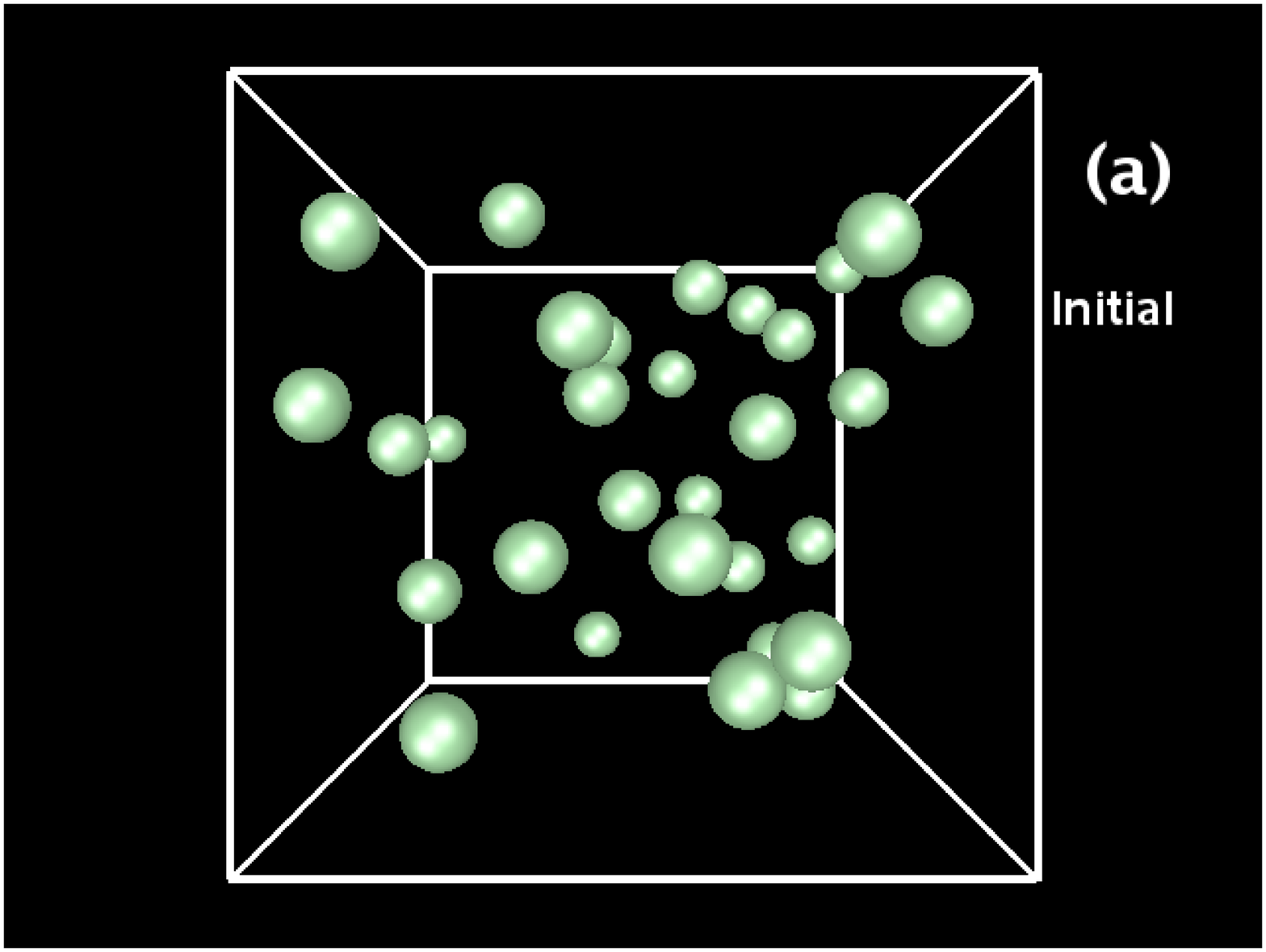}
        \includegraphics[width=\halfwidth]{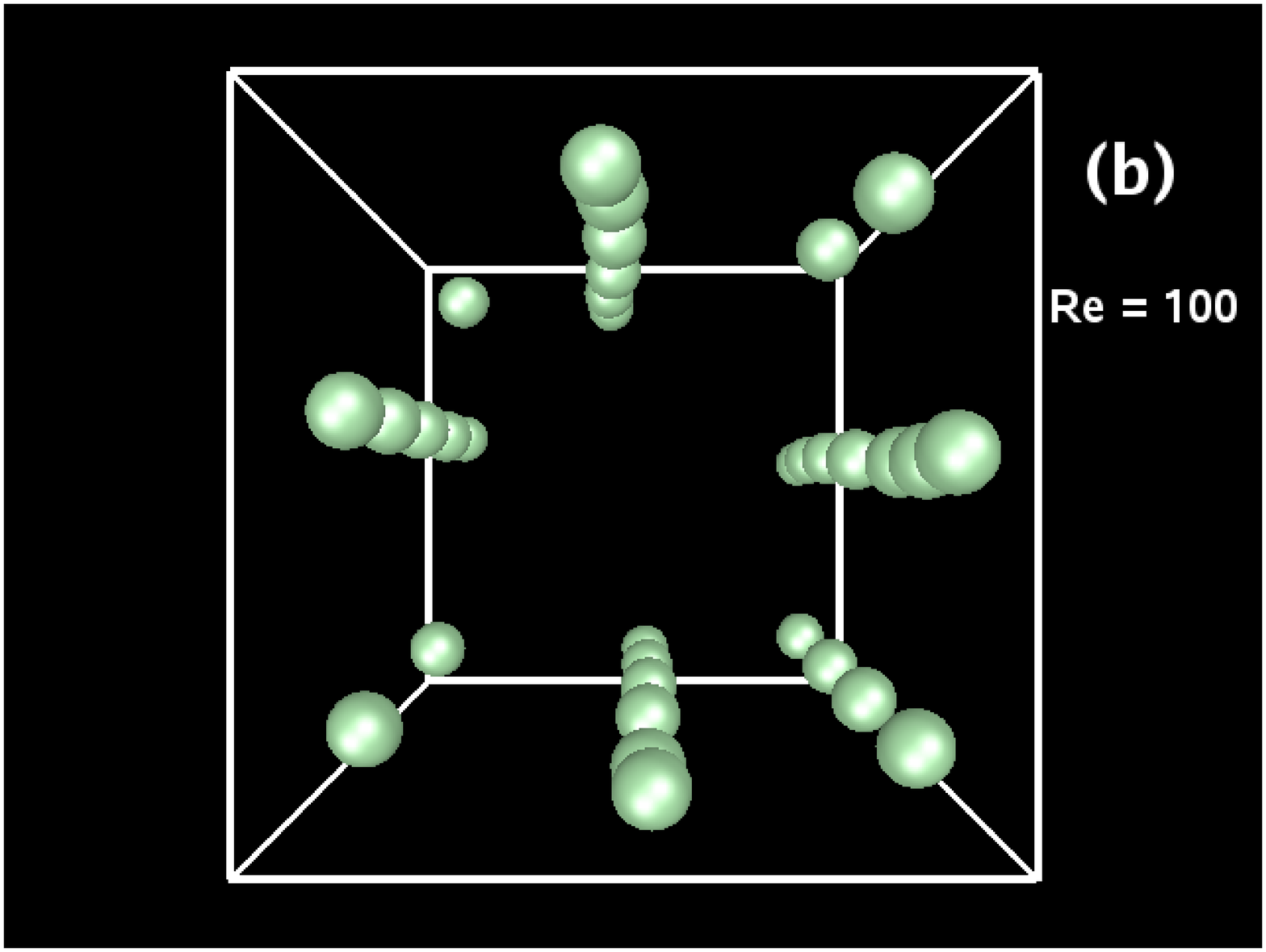}}
\centering \mbox{
        \includegraphics[width=\halfwidth]{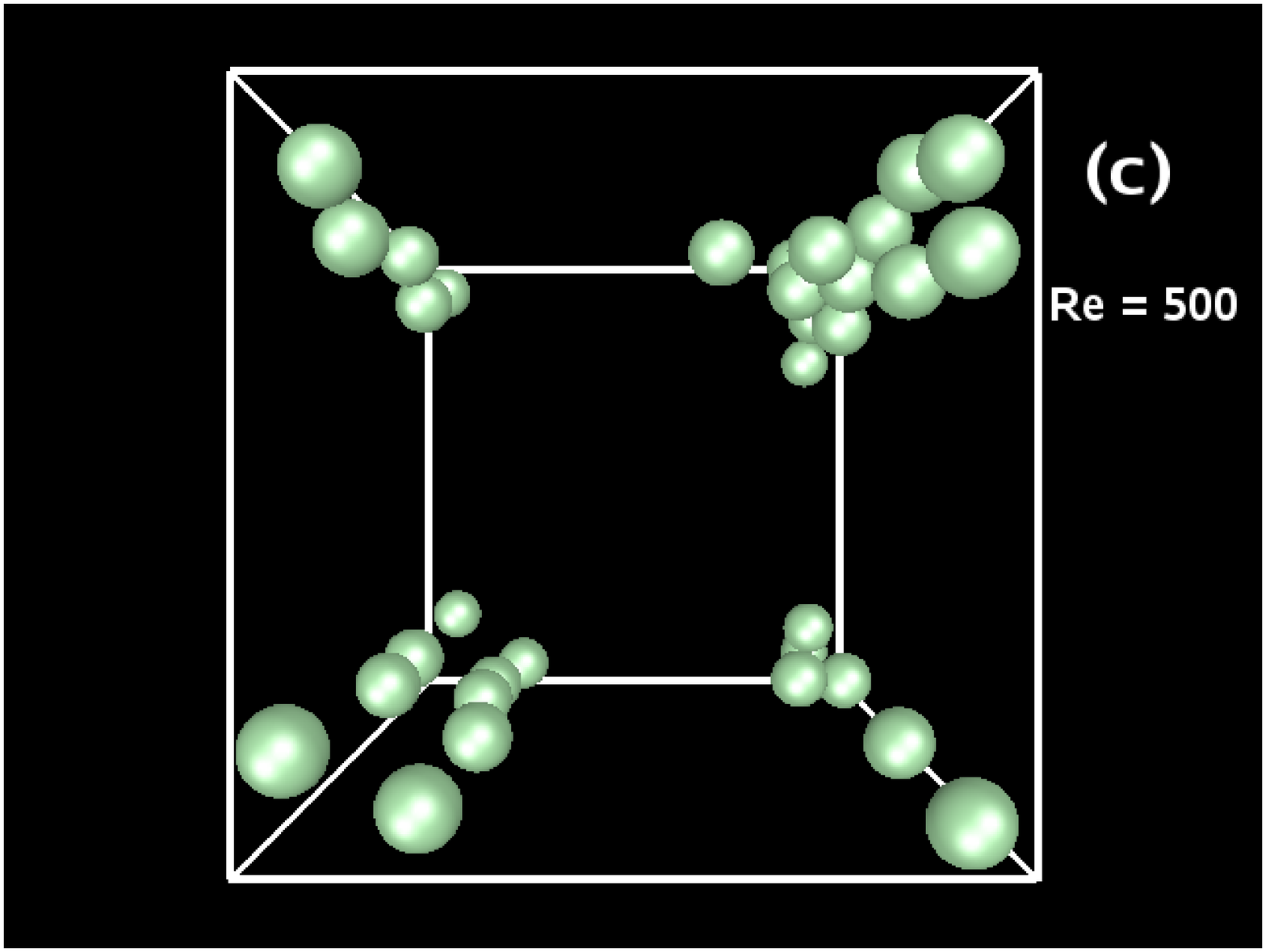}
        \includegraphics[width=\halfwidth]{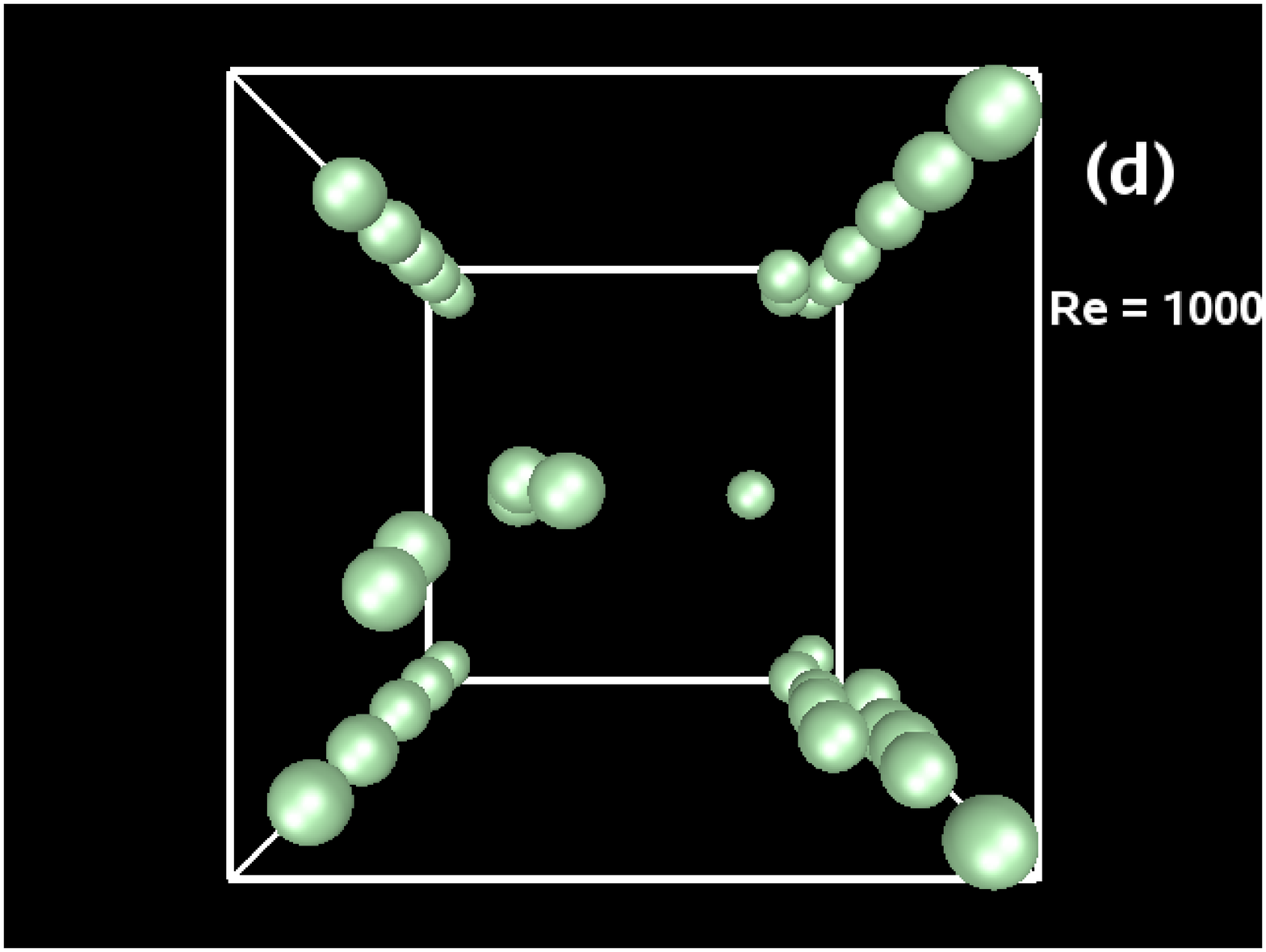}}
\caption{Snapshots of particle configurations in a duct flow at
different Reynolds numbers; the flow is into the plane of the
paper: (a) Initial configuration (b) $Re = 100$ (c) $Re = 500$
(d) $Re = 1000$. The ratio $H/d = 9.1$, the number of particles $N = 32$
and the volume fraction $\phi = 1\%$.}\label{fig:migration}
\end{figure}

To investigate multiparticle suspensions,
random configurations of particles were prepared at a volume
fraction $\phi=1\%$ and size ratio $H/d=9.1$. The Reynolds number in
the simulations varied between 100 and 1000. An initially uniform
distribution, shown in Fig.~\ref{fig:migration}a, evolves into three
different steady-state distributions depending on $Re$. At $Re=100$
(Fig.~\ref{fig:migration}b) particles are gathered around the eight
equilibrium positions and strongly aligned in the direction of the
flow, making linear chains of more or less uniformly spaced
particles. Similar trains of particles were observed in laboratory
experiments \cite{Mat04a}. At $Re=500$ (Fig.~\ref{fig:migration}c) the
particles are gathered in one of the four most stable positions, near each
corner. By a Reynolds number of 500, the trains are unstable and the
spacing between the particles is no longer uniform. Instead transient
aggregates of closely-spaced particles are formed, again near the
corners of the duct. However at still higher Reynolds number, $Re =
1000$, there is another change in particle configuration
(Fig.~\ref{fig:migration}d), and particles appear in the center of
the duct. A central band was first observed in experiments in a
cylindrical pipe \cite{Mat04}, but its origin remains unclear. We
observe that the central particles have a substantial diffusive motion
in the velocity-gradient plane, whereas the particle trains exhibit
little transverse diffusion. Since there are no single-particle
equilibrium positions at the duct center, the presence of particles in
the inner region is clearly due to multi-body
interactions. Nevertheless this migration cannot be a shear-induced
migration of the kind that occurs in low-Reynolds number flows
\cite{Fra03}.

Our simulations suggest that the inner band of particles is the
result of the formation of transient clusters of particles. We
proposed~\cite{Chu06}
that at higher Reynolds numbers ($Re > 500$) the trains become unstable
and clusters of closely-spaced particles arise, as can be seen in
Fig.~\ref{fig:migration}c. Simulations of tethered pairs of particles
have shown that additional equilibrium positions arise for pairs of
particles at Reynolds numbers in excess of 750~\cite{Chu06}. Thus
transient clusters are formed at higher Reynolds numbers, which drift
towards the center of the channel making the additional ring observed in
experiments~\cite{Mat04} and simulations (Fig.~\ref{fig:migration}d).
Eventually, the cluster disintegrates from hydrodynamic dispersion
and the particles return to the walls. At steady state, there is
a flux of pairs and triplets of particles moving towards the center,
balanced by individual particles moving towards the walls. This
also explains why the particles in the inner region are highly mobile,
while those near the walls have a very small diffusivity.

%% file: adv_poly_article.bbl
\begin{thebibliography}{100}
\providecommand{\url}[1]{\texttt{#1}}
\providecommand{\urlprefix}{URL }
\providecommand{\eprint}[2][]{\url{#2}}

\bibitem{chaikin_lubensky}
Chaikin PM, Lubensky TC (1997) Principles of Condensed Matter Physics.
\newblock Cambridge University Press, Cambridge

\bibitem{russel_saville_schowalter}
Russel WB, Saville DA, Schowalter WR (1995) Colloidal Dispersions.
\newblock Cambridge University Press, Cambridge

\bibitem{lyklema}
Lyklema J (1991) Fundamentals of Interface and Colloid Science.
\newblock Academic Press, London

\bibitem{evans_wennerstrom}
Evans DF, Wennerstr{\"o}m H (1999) The Colloidal Domain.
\newblock Wiley-VCH, New York, 2nd edn.

\bibitem{dGbook}
de~Gennes PG (1979) Scaling Concepts in Polymer Physics.
\newblock Cornell University Press, Ithaca

\bibitem{doi:86}
Doi M, Edwards SF (1986) The Theory of Polymer Dynamics.
\newblock Oxford University Press, Oxford

\bibitem{grosberg}
Grosberg AY, Khokhlov AR (1994) Statistical Physics of Macromolecules.
\newblock AIP Press, New York

\bibitem{phil_nelson}
Nelson P (2007) Biological Physics.
\newblock W. H. Freeman, New York

\bibitem{Ald70}
Alder BJ, Wainwright TE (1970) Phys Rev A 1:18

\bibitem{Mon89}
Montfrooij W, de~Schepper I (1989) Phys Rev A 39:2731

\bibitem{bird}
Bird RB, Curtiss CF, Armstrong RC, Hassager O (1987) Dynamics of Polymeric
  Liquids.
\newblock Wiley, New York

\bibitem{grest:1986}
Grest GS, Kremer K (1986) Phys Rev A 33:3628

\bibitem{wca}
Weeks JD, Chandler D, Andersen HC (1971) J Chem Phys 54:5237

\bibitem{Han86}
Hansen JP, McDonald IR (1986) {Theory of Simple Liquids}.
\newblock Academic Press, London

\bibitem{risken}
Risken H (1984) The Fokker--Planck Equation.
\newblock Sprin\-ger--Ver\-lag, Berlin

\bibitem{gardiner:1985}
Gardiner CW (1985) {Handbook of Stochastic Methods for Physics, Chemistry, and
  the Natural Sciences}.
\newblock Springer-Verlag, Berlin

\bibitem{Maz82}
Mazur P, Saarloos WV (1982) Physica A 115:21

\bibitem{cichocki_felderhof_multipole}
Cichocki B, Felderhof BU (1988) J Chem Phys 89:3705

\bibitem{Lad90b}
Ladd AJC (1990) J Chem Phys 93:3484

\bibitem{sangani_mo}
Sangani AS, Mo G (1996) Phys Fluids 8:1990

\bibitem{Cic00}
Cichocki B, Jones RB, Kutteh R, Wajnryb E (2000) J Chem Phys 112:2548

\bibitem{Hap86}
Happel J, Brenner H (1986) {Low-Reynolds Number Hydrodynamics}.
\newblock Martinus Nijhoff, Dordrecht

\bibitem{Rot69}
Rotne J, Prager S (1969) J Chem Phys 50:4831

\bibitem{wajnryb_divergence}
Wajnryb E, Szymczak P, Cichocki B (2004) Physica A 335:339

\bibitem{Ott96}
{\"{O}}ttinger HC (1996) {Stochastic Processes in Polymeric Fluids}.
\newblock Springer-Verlag, Berlin

\bibitem{kroger_review}
Kr{\"o}ger M (2004) Phys Rep 390:453

\bibitem{Erm78}
Ermak DL, McCammon JA (1978) J Chem Phys 69:1352

\bibitem{boliu}
Liu B, D{\"u}nweg B (2003) J Chem Phys 118:8061

\bibitem{Bra88b}
Brady JF, Bossis G (1988) Ann Rev Fluid Mech 20:111

\bibitem{Fix86}
Fixman M (1986) Macromolecules 19:1204

\bibitem{jendrejack00}
Jendrejack RM, Graham MD, dePablo JJ (2000) J Chem Phys 113:2894

\bibitem{Sie01}
Sierou A, Brady JF (2001) J Fluid Mech 448:115

\bibitem{BanchioBrady03}
Banchio AJ, Brady JF (2003) J Chem Phys 118:10323

\bibitem{Saintillanetal05}
Saintillan D, Darve E, Shaqfeh ESG (2005) Macromolecules 17:33301

\bibitem{Hernandez-Ortizetal07}
Hernandez-Ortiz JP, de~Pablo JJ, Graham MD (2007) Phys Rev Lett 98:140602

\bibitem{menghigdon1}
Meng Q, Higdon JJL (2008) J Rheol 52:1

\bibitem{menghigdon2}
Meng Q, Higdon JJL (2008) J Rheol 52:37

\bibitem{Bla71}
Blake JR (1971) Proc Camb Phil Soc 70:303

\bibitem{Lir76}
Liron N, Mochon S (1976) J Eng Math 10:287

\bibitem{cichocki_jones_physica}
Cichocki B, Jones RB (1998) Physica A 258:273

\bibitem{jendrejack03}
Jendrejack RM, Graham MD, dePablo JJ (2003) J Chem Phys 119:1165

\bibitem{Lan59}
Landau LD, Lifshitz EM (1959) {Fluid Mechanics}.
\newblock Addison-Wesley, London

\bibitem{ahlrichs:01}
Ahlrichs P, Everaers R, D{\"u}nweg B (2001) Phys Rev E 64:040501 (R)

\bibitem{koelman:1992}
Hoogerbrugge PJ, Koelman JMVA (1992) Europhys Lett 19:155

\bibitem{koelman:1993}
Koelman JMVA, Hoogerbrugge PJ (1993) Europhys Lett 21:369

\bibitem{pep:1995}
Espa{\~{n}}ol P, Warren P (1995) Europhys Lett 30:191

\bibitem{pep:1995:2}
Espa{\~{n}}ol P (1995) Phys Rev E 52:1734

\bibitem{groot:1997}
Groot R, Warren P (1997) J Chem Phys 107:4423

\bibitem{pep:1998}
Espa{\~{n}}ol P (1998) Phys Rev E 57:2930

\bibitem{ignacio}
Pagonabarraga I, Hagen MJH, Frenkel D (1998) Europhys Lett 42:377

\bibitem{gibson}
Gibson JB, Chen K, Chynoweth S (1999) Int J Mod Phys C 10:241

\bibitem{besold}
Besold G, Vattulainen I, Karttunen M, Polson JM (2000) Phys Rev E 62:R7611

\bibitem{besold2}
Vattulainen I, Karttunen M, Besold G, Polson JM (2002) J Chem Phys 116:3967

\bibitem{mikko}
Nikunen P, Karttunen M, Vattulainen I (2003) Comp Phys Comm 153:407

\bibitem{shardlow}
Shardlow T (2003) SIAM J Sci Comp 24:1267

\bibitem{thoso}
Soddemann T, D{\"u}nweg B, Kremer K (2003) Phys Rev E 68:046702

\bibitem{junghans}
Junghans C, Praprotnik M, Kremer K (2008) Soft Matter 4:156

\bibitem{malevanets_kapral}
Malevanets A, Kapral R (1999) J Chem Phys 110:8605

\bibitem{ihle_mpcd1}
Ihle T, Kroll DM (2003) Phys Rev E 67:066705

\bibitem{ihle_mpcd2}
Ihle T, Kroll DM (2003) Phys Rev E 67:066706

\bibitem{kikuchi}
Kikuchi N, Pooley CM, Ryder JF, Yeomans JM (2003) J Chem Phys 119:6388

\bibitem{ihle_mpcd3}
Ihle T, T{\"u}zel E, Kroll DM (2004) Phys Rev E 70:035701(R)

\bibitem{Sha04}
Sharma N, Patankar NA (2004) J Comp Phys 201:466

\bibitem{carparrinello}
Car R, Parrinello M (1985) Phys Rev Lett 55:2471

\bibitem{hofler_schwarzer}
H{\"o}fler K, Schwarzer S (2000) Phys Rev E 61:7146

\bibitem{schwarzer}
Schwarzer S (1995) Phys Rev E 52:6461

\bibitem{kalthoff_schwarzer}
Kalthoff W, Schwarzer S, Herrmann HJ (1997) Phys Rev E 56:2234

\bibitem{wachmann_granular}
Wachmann B, Kalthoff W, Schwarzer S, Herrmann HJ (1998) Granular Matter 1:75

\bibitem{delgado1}
Delgado-Buscalioni R, Coveney PV (2003) Phys Rev E 67:046704

\bibitem{delgado2}
Delgado-Buscalioni R, Flekkoy EG, Coveney PV (2005) Europhys Lett 69:959

\bibitem{delgado3}
Delgado-Buscalioni R, Coveney PV, Riley GD, Ford RW (2005) Phil Trans: Math,
  Phys Eng Sci 363:1975

\bibitem{defabritiis}
Fabritiis GD, Serrano M, Delgado-Buscalioni R, Coveney PV (2007) Phys Rev E
  75:026307

\bibitem{giupponi}
Giupponi G, Fabritiis GD, Coveney PV (2007) J Chem Phys 126:154903

\bibitem{delgado4}
Delgado-Buscalioni R, Fabritiis GD (2007) Phys Rev E 76:036709

\bibitem{Fri86}
Frisch U, Hasslacher B, Pomeau Y (1986) Phys Rev Lett 56:1505

\bibitem{Fri87}
Frisch U, d'Humi{\`{e}}res D, Hasslacher B, Lallemand P, Pomeau Y, Rivet JP
  (1987) Complex Systems 1:649

\bibitem{sauro}
Succi S (2001) The Lattice Boltzmann Equation for Fluid Dynamics and Beyond.
\newblock Oxford University Press, Oxford

\bibitem{McN88}
McNamara GR, Zanetti G (1988) Phys Rev Lett 61:2332

\bibitem{Hig89}
Higuera F, Succi S, Benzi R (1989) Europhys Lett 9:345

\bibitem{Ben92}
Benzi R, Succi S, Vergassola M (1992) Phys Rep 222:145

\bibitem{alexander:93}
Alexander FJ, Chen S, Sterling JD (1993) Phys Rev E 47:R2249

\bibitem{ihle_thermal:00}
Ihle T, Kroll D (2000) Comp Phys Comm 129:1

\bibitem{lallemand_thermal:03}
Lallemand P, Luo LS (2003) Phys Rev E 68:036706

\bibitem{guo:07}
Guo Z, Zheng C, Shi B, Zhao TS (2007) Phys Rev E 75:036704

\bibitem{ansumali:05}
Ansumali S, Karlin IV (2005) Phys Rev Lett 95:260605

\bibitem{prasianakis:07}
Prasianakis NI, Karlin IV (2007) Phys Rev E 76:016702

\bibitem{Gun91}
Gunstensen AK, Rothman DH, Zaleski S, Zanetti G (1991) Phys Rev A 43:4320

\bibitem{shanchen:93}
Shan X, Chen H (1993) Phys Rev E 47:1815

\bibitem{julia}
Swift MR, Orlandini E, Osborn WR, Yeomans JM (1996) Phys Rev E 54:5041

\bibitem{julia_noise:99}
Gonnella G, Orlandini E, Yeomans JM (1999) Phys Rev E 59:R4741

\bibitem{lishinonideal}
Luo LS (2000) Phys Rev E 62:4982

\bibitem{lishimixture}
Luo LS, Girimaji SS (2003) Phys Rev E 67:036302

\bibitem{guo_zhao:05}
Guo Z, Zhao TS (2005) Phys Rev E 71:026701

\bibitem{arcidiacono:06}
Arcidiacono S, Mantzaras J, Ansumali S, Karlin IV, Frouzakis C, Boulouchos KB
  (2006) Phys Rev E 74:056707

\bibitem{halliday:07}
Halliday I, Hollis AP, Care CM (2007) Phys Rev E 76:026708

\bibitem{li_wagner:07}
Li Q, Wagner AJ (2007) Phys Rev E 76:036701

\bibitem{qian}
Qian YH, D'Humieres D, Lallemand P (1992) Europhys Lett 17:479

\bibitem{Lad94}
Ladd AJC (1994) J Fluid Mech 271:285

\bibitem{Lad94a}
Ladd AJC (1994) J Fluid Mech 271:311

\bibitem{Lad01}
Ladd AJC, Verberg R (2001) J Stat Phys 104:1191

\bibitem{Adh05}
Adhikari R, Stratford K, Cates ME, Wagner AJ (2005) Europhys Lett 3:473

\bibitem{DSL}
D{\"u}nweg B, Schiller UD, Ladd AJC (2007) Phys Rev E 76:036704

\bibitem{hinch}
Hinch EJ (1991) Perturbation Methods.
\newblock Cambridge University Press, Cambridge

\bibitem{Ngu02}
Nguyen NQ, Ladd AJC (2002) Phys Rev E 66:046708

\bibitem{cha60}
Chapman S, Cowling TG (1960) {The Mathematical Theory of Non-Uniform Gases}.
\newblock Cambridge University Press, Cambridge

\bibitem{McN92}
McNamara GR, Alder BJ (1992) In: M~Mareschal, BL~Holian (eds.) Microscopic
  Simulations of Complex Hydrodynamic Phenomena. Plenum, New York

\bibitem{Wag98}
Wagner AJ (1998) Europhys Lett 44:144

\bibitem{Kar99}
Karlin IV, Ferrante A, \"Ottinger HC (1999) Europhys Lett 47:182

\bibitem{Bog03}
Boghosian BM, Love PJ, Coveney PV, Karlin IV, Succi S, Yepez J (2003) Phys Rev
  E 68:025103(R)

\bibitem{Dhu02}
D'Humi{\`{e}}res D, Ginzburg I, Krafczyk M, Lallemand P, Luo LS (2002) Phil
  Trans Royal Soc London A 360:437

\bibitem{Lal00}
Lallemand P, Luo LS (2000) Phys Rev E 61:6546

\bibitem{Dhu92}
D'Humi{\`{e}}res D (1992) Progress in Astronautics and Aeronautics 159:450

\bibitem{Gin03}
Ginzburg I, d'Humi{\`{e}}res D (2003) Phys Rev E 68:066614

\bibitem{Chu07}
Chun B, Ladd AJC (2007) Phys Rev E 75:066705

\bibitem{McN93}
McNamara GR, Alder BJ (1993) Physica A 194:218

\bibitem{landbind}
Landau DP, Binder K (2000) A Guide to Monte Carlo Simulations in Statistical
  Physics.
\newblock Cambridge University Press, Cambridge

\bibitem{Gin94}
Ginzburg I, Adler PM (1994) J Phys II France 4:191

\bibitem{Guo02}
Guo Z, Zheng C, B~Shi B (2002) Phys Rev E 65:046308

\bibitem{Lad89c}
Ladd AJC, Frenkel D (1989) In: P~Manneville, N~Boccara, GY~Vichniac, R~Bidaux
  (eds.) Cellular Automata and Modeling of Complex Physical Systems.
  Springer-Verlag, Berlin-Heidelberg, no.~46 in Springer Proceedings in
  Physics, pp. 242--245

\bibitem{Lad90a}
Ladd AJC, Frenkel D (1990) Physics of Fluids A 2:1921

\bibitem{ahlrichs:98}
Ahlrichs P, D{\"u}nweg B (1998) Int J Mod Phys C 9:1429

\bibitem{Ahl99}
Ahlrichs P, D{\"{u}}nweg B (1999) J Chem Phys 111:8225

\bibitem{fyta}
Fyta MG, Melchionna S, Kaxiras E, Succi S (2006) Multiscale Model Simul 5:1156

\bibitem{vladimir}
Lobaskin V, D{\"u}nweg B (2004) New J Phys 6:54

\bibitem{vladimir2}
Lobaskin V, D{\"u}nweg B, Holm C (2004) J Physics: Cond Matt 16:S4063

\bibitem{vladimir3}
Lobaskin V, D\"{u}nweg B, Medebach M, Palberg T, Holm C (2007) Phys Rev Lett
  98:176105

\bibitem{chatterji2005cmd}
Chatterji A, Horbach J (2005) J Chem Phys 122:184903

\bibitem{chatterji2007eph}
Chatterji A, Horbach J (2007) J Chem Phys 126:064907

\bibitem{Pes02}
Peskin CS (2002) Acta Numerica 11:479

\bibitem{nash07}
Nash RW, Adhikari R, Cates ME (2008) Phys Rev E 77:026709

\bibitem{Fen03}
Feng ZG, Michaelides E (2003) J Comp Phys 195:602

\bibitem{Shi05}
Shi X, Phan-Thien N (2005) J Comp Phys 206:81

\bibitem{Aid98}
Aidun CK, Lu YN, Ding E (1998) J Fluid Mech 373:287

\bibitem{Low95}
Lowe CP, Frenkel D, Masters AJ (1995) J Chem Phys 103:1582

\bibitem{ricci_ciccotti}
Ricci A, Ciccotti G (2003) Mol Phys 101:1927

\bibitem{bussiparrinello}
Bussi G, Parrinello M (2007) Phys Rev E 75:056707

\bibitem{thalmann}
Thalmann F, Farago J (2007) J Chem Phys 127:124109

\bibitem{DeF06}
Fabritiis GD, Serrano M, Espa{\~{n}}ol P, Coveney P (2006) Physica A 361:429

\bibitem{Ser06}
Serrano M, Fabritiis GD, Espa{\~{n}}ol P, Coveney P (2006) Math Comput Simul
  72:190

\bibitem{Ust05}
Usta OB, Ladd AJC, Butler JE (2005) J Chem Phys 122:094902

\bibitem{Gha95}
Ghadder CK (1995) Phys Fluids 7:2563

\bibitem{Din03}
Ding EJ, Aidun CK (2003) J Stat Phys 112:685

\bibitem{Cla89}
Claeys IL, Brady JF (1989) PhysicoChem Hydrodyn II:261

\bibitem{Lad97b}
Ladd AJC (1997) Phys Fluids 9:491

\bibitem{Bra88}
Brady JF, Durlofsky LJ (1988) Phys Fluids 31:717

\bibitem{Bux05}
Buxton GA, Verberg R, Jasnow D, Balazs AC (2005) Phys Rev E 71:056707

\bibitem{Ale06}
Alexeev A, Verberg R, Balazs AC (2006) Soft Matter 2:499

\bibitem{Jun05b}
Junk M, Yang Z (2005) J Stat Phys 121:3

\bibitem{Che98a}
Chen HD (1998) Phys Rev E 58:3955

\bibitem{Che98b}
Chen HD, Teixeira C, Molvig K (1998) Int J Mod Phys C 9:1281

\bibitem{Bou01}
Bouzidi M, Firdaouss M, Lallemand P (2001) Phys Fluids 13:3452

\bibitem{Fil98}
Filippova O, H{\"{a}}nel D (1998) J Comput Phys 147:219

\bibitem{Mei99}
Mei RW, Luo LS, Shyy W (1999) J Comput Phys 155:307

\bibitem{Lal03}
Lallemand P, Luo LS (2003) J Comp Phys 184:406

\bibitem{Max01}
Maxey MR, Patel BK (2001) Int J Multiphase Flow 27:1603

\bibitem{Lom02}
Lomholt S, Stenum B, Maxey MR (2002) Int J Multiphase Flow 28:225

\bibitem{zakharov:97}
Zakharov VE, Kuznetsov EA (1997) Physics--Uspekhi 40:1087

\bibitem{chandra}
Chandrasekhar S (1943) Rev Mod Phys 15:1

\bibitem{bdlangevin}
D{\"u}nweg B (2003) In: B~D{\"u}nweg, DP~Landau, AI~Milchev (eds.) Computer
  Simulations of Surfaces and Interfaces, Kluwer, Dordrecht

\bibitem{foxuhlenbeck}
Fox RF, Uhlenbeck GE (1970) Phys Fluids 18:1893

\bibitem{functional_derivative}
See, e.g., http://en.wikipedia.org

\bibitem{vangunsteren_berendsen}
van Gunsteren W, Berendsen HJC (1988) Molec Simul 1:173

\bibitem{mclachlan}
McLachlan RI, Quispel GRW (2002) Acta Numer 11:341

\bibitem{forbert_chin}
Forbert HA, Chin SA (2000) Phys Rev E 63:016703

\bibitem{Has59}
Hasimoto H (1959) J Fluid Mech 5:317

\bibitem{Lad00}
Ladd AJC (2000) In: H~van Beijeren, J~Karkheck (eds.) Dynamics: Models and
  Kinetic Methods for Non-equilibrium Many Body Systems. Kluwer Academic
  Publishers, Dordrecht, pp. 17--30

\bibitem{beenakker:86}
Beenakker CWJ (1986) J Chem Phys 85:1581

\bibitem{Lad88}
Ladd AJC (1988) J Chem Phys 88:5051

\bibitem{Wei89}
Weitz DA, Pine DJ, Pusey PN, Tough RJA (1989) Phys Rev Lett 63:1747

\bibitem{Ern70}
Ernst MH, Hauge EH, van Leeuwen JMJ (1970) Phys Rev Lett 25:1254

\bibitem{Dor75}
Dorfman JR, Cohen EGD (1975) Phys Rev A 12:292

\bibitem{Hau73}
Hauge EH, Martin-L{\"{o}}f A (1973) J Stat Phys 7:259

\bibitem{Hoe91}
van~der Hoef MA, Frenkel D, Ladd AJC (1991) Phys Rev Lett 67:3459

\bibitem{Hoe91a}
van~der Hoef MA, Frenkel D (1991) Phys Rev Lett 66:1591

\bibitem{Lad93a}
Ladd AJC (1993) Phys Rev Lett 70:1339

\bibitem{Zhu92}
Zhu JX, Durian DJ, M{\"{u}}ller J, Weitz DA, Pine DJ (1992) Phys Rev Lett
  68:2559

\bibitem{Kao93}
Kao MH, Yodh AG, Pine DJ (1993) Phys Rev Lett 70:242

\bibitem{zimm}
Zimm BH (1956) J Chem Phys 24:269

\bibitem{rouse}
Rouse PE (1953) J Chem Phys 21:1272

\bibitem{pierleoni:92}
Pierleoni C, Ryckaert JP (1992) J Chem Phys 96:8539

\bibitem{smith:92}
Smith W, Rapaport DC (1992) Mol Sim 9:25

\bibitem{duenweg:93}
D{\"u}nweg B, Kremer K (1993) J Chem Phys 99:6983

\bibitem{schlijper:95}
Schlijper AG, Hoogerbrugge PJ, Manke CW (1995) J Rheol 39:567

\bibitem{spenley:00}
Spenley NA (2000) Europhys Lett 49:534

\bibitem{malevanets_yeomans:00}
Malevanets A, Yeomans JM (2000) Europhys Lett 52:231

\bibitem{mussawisade:05}
Mussawisade K, Ripoll M, Winkler RG, Gompper G (2005) J Chem Phys 123:144905

\bibitem{Lad92}
Ladd AJC, Frenkel D (1992) Macromolecules 25:3435

\bibitem{duenweg:02}
D{\"u}nweg B, Reith D, Steinhauser M, Kremer K (2002) J Chem Phys 117:914

\bibitem{degennes:76}
de~Gennes PG (1976) Macromolecules 9:594

\bibitem{richter:84}
Richter D, Binder K, Ewen B, St{\"u}hn B (1984) J Phys Chem 88:6618

\bibitem{Fan05}
Fang L, Hu H, Larson RG (2005) J Rheol 49:127

\bibitem{Jen04}
Jendrejack RM, Schwartz DC, de~Pablo JJ, Graham MD (2004) J Chem Phys 120:2513

\bibitem{Ma05}
Ma H, Graham M (2005) Phys Fluids 17:083103

\bibitem{Ust06}
Usta OB, Butler JE, Ladd AJC (2006) Phys Fluids 18:031703

\bibitem{Ust07}
Usta OB, Butler JE, Ladd AJC (2007) Phys Rev Lett 98:090831

\bibitem{Zhe02}
Zheng J, Yeung ES (2002) Anal Chem 74:4536

\bibitem{Zhe03}
Zheng J, Yeung ES (2003) Anal Chem 75:3675

\bibitem{Lon01}
Long D, Ajdari A (2001) Euro Phys J E 4:29

\bibitem{But07}
Butler JE, Usta OB, Kekre R, Ladd AJC (2007) Phys Fluids 19:113101

\bibitem{Caf85}
Caflisch RE, Luke JHC (1985) Phys Fluids 28:759

\bibitem{Hin88}
Hinch EJ (1988) In: E~Guyon, Y~Pomeau, JP~Nadal (eds.) Disorder and Mixing.
  Kluwer Academic, Dordrecht, pp. 153--161

\bibitem{Nic95}
Nicolai H, Guazzelli E (1995) Phys Fluids 7:3

\bibitem{Seg97}
Segr\'{e} PN, Herbolzheimer E, Chaikin PM (1997) Phys Rev Lett 79:2574

\bibitem{Koc91}
Koch DL, Shaqfeh ESG (1991) J Fluid Mech 224:275

\bibitem{Lad96}
Ladd AJC (1996) Phys Rev Lett 76:1392

\bibitem{Ton98}
Tong P, Ackerson BJ (1998) Phys Rev E 58:R6931

\bibitem{Lev98}
Levine A, Ramaswamy S, Frey E, Bruinsma R (1998) Phys Rev Lett 81:5944

\bibitem{Bre99}
Brenner MP (1999) Phys Fluids 11:754

\bibitem{Luk00}
Luke JHC (2000) Phys Fluids 12:1619

\bibitem{Muc03}
Mucha PJ, Brenner MP (2003) Phys of Fluids 15:1305

\bibitem{Lad02}
Ladd AJC (2002) Phys Rev Lett 88:048301

\bibitem{Gua01}
Guazzelli E (2001) Phys Fluids 13:1537

\bibitem{Saf65}
Saffman PG (1965) J Fluid Mech 22:385

\bibitem{Ho74}
Ho BP, Leal LG (1974) J Fluid Mech 65:365

\bibitem{Seg62}
Segr\'{e} G, Silberberg A (1962) J Fluid Mech 14:136

\bibitem{Sch89}
Schonberg JA, Hinch EJ (1989) J Fluid Mech 203:517

\bibitem{Asm99}
Asmolov ES (1999) J Fluid Mech 381:63

\bibitem{Mat04a}
Matas J, Glezer V, Guazzelli E, Morris J (2004) Phys Fluids 16:4192

\bibitem{Mat04}
Matas JP, Morris JF, Guazzelli E (2004) J Fluid Mech 515:171

\bibitem{Chu06}
Chun B, Ladd AJC (2006) Phys Fluids 18:031704

\bibitem{Fra03}
Frank M, Anderson D, Weeks ER, Morris JF (2003) J Fluid Mech 493:363

\end{thebibliography}
